\RequirePackage{lineno}
 \documentclass[prd,twocolumn,preprintnumbers,superscriptaddress,nofootinbib]{revtex4}
\usepackage [english]{babel}
\usepackage [autostyle, english = american]{csquotes}
\MakeOuterQuote{"}
\usepackage{tabularx} 
\usepackage[title]{appendix}
\usepackage{graphicx}  
\usepackage{dcolumn}   
\usepackage{bm}        
\usepackage{amssymb}   
\usepackage{multirow}
\usepackage{lineno}
\usepackage[hyperfootnotes=false]{hyperref}

\usepackage[title]{appendix}
\usepackage{graphicx}  
\usepackage{dcolumn}   
\usepackage{bm}        
\usepackage{amssymb}   
\usepackage{multirow}
\usepackage{lineno}
\usepackage{pdfpages}
\usepackage[hyperfootnotes=false]{hyperref}
\usepackage{capt-of}
\usepackage{float}
\usepackage{pdfpages}

\begin{document}
\newcommand{\beq}{\begin{equation}}
\newcommand{\eeq}{\end{equation}}
\newcommand{\de}{\delta}
\newcommand{\di}{\displaystyle}
\newcommand{\ep}{\epsilon}
\newcommand{\ga}{\gamma}
\newcommand{\Ga}{\Gamma}
\newcommand{\la}{\lambda}
\newcommand{\om}{\omega}
\newcommand{\si}{\sigma}
\newcommand{\ve}{\varepsilon}
\newcommand{\vp}{\varphi}
\newcommand{\dmbarbest} {{(3.36^{+0.46}_{-0.40} {(stat.)}\pm0.06{(syst.)})\times 10^{-3} eV^{2}}}
\newcommand{\dmbest}  {(2.32^{+0.12}_{-0.08})\times10^{-3} eV^{2} }
\newcommand{\pk}{{k \cdot p}}
\newcommand{\pkprime}{{k' \cdot p}}
\newcommand{\kq}{{k \cdot q}}
\newcommand{\pq}{{q\cdot p}}
\newcommand{\W}{{p'}}
\newcommand{\qW}{{q \cdot p'}}
\newcommand{\Wq}{{q \cdot p'}}
\newcommand{\pW}{{p\cdot p'}}
\newcommand{\GeV}{\; {\mathrm{GeV}}}
\newcommand{\cm}{\; {\mathrm{cm}}}
\newcommand{\D}{\displaystyle}
\newcommand{\nuwro}{\textsc{NuWro}}
\newcommand{\pythia}    {{\sc{pythia}}}
\newcommand{\achilles}  {{\sc{achilles}}}
\newcommand{\minerva}{MINER$\nu$A}
\newcommand{\ttbs}{\char'134}
\newcommand{\integ}{\iint}
\newcommand{\FA}{${\cal F}_A$}
\newcommand{\fa}{${\cal F}_A(q^2)$}
\newcommand{\gepQ}{$G_{Ep}(q^2)$}
\newcommand{\nubar}[0]{$\overline{\nu}$}
\newcommand{\gep}{G_{Ep}}
\newcommand{\gmp}{G_{Mp}}
\newcommand{\gmn}{G_{Mn}}
\newcommand{\gmpmu}{G_{Mp}/\mu_{p}}
\newcommand{\gepK}{G^{Kelly}_{Ep}}
\newcommand{\gmpK}{G^{Kelly-upd}_{Mp}}
\newcommand{\gen}{G_{En}}
\newcommand{\gmnmu}{G_{Mn}/\mu_{n}}
\newcommand{\gepnew}{G_{Ep}^{new}}
\newcommand{\gmpnewmu}{G_{Mp}^{new}/\mu_{p}}
\newcommand{\gmpKmu}{G^{Kelly-upd}_{Mp}/\mu_{p}}
\newcommand{\gennew}{G_{En}^{new}}
\newcommand{\gmnnewmu}{G_{Mn}^{new}/\mu_{n}}
\newcommand{\numu}{\nu_{\mu}}
\newcommand{\muminus}{\mu^{-}}
\newcommand{\muplus}{\mu^{+}}
\newcommand{\numubar}{\overline{\nu}_{\mu}}
%
%
\newcommand{\carbon}{\rm ^{12}C}
\newcommand{\oxygen}{\rm ^{16}O}
\newcommand{\deuteron}{\rm ^{2}H}
\newcommand{\hydrogen}{\rm ^{1}H}
\newcommand{\Hefour}{\rm ^{4}He}
\newcommand{\lead}{\rm^{208}Pb}
\newcommand{\Hethree}{\rm ^{3}He}
\newcommand{\neon}{\rm ^{20}Ne}
\newcommand{\aluminum}{\rm^ {27}Al}
\newcommand{\argon}{\rm ^{40}Ar}
\newcommand{\iron}{\rm ^{56}Fe}
\newcommand{\genie}{$\textsc{genie}$}
\newcommand{\qv}{$\bf |\vec q|$}
\newcommand{\rlqe}{${\cal R}_L^{QE}(\bf q,\nu)$ }
\newcommand{\rtqe}{${\cal R}_T^{QE}(\bf q,\nu)$ }
\newcommand{\rltot}{${\cal R}_L(\bf q, \nu)$ }
\newcommand{\rttot}{${\cal R}_T(\bf q, \nu)$ }

\newcommand{\Rochester}{Department of Physics and Astronomy, University of Rochester, Rochester, NY  14627, USA}
\newcommand{\JLAB}{Thomas Jefferson National Accelerator Facility, Newport News, VA 23606, USA}
\newcommand{\Poland}{Institute of Theoretical Physics, University of Wroc\l aw, plac Maxa Borna 9, 50-204, Wroc\l aw, Poland}
\newcommand{\Israel}{School of Physics and Astronomy, Tel Aviv University, Israel}
\newcommand{\Sevile}{Departamento de F\'isica At\'omica, Molecular y Nuclear, Universidad de Sevilla, 41080 Sevilla, Spain}
%
%
\title{Global Extraction of the  $\rm^{12}C$ Nuclear Electromagnetic Response Functions (${\cal R}_L$ and ${\cal R}_T$) and Comparisons to Nuclear Theory and Neutrino/Electron Monte Carlo Generators}
%
%

\author{Arie~Bodek}
\affiliation{\Rochester}
\email{bodek@pas.rochester.edu}
   \author{M.~E.~Christy}
\affiliation{\JLAB}
\email{christy@jlab.org}
\author{Zihao Lin}
\affiliation{\Rochester}
\email{zlin22@ur.rochester.edu}
\author{Giulia-Maria  Bulugean}
\affiliation{\Rochester}
\email{gbulugea@ur.rochester.edu}
\author{Amii Daniela Matamoros Delgado }
\affiliation{\Rochester}
\email{amatamor@u.rochester.edu}
\author{Artur M. Ankowski}
\affiliation{\Poland}
\author{G. D. Megias}
 \affiliation{Departamento de F\'isica At\'omica, Molecular y Nuclear, Universidad de Sevilla, 41080 Sevilla, Spain}
\author{Julia Tena Vidal} 
\affiliation{\Israel}
%
\date{\today}


\begin{abstract}
We have performed a global extraction of the ${\rm ^{12}C}$ longitudinal  (${\cal R}_L$) and transverse (${\cal R}_T$) nuclear electromagnetic response functions from  an analysis of all available electron scattering data on carbon. The response functions are extracted for energy transfer $\nu$,  spanning the nuclear excitation, quasielastic (QE),  resonance and inelastic continuum  over a large range of  the square of the four-momentum transfer, $Q^2.$  In addition, we perform a universal fit to all ${\rm ^{12}C}$ electron scattering data which also provides parmeterizations of ${\cal R}_L$ and ${\cal R}_T$ over a larger kinematic range. 
 Given the nuclear physics common to both electron and neutrino scattering from nuclei, extracted response functions from electron scattering spanning a large range of  $Q^2$ and $\nu$  also provide a powerful tool for validation and tuning of neutrino  Monte Carlo (MC) generators. In this paper  we focus on the nuclear excitation, single nucleon (QE-1p1h) and two nucleon (2p2h) final state regions and compare the measurements to theoretical predictions including   ``Energy Dependent-Relativistic Mean Field'' (ED-RMF),  ``Green's Function Monte Carlo'' (GFMC),   "Short Time Approximation Quantum Monte Carlo" (STA-QMC), an improved superscaling model  (SuSAv2), "Correlated Fermi Gas" (CFG), as well as the \nuwro{}, and  \achilles~ generators. Combining the ED-RMF-QE-1p1h  predictions with the SuSAv2-MEC-2p2h  predictions provides a good  description of ${\cal R}_L$ and ${\cal R}_T$ for  both single nucleon (from QE and nuclear excitations) and two nucleon final states over the  entire kinematic range.
\end{abstract}


\maketitle

%
\section{Introduction}
Electron scattering cross sections on nuclear targets are completely  described by longitudinal (${\cal R}_L$)  and transverse (${\cal R}_T$) nuclear electromagnetic response functions. Here  ${\cal R}_L$  and  ${\cal R}_T$  are functions of the energy transfer $\nu$ (or excitation energy $E_x$) and the  square of the 4-momentum transfer $Q^2$ (or  alternatively the  3-momentum transfer $\bf q$). Recent  theoretical ~\cite{Mihaila:1999nn,Lovato:2016gkq,Lovato:2023raf,Cloet:2015tha,Franco-Munoz:2022jcl,Franco-Munoz:2023zoa,Sobczyk:2020qtw,Carlson:2014vla,Carlson:2001mp} 
calculations of ${\cal R}_L({\bf q}, \nu)$  and  ${\cal R}_T({\bf q}, \nu)$  can be tested by comparing the predictions to experimental data.   We perform  extraction of both response functions as functions of (${\bf q}, \nu$), as well as  ($Q^2, \nu)$ from an analysis of all available inclusive cross section data for $\rm^{12}C$, and compare them to theoretical predictions.

 Given the nuclear physics common to both electron and neutrino scattering from nuclei, extracted response functions from electron scattering spanning a large range of  $Q^2$ and $\nu$  also provide a powerful tool for validation and tuning of neutrino  Monte Carlo (MC) generators\cite{generators:2022A, generators:2022}.
    Early MC generators include   {{\sc{neugen}}~\cite{neugen:2002} and {{\sc{nuance}}}~\cite{nuance:2002}. More  recent calculations and MC generators include {{\sc{neut}}}~\cite{neut:2002,Jake:2024,Dolan:2023iik}, {{\sc{gibuu}}} (Giessen Boltzmann-Uehling-Uhlenbeck) \cite{gibuu:2009,gibuu:2012},  \nuwro{}~\cite{NuWro}, \achilles~ (A CHIcago Land Lepton Event Simulator)~\cite{Isaacson:2023} and {{\sc{genie}}}~\cite{Andreopoulos:2009rq}.
  With  the advent of  DUNE\cite{DUNE:2020jqi}(Deep Underground Neutrino Experiment), the next generation neutrino oscillation experiments aim to search for  CP violations in neutrino oscillations.  Therefore, current neutrino MC generators need to be validated and tuned over the complete range of relevance in $Q^2$ and $\nu$ to provide a better description of  the cross sections for electron and neutrino interactions. 
  
  In this paper we  focus on the nuclear excitation, quasielastic (QE) and  two nucleon final state (2p2h) regions and  
 compare the extracted $\rm^{12}C$  ${\cal R}_L({\bf q}, \nu)$  and  ${\cal R}_T({\bf q}, \nu)$   to nuclear theory predictions
  of the  ``Green's Function Monte Carlo'' (GFMC)~\cite{Lovato:2016gkq,Lovato:2023raf}, the ``Energy Dependent-Relativistic Mean Field'' (ED-RMF)~\cite{Franco-Munoz:2022jcl,Franco-Munoz:2023zoa},  the  "Short Time Approximation Quantum Monte Carlo" (STA-QMC)~\cite{Andreoli:2024,Pastore:2019urn}, the "Correlated Fermi Gas" (CFG)~\cite{Bhattacharya:2024win}, the  \nuwro{} theoretical approach~\cite{NuWro}, the \achilles~\cite{Isaacson:2023} theoretical approach, and an improved superscaling model  (SuSAv2)\cite{Gonzalez-Jimenez:2014eqa, Megias:2016lke,Gonzalez-Rosa:2022ltp,Gonzalez-Rosa:2023aim}.  We have not done comparisons to the {{\sc{gibuu}}}  theoretical approach because nuclear response function predictions by  
  {{\sc{gibuu}}} are not  readily available.

The theoretical approaches of    \nuwro{} and  \achilles~  have been implemented in the corresponding  neutrino MC generators. Recently,  ED-RMF has been  implemented\cite{Jake:2024}  in an update of the  {{\sc{neut}}} neutrino MC generator.  The  improved SuSAv2 predictions for the  1p1h and 2p2h channels has been  implemented in  {{\sc{genie}}}. SuSAv2-inelastic is being implemented in {{\sc{genie}}}, and  STA-QMC for $\rm^4He$ has been implemented in   {{\sc{genie}}}.  In this paper we focus on testing  theoretical predictions.  The testing of the implementation of  the theoretical approaches into the  {{\sc{genie}}} neutrino MC generator (run in electron scattering mode) will be presented in a future communication.   Similar investigation of electron scattering data on $\rm ^{40}Ca$ and  $\rm ^{56}Fe$ (as well as hydrogen and deuterium data)    are currently under way.

\begin{figure*}[ht]
\begin{center}
\includegraphics[width=3.5in,height=4.5 in]{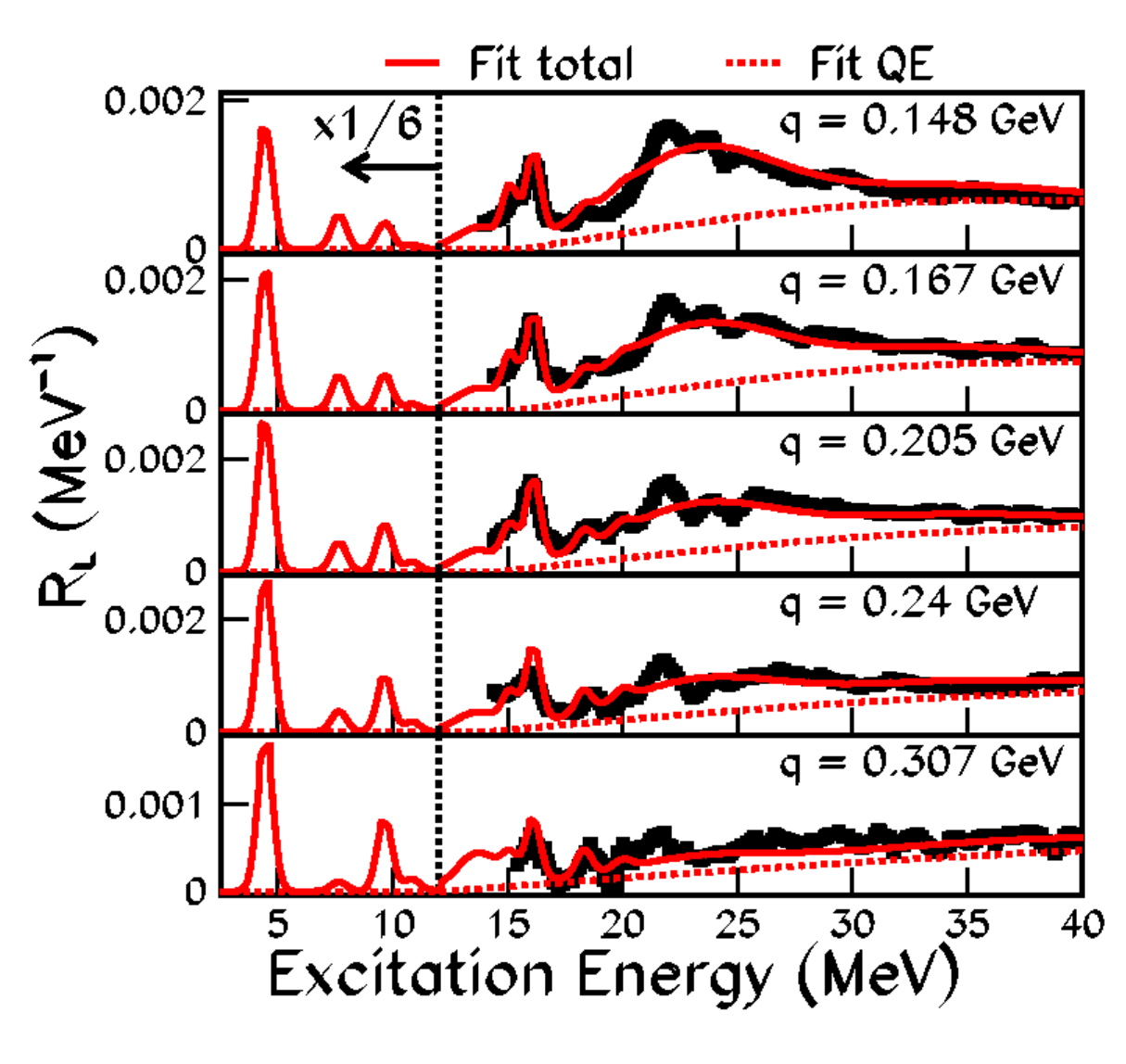}
\includegraphics[width=3.5in,height=4.5 in]{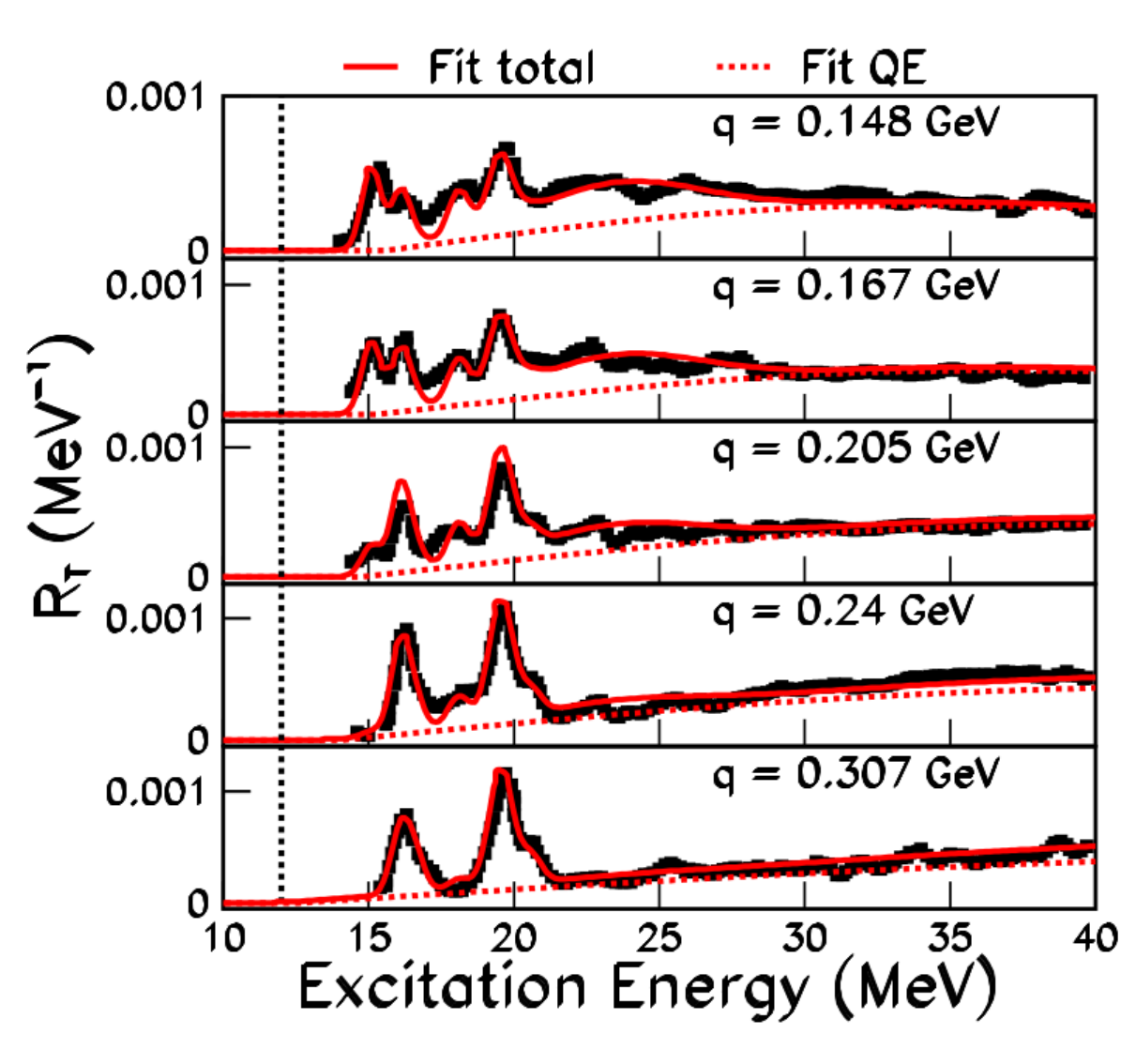}
\caption{{\bf Nuclear excitation region}: Comparison of the  ${\rm ^{12}C}$  longitudinal ${\cal R}_L/Z^2$ (left) and transverse ${\cal R}_T/Z^2$ (right) response functions extracted by Yamaguchi:71~\cite{Yamaguchi:1971ua} (black squares)  versus excitation energy $E_x$,  to the response functions extracted from the Christy-Bodek universal fit to all available electron scattering cross section data on ${\rm ^{12}C}$ (solid red line). The  contributions from nuclear excitations with $E_x<12$ MeV are multiplied by (1/6).  The QE contribution to the total response functions is represented by the red dashed line.  The response functions for all states in the  region of the Giant Dipole Resonance (20-30 MeV) region  are modeled as one average broad excitation. Note there are no transverse nuclear excitations with $E_x<10$ MeV. (Figure adapted from ~\cite{Bodek:2023dsr}). }
\label{Yamaguchi_RL_RT}
\end{center}
\end{figure*} 
%
\begin{figure*}
\begin{center}
\includegraphics[width=7.0in,height=8.0in]{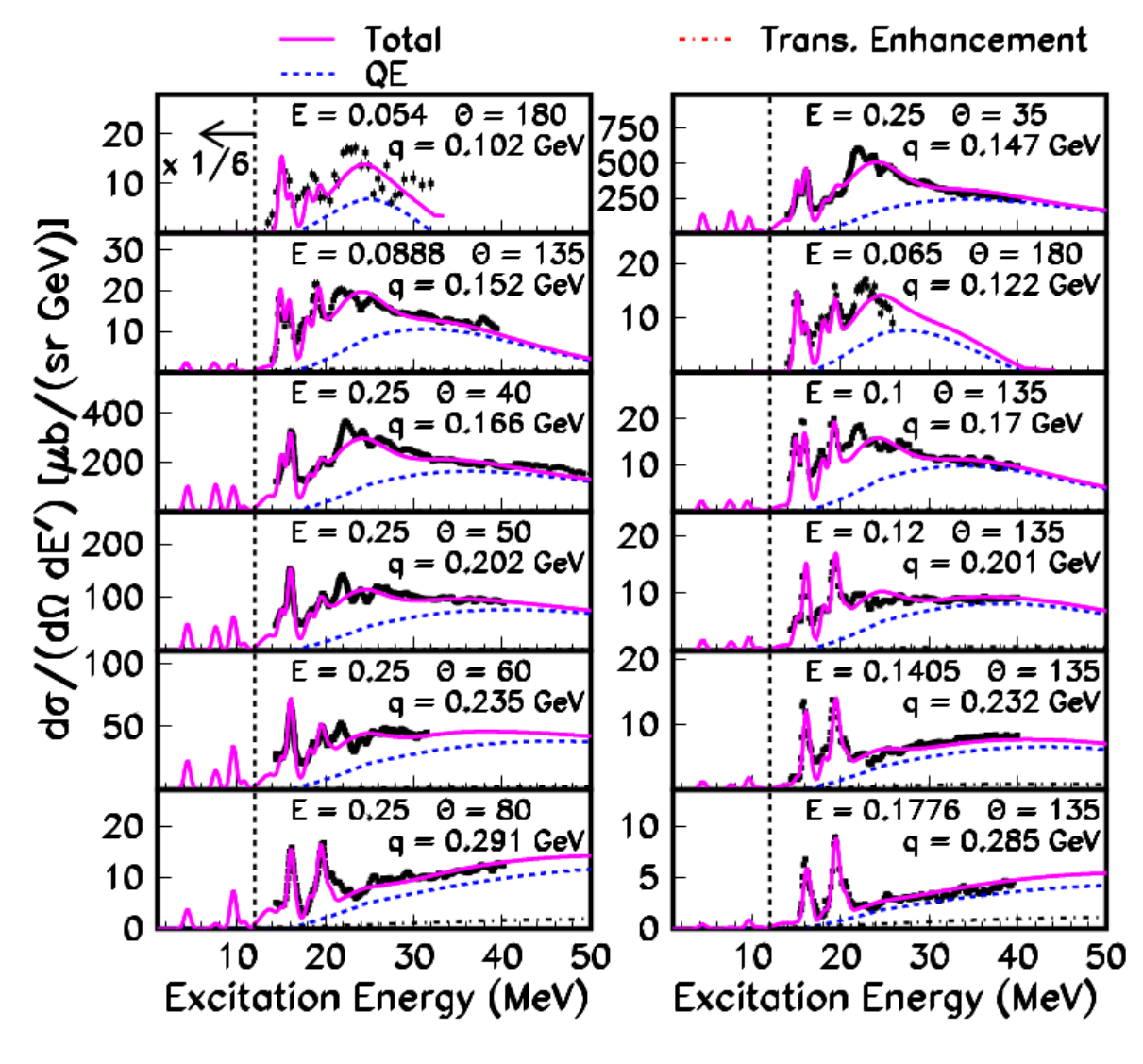}
\caption{{\bf Nuclear excitation region}:  Comparison of our fits to published radiatively corrected inelastic electron scattering cross sections on ${\rm ^{12}C}$ for excitation energies less than 50 MeV.  The  cross sections for excitation energies less than 12 MeV are multiplied by (1/6). The pink solid line is the predicted total cross section from the Christy-Bodek universal fit~\cite{Bodek:2022gli,Bodek:2023dsr} to all electron scattering data on ${\rm ^{12}C}$.  The fit includes  nuclear excitations, a superscaling QE model~\cite{Maieron:2001it,Amaro:2004bs,Amaro:2019zos,Megias:2014kia}   with Rosenfelder Pauli suppression~\cite{Rosenfelder:1980nd} (dashed blue line), "Transverse Enhancement/Meson Exchange Currents" (dot-dashed line) and pion production processes (at higher excitation energies). 
The data are from Yamaguchi:71~\cite{Yamaguchi:1971ua} except for  the cross sections  for $E_0=54$ MeV at 180$^{\circ}$ (from Goldemberg:64~\cite{PhysRev.134.B963}) and the cross sections for  $E_0=65$ MeV at  180$^{\circ}$  (from Deforest:65~\cite{DEFOREST1965311}). The measurements at 180$^{\circ}$ are only sensitive to the transverse form factors.  (Figure adapted from ~\cite{Bodek:2023dsr}).  }
\label{Yamaguchi_states}
\end{center}
\end{figure*}
%
\begin{figure*}
\begin{center}
\includegraphics[width=7.0in, height=4.5in] {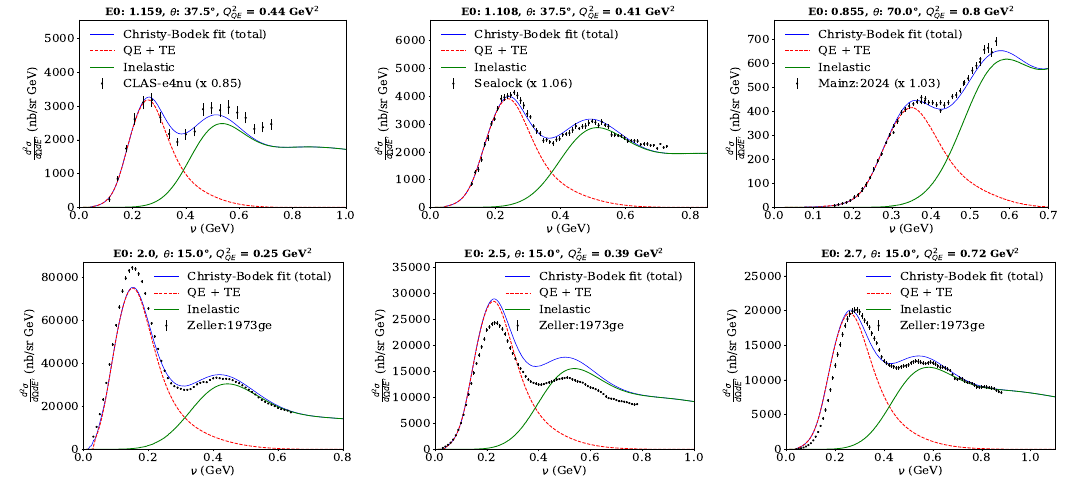}
\includegraphics[width=3.5in,height=2.5in]{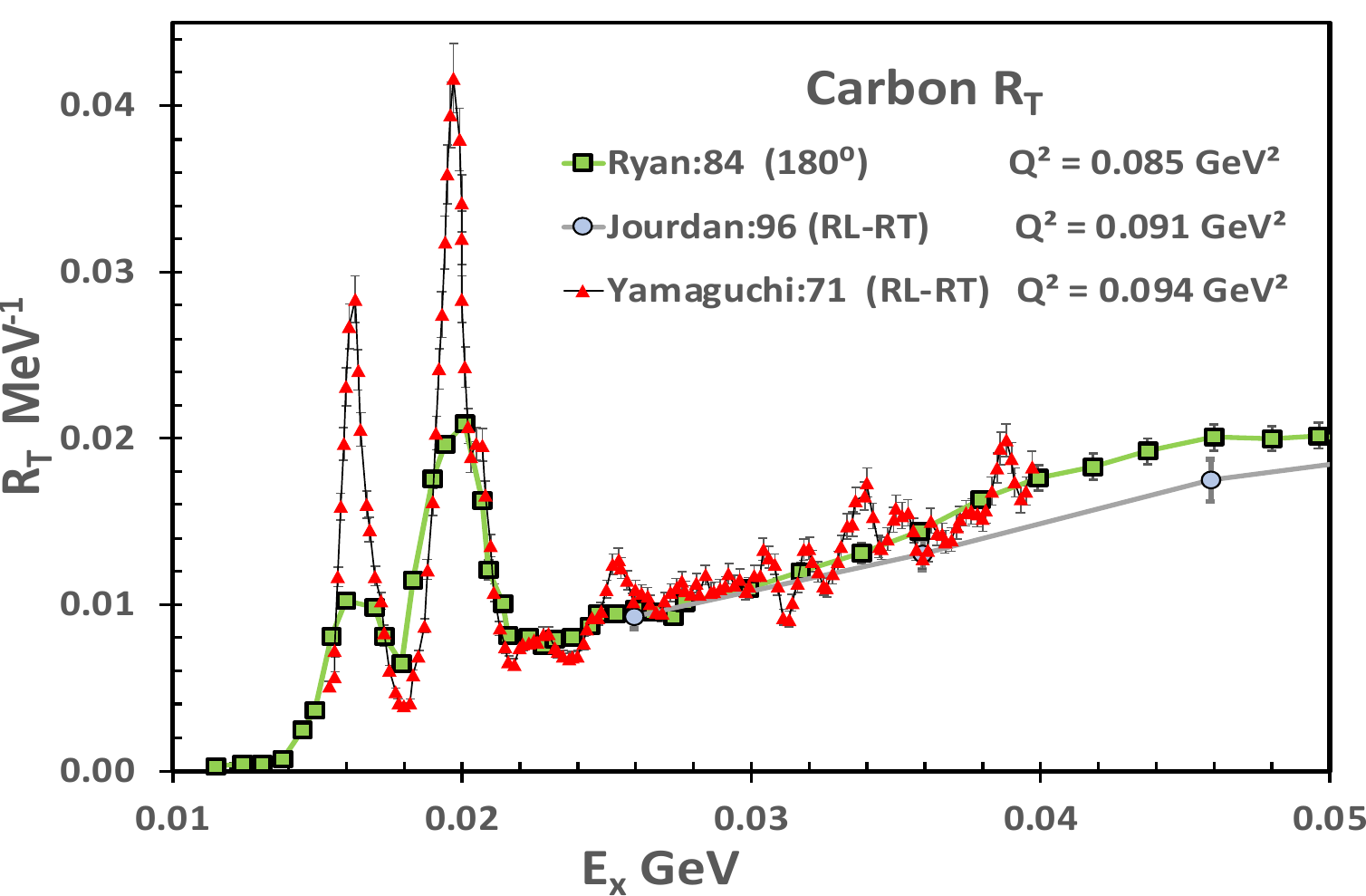}
\includegraphics[width=3.5in, height=2.5in] {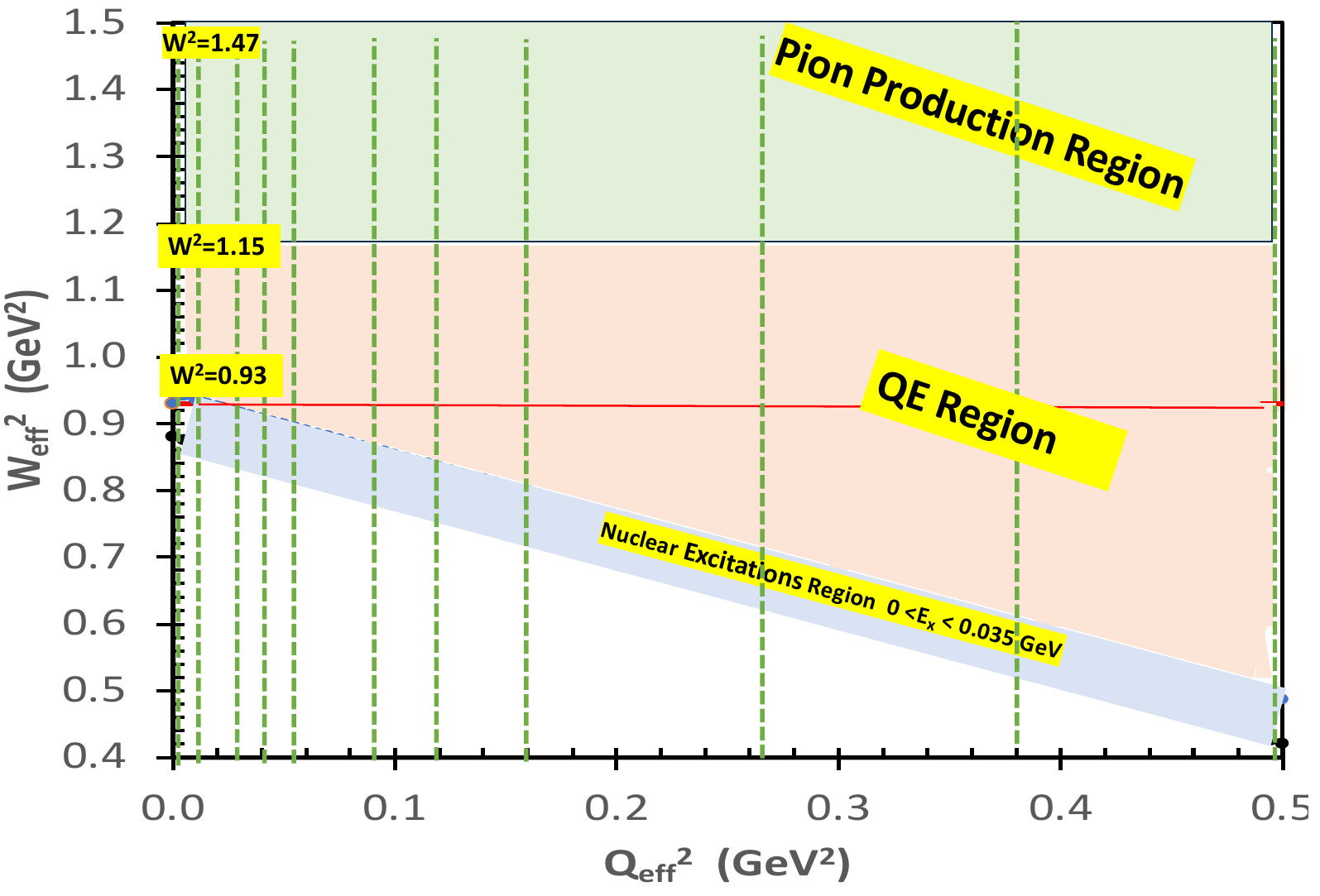}
\caption{ {\bf Top:} Comparison of the Christy-Bodek  fit to measurements. {\bf (a)} A recent (not included in the fit) CLAS-e4nu:2023 cross section\cite{Isaacson:2023} measurement  at E$_0$=1.159 GeV and $\theta$=37.45$^0$ (multiplied by 0.85). {\bf (b) }  A Sealock:1989 measurement \cite{Sealock:1989nx}  at 1.108 GeV and $\theta$=37.5$^0$ (multiplied by 1.06) {\bf  (c) } A  recent  (not included in the fit) Mainz:2024\cite{Mihovilovic:2024ymj} measurement at 0.855 GeV and $\theta$=70$^0$ (multiplied by 1.03). {\bf  (d-f)}  Zeller:1973 measurements\cite{Zeller:short} (not included in the fit) for 2.0, 2.5 and 2.7 GeV at $\theta$=37.5$^0$ (found to be  inconsistent with world data).
{\bf Bottom Left:} Comparison of the high resolution measurement of  ${\cal R}_T$ in Yamaguchi:71~\cite{Yamaguchi:1971ua} for $Q^2=0.085$ GeV$^2$  (versus  $E_x$) to  values of  ${\cal R}_T$ extracted from the lower resolution 180$^{\circ}$  electron scattering cross-section measurements  published in  Ryan:84~\cite{Ryan:1984yp} (multiplied by 1.05). {\bf Bottom Right: }  $W^2$  range for the fixed $Q^2$ values investigated in this analysis.  At low $Q^2$ (or $\bf q$) the contribution from nuclear excitation is significant. 
}
\label{Ryan_vs_Yama_plot}
\end{center}
\end{figure*} 

%

Previous extractions of ${\cal R}_L({\bf q}, \nu)$ and ${\cal R}_T({\bf q}, \nu)$ for $\rm^{12}C$ were performed with a limited set of cross section data.  Consequently the extractions are only available for a very limited range of $\bf q$ and  $\nu$.  The first  extraction  of  ${\cal R}_T({\bf q}, \nu)$ and ${\cal R}_L({\bf q}, \nu)$ for $\rm^{12}C$ was performed in 1983 and  used experimental cross sections data from only one experiment at Saclay (Barreau:83\cite{Barreau:1981rr,Barreau:1983ht,Barreau:1983A}). 
The extractions were performed for $\bf q$ values of 0.30, 0.40 and 0.55 GeV and values of $\nu$ in the quasielastic (QE) region. 
 A later extraction using the same Saclay data in combination with cross sections measured at the Stanford Linear Accelerator Center (SLAC) was published in 1996 (Jourdan:96\cite{Jourdan:1996np, Jourdan:1995np}) also  for three similar values of $\bf q$ (0.30, 0.38 and 0.57 GeV) and values of $\nu$ in the QE region. The most recent extraction was published in 2021 (Buki:21\cite{Buki:2021mjo}) used a limited set of cross section data  measured at Kharkov.  That extraction was only performed for ${\bf q}=0.3$ GeV in the QE region and has larger uncertainties than the two earlier analyses.  An extraction of ${\cal R}_L({\bf q}, \nu)$ and ${\cal R}_T({\bf q}, \nu)$ for $\rm^{12}C$  at  $Q^2$=0.1 GeV$^2$ in the $\Delta$(1232) nucleon resonance region was published in 1998 by Baran and collaborators~\cite{Baran:1988tw}.
 

 
  In this communication we  report on the  extraction of  ${\cal R}_L$ and  ${\cal R}_T$ for  ${\rm ^{12}C}$ at 18 distinct values of both {\bf q} and $Q^2$ by including all available electron scattering cross section measurements on carbon.  We note that a significant number of cross section measurements were not available in tabular form and required digitization of the cross sections from figures in the publications.  As summarized  in Table \ref{tab:Q2bins}, the response functions are extracted  for a large range of energy transfer $\nu$ at 18 fixed values of $\bf q$ in the range $0.1<{\bf q}<3.75$ ~GeV and at 18 fixed values of $Q^2$  in the range $0<Q^2<3.5$ GeV$^2$.  The range of  $\nu$ spans  the nuclear excitation, QE, resonance and inelastic continuum.    The complete data set consists of about 10,000 ${\rm ^{12}C}$ differential cross section measurements points as well as  photo-absorption data ($Q^2=0$).
We also include  high precision cross section measurements from   Jefferson Lab Hall C experiment E-04-001~\cite{JUPITER:2025uny,Alsalmi:2019sie}.  

Relative normalization factors for each experiment are determined from a global fit to the cross section data.  This is done by including the normalizations as free parameters in the fit (including the normalization uncertainty quoted for each experiment). 
Leveraging both the expanded data set and the global fit for centering the data to fixed {\bf q} ($Q^2$) allowed for more precise extractions of ${\cal R}_L$ and ${\cal R}_T$ for ${\rm ^{12}C}$. In the analysis, the global fit used all the cross section points measurements, and the  individual ${\cal R}_L$ and ${\cal R}_T$ extractions used 8,500 cross section points measurements.  This is because individual ${\cal R}_L$ and ${\cal R}_T$ extractions require data at both small angles and large angles for the same $\bf q$ and $\nu$ bin.  


 In the  nuclear excitation region (excitation energy less than 50 MeV) we extract  \rltot and \rttot  from fits to ${\rm ^{12}C}$ nuclear excitation form factors by Bodek and Christy~\cite{Bodek:2023dsr}.  In addition, for $16< E_x <40$ MeV, values of  ${\cal R}_L({\bf q},E_x)$ 
and ${\cal R}_T ({\bf q},E_x)$  are available from the analysis of  Yamaguchi:71~\cite{Yamaguchi:1971ua} for  five values of  $\bf q$ (0.148, 0.167, 0.205, 0.240 and 0.307 GeV).  Comparisons of the fit to the Yamaguchi response functions and cross sections are shown in 
Figures~\ref{Yamaguchi_RL_RT} and~\ref{Yamaguchi_states}, respectively.
 
 \subsection{Overview of this analysis}
The Christy-Bodek  2024 universal fit to all available cross section measurements for ${\rm ^{12}C}$ (and Hydrogen and Deuterium)   is  described in two recent  publications~\cite{Bodek:2022gli,Bodek:2023dsr} (similar fits were also done for  other nuclei). The fit includes nuclear elastic form factors, nuclear-excitations, quasielastic scattering, resonance  production and  inelastic continuum.   In the quasielastic region the fit is based on the $\psi^\prime$ SuSA (Super Scaling Approximation) formalism~\cite{Maieron:2001it,Amaro:2004bs,Amaro:2019zos,Megias:2014kia}  with Rosenfelder~\cite{Rosenfelder:1980nd} Pauli suppression. For the QE region the fit also includes  parametrizations of a  $multiplicative$  ``Longitudinal Quenching Factor'' at low $\bf q$ and an $additive$ ``Transverse Enhancement'' contribution, which is significant at intermediate $\bf q$. In the inelastic resonance and continuum region the fit is based on a Gaussian Fermi motion smeared free proton and neutron cross sections with a multiplicative medium modification factor.  The relative normalizations of different  ${\rm ^{12}C}$ cross section data sets are also extracted from the fit.   In summary the fit includes:
 \begin{enumerate} 
\item
   All available electron scattering data on $\hydrogen$, $\deuteron$, $\carbon$ 
   (for $\carbon$ we digitized additional data
   to supplement the data  included in  the QE~\cite{QEarchive}  and resonance~\cite{ResArchive} archives as
   summarized in Table  \ref{datasets}).
\item 
 Coulomb corrections~\cite{Gueye:1999mm} using the Effective Momentum Approximation (EMA) in modeling scattering from nuclear targets. We use
 {\bf $V_{eff}$= 3.10$\pm$ 0.25 MeV} for $\rm^{12}C$
\item
Updated $\rm^{12}C$ nuclear elastic form factor\cite{Bodek:2023dsr}.
\item  Parameterizations of $\rm^{12}C$ nuclear excitations  form factors\cite{Bodek:2023dsr}.
\item  
 Superscaling  function $FN(\psi^\prime)$ parameters are re-extracted including the Fermi broadening parameter $K_F$.
\item 
 Parameterizations of the free nucleon form factors~\cite{Bosted:1994tm} are re-derived from all  $\hydrogen$  and $\deuteron$ data.
\item 
  Rosenfelder Pauli suppression~\cite{Megias:2014kia, Rosenfelder:1980nd} which  reduces and modifies the QE distribution at low $\bf{q}$ and $\nu$. 
\item 
Updates of fits~\cite{Bosted:2012qc} to inelastic electron scattering data (in the nucleon resonance region and inelastic continuum) for $\hydrogen$ and $\deuteron$, providing the structure functions for the proton and neutron.
\item  
 A $\bf q$ dependent  $E_{\mathrm{shift}}^{QE}(\bf q)$ parameter  for the QE process to account for the optical potential~\cite{Bodek:2018lmc} of final state nucleons.   The separation energies of a proton and a neutron  from $\carbon$  are 16 MeV and  18.7  MeV, respectively.
Therefore, we  ensure that the QE cross section  on protons is zero for excitation energy below 16 MeV,  and the QE cross section  on neutrons is zero  for excitation energy less than 18.7 MeV.
 %
\item 
  Photo-production data in the  nuclear excitation region,  nucleon resonance and inelastic  continuum~\cite{Bosted:2007xd}.
\item  
 Gaussian Fermi  motion smeared nucleon resonance and inelastic continuum~\cite{Christy:2007ve,Bosted:2007xd}. The $K_F$ parameters for pion production and QE can be different.
\item 
 Parametrizations of the medium modifications of both the inelastic  \rltot and \rttot~structure functions responsible for the EMC effect (nuclear dependence of inelastic structure functions).  These are applied as multiplicative factors to the free nucleon cross sections prior to application of the Fermi smearing.    
\item For QE scattering we include parameterizations of   additive Transverse Enhancement  TE(${\bf q},\nu$)  and multiplicative  Longitudinal Quenching factor   $F_{quench}^L({\bf q})$.
\item The TE(${\bf q},\nu$)  is composed of three independent contributions. (a) An enhancement in the region of the QE peak accounting for
enhancement of \rtqe from the interference of 1-body and 2-body currents. (b)  An enhancement in the dip region between the QE peak and the $\Delta$(1232) resonance accounting for the contribution of 2-body currents leading to two nucleons in the final state.
(c) We add the contribution of scattering from Quasi-deuterons (scattering from short-range correlation neutron proton pairs), by using a fit  of  the contribution of Quasi-deuterons to the  photoproduction ($Q^2=0$)
cross section ($\sigma_{Quasi-D}$) as a function of $\nu$.  The fit, which is obtained from  \cite{Plujko:2018uum},  is given in Appendix B and shown in Fig.\ref{photo_cross_figure}).  For higher $Q^2$ values it is multiplied  by a suppression  factor given by
 [1/(1+$Q^2$/0.5)]$^5$, where $Q^2$ is in GeV$^2$.  In addition, we limit the this contribution to the region below the nucleon resonance
 region by multiplying by $1/[e^{(\nu-a)/\tau)}+1)]$ where a = 0.12 GeV and $\tau$=0.005 GeV.
\item  There is an  apparent shift of the
peak of the  $\Delta$(1232) to smaller values of $\nu$ which in the 2024 version of the fit is  accounted for by an "effective optical  potential"\cite{Bodek:2020wbk}. 

\item  We  include 
QE data at {\it all values} of $Q^2$ down to  $Q^2=0.01$ GeV$^2$ ($\bf q$=0.1 GeV) (which were not included in the Bosted-Mamyan fit~\cite{Bosted:2012qc}.)
\item The relative normalizations between different experiments are extracted from the fit. 
\item Data sets which are inconsistent~\cite{Zeller:1973ge,Heimlich:1974rk} with the world's data are identified and not included in the analysis (see 
Table \ref{datasets}).
 \end{enumerate}
 %

 The primary purpose of the fit is  to model cross sections used in calculation of radiative corrections in electron scattering experiments. The  fit  describes all electron scattering data on $\carbon$ and the fit's \rltot and \rttot are valid for a larger kinematic range than the individual \rltot and \rttot measurements. Therefore,  the fit can also be  used to validate  nuclear models  and tune  MC generators for electron and neutrino scattering experiments over a large kinematic range.   
 
The use of the   Christy-Bodek 2024 universal fit~\cite{Bodek:2022gli,Bodek:2023dsr}  {\it  in  identifying   the few data sets that are inconsistent with all  the other measurements} is illustrated in the top six sub-figures of Fig. \ref{Ryan_vs_Yama_plot}.
Here, the comparison of the fit to recent  CLAS-e4nu:2023 cross section\cite{Isaacson:2023} measurement (which was not included in the universal  fit) at E$_0$=1.159 GeV and $\theta$=37.5$^0$ shown  in sub-figure (a) indicates that normalization of this data relative to all  previous data is 0.85.   A comparison to the  Sealock:1989 measurement \cite{Sealock:1989nx}  at 1.108 GeV and $\theta$=37.5$^0$  (which was included in the universal  fit)  shown in sub-figure (b)  indicates a normalization of 1.06.  A comparison to  a recent Mainz:2024\cite{Mihovilovic:2024ymj}  measurement at  0.855 GeV and $\theta$=70$^0$  (which was not included in the universal  fit)   shown in sub-figure (c)  indicates a normalization of 1.03.   Comparisons to   Zeller:1973 measurements\cite{Zeller:short,Heimlich:1974rk}
for incident energies of  2.0, 2.5 and 2.7 GeV at $\theta$=37.5$^0$ (without any normalizations)   are  shown in sub-figures} (d-f).
 The Zeller:1973 measurements are excluded from the universal fit because of inconsistent normalizations between the  three energies,  and an unexplained  shift in $\nu$ in the 2.7 GeV  data.

\begin{figure}
\includegraphics[width=3.55in, height=1.5in] {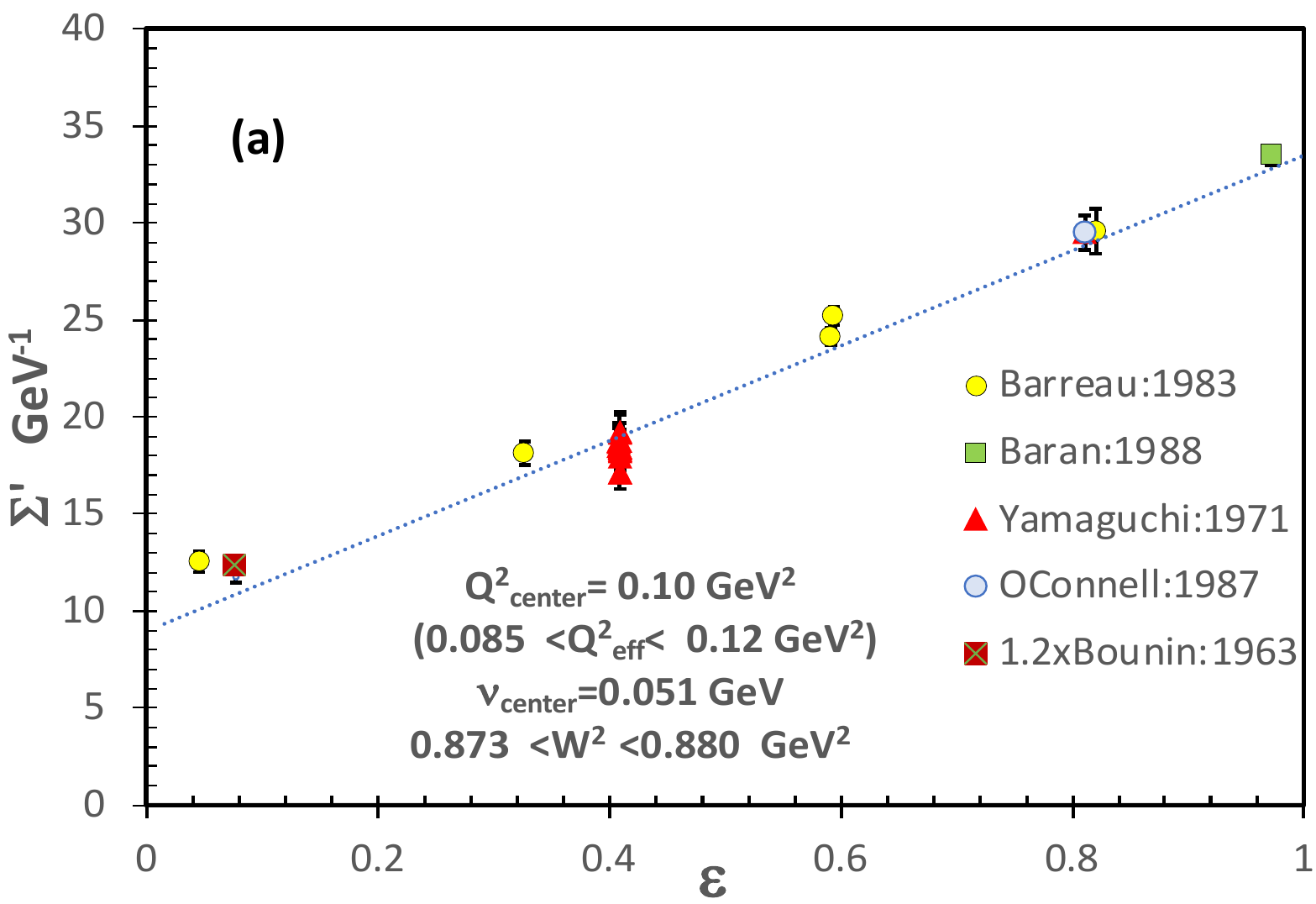}
\includegraphics[width=3.55in, height=1.5in] {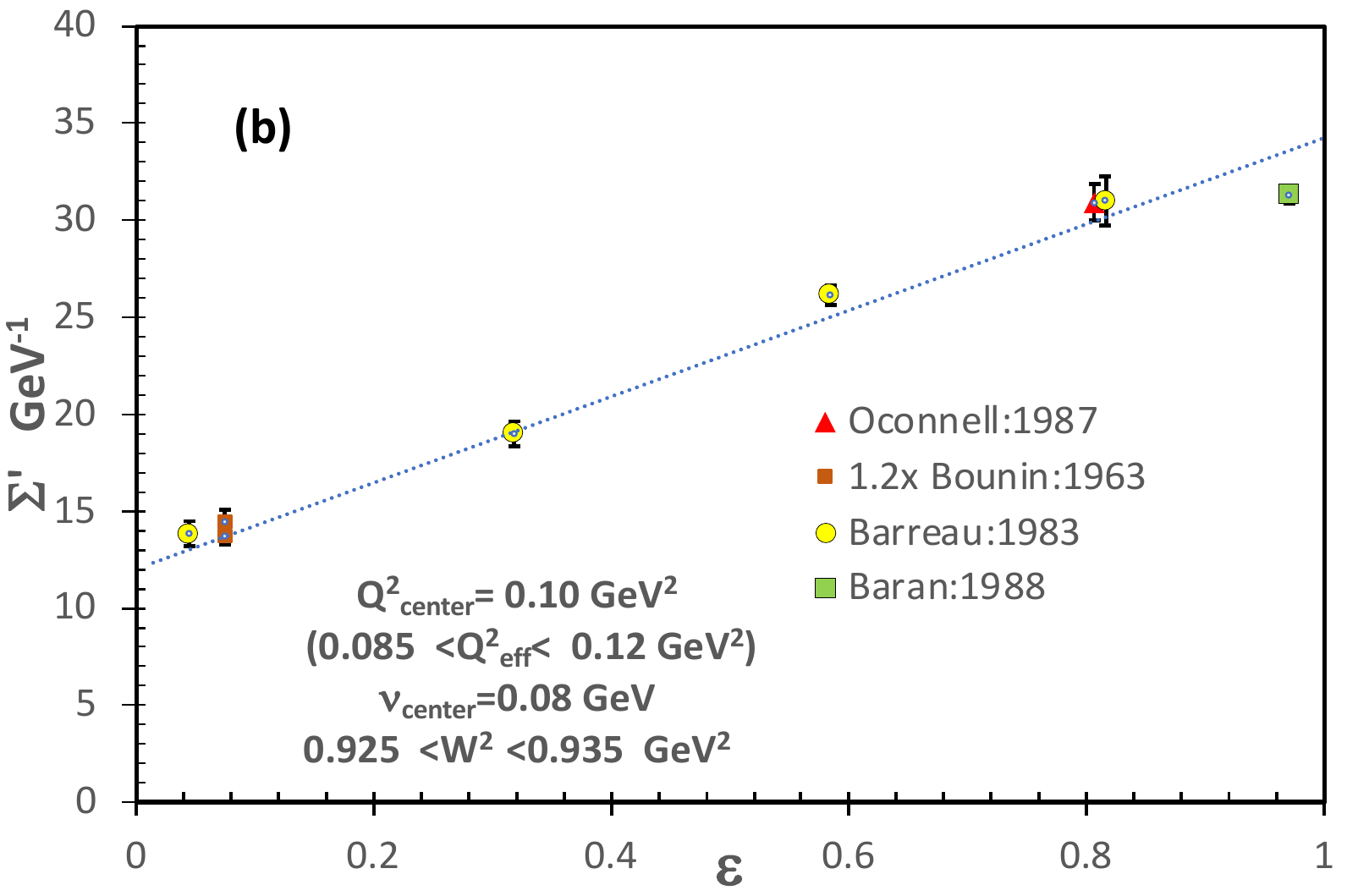}
\includegraphics[width=3.55in, height=1.5in] {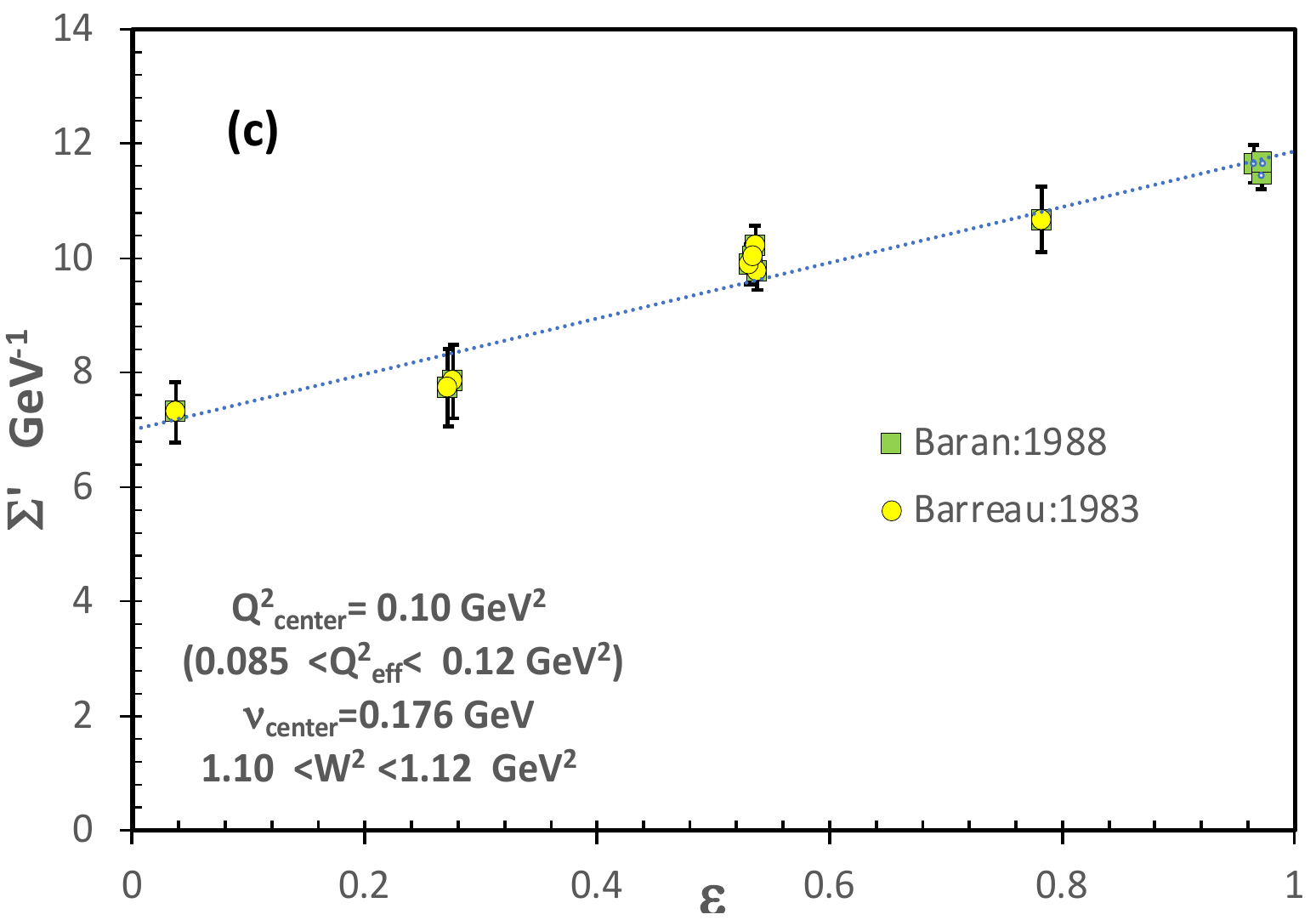}
\includegraphics[width=3.55in, height=1.5in] {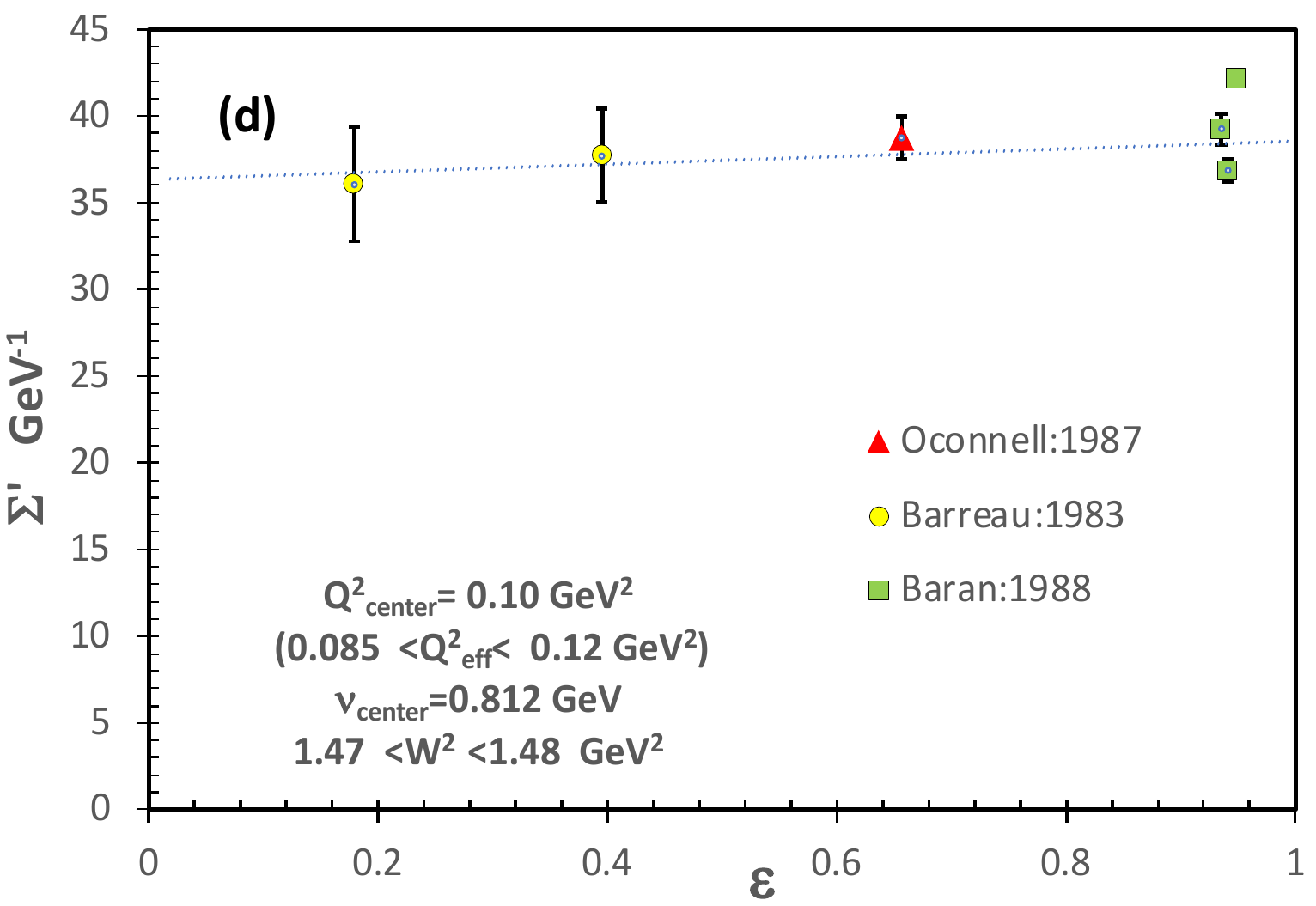}
\caption{ Sample ${\cal R}_L$  ${\cal R}_T$  Rosenbluth plots (before bin centering corrections).
 }
\label{Rosen_figure}
\end{figure}

 In this analysis the  fit is  primarily used to calculate "bin-centering"  corrections as described below.
Rosenbluth \cite{Rosenbluth:1950} extractions of ${\cal R}_T(Q^2, \nu)$ and  ${\cal R}_L(Q^2, \nu)$ require cross section measurements at different angles for the same values of ${\bf q}$ and $\nu$.  We bin the cross sections in fixed bins of ${\bf q}$ and use the fit for determining "bin centering" corrections  to account for the small differences in ${\bf q}$  and $\nu$  of the binned  cross sections measurements from different experiments.  In addition to  bins in  fixed values of ${\bf q}$ we also perform the same analysis in bins of fixed values of   $Q^2$.  Sample Rosenbluth plots before the application of  "bin centering" corrections are shown in  Fig. \ref{Rosen_figure}.

 In order to minimize the "bin centering" corrections it is critical to bin in a kinematic variable for which features (i.e. peaks) in the cross section remain fixed independent of the angle.  For this reason we perform the analysis in bins of both the excitation energy and bins of the square of the  final state mass ($W^2=M^2+2M\nu-Q^2$). We  use the results of the analysis in bins of excitation energy for $E_x$ less than 50~MeV (a region dominated by nuclear excitations) and the results of the analysis in bins of $W^2$ for $E_x$  greater than 50  MeV (a region dominated by QE scattering and pion production). Afterwards, we  convert the $E_x$ and $W^2$ values in the center of each bin to the corresponding values of $\nu$ in the center of the bin.

The  longitudinal and transverse response functions  (defined in section \ref{RLRTdescription}) in  the nuclear excitation region  are well  described by the universal  fit~\cite{Bodek:2023dsr}  to measured  transverse  and longitudinal form factors for each nuclear excitation. The ${\rm ^{12}C}$ ${\cal R}_L/Z^2$ and ${\cal R}_T/Z^2$ measurements published in  Yamaguchi:71~\cite{Yamaguchi:1971ua} in the nuclear  excitation region above the proton separation energy (16$<E_x<40$ MeV) are compared to the  ${\cal R}_L/Z^2$ and ${\cal R}_T/Z^2$ predictions from the  fit  in Figure \ref{Yamaguchi_RL_RT}. 

Rosenbluth extractions of  ${\cal R}_L$ and ${\cal R}_T$  using  data spanning a range of angles from different experiments are valid in the QE and pion production regions. However, in the nuclear excitation region it is not valid to combine experiments with different experimental resolutions, as this will lead to structural artifacts in the extracted response functions.   For example, in the bottom left panel of  Fig. \ref{Ryan_vs_Yama_plot} we show a comparison of high resolution extraction of  ${\cal R}_T$ in Yamaguchi:71~\cite{Yamaguchi:1971ua} for $Q^2$=0.085 GeV$^2$ to ${\cal R}_T$ extracted from the lower resolution 180$^{\circ}$  cross section data published in Ryan:84~\cite{Ryan:1984yp}.

In contrast, at  higher values of excitation energy (for $E_x>30$  MeV) the cross sections are relatively smooth on the scale of the experimental resolutions and a Rosenbluth analysis using  cross sections from different experiments can be performed.

\begin{figure*}
\includegraphics[width=3.5 in, height=2.6in] {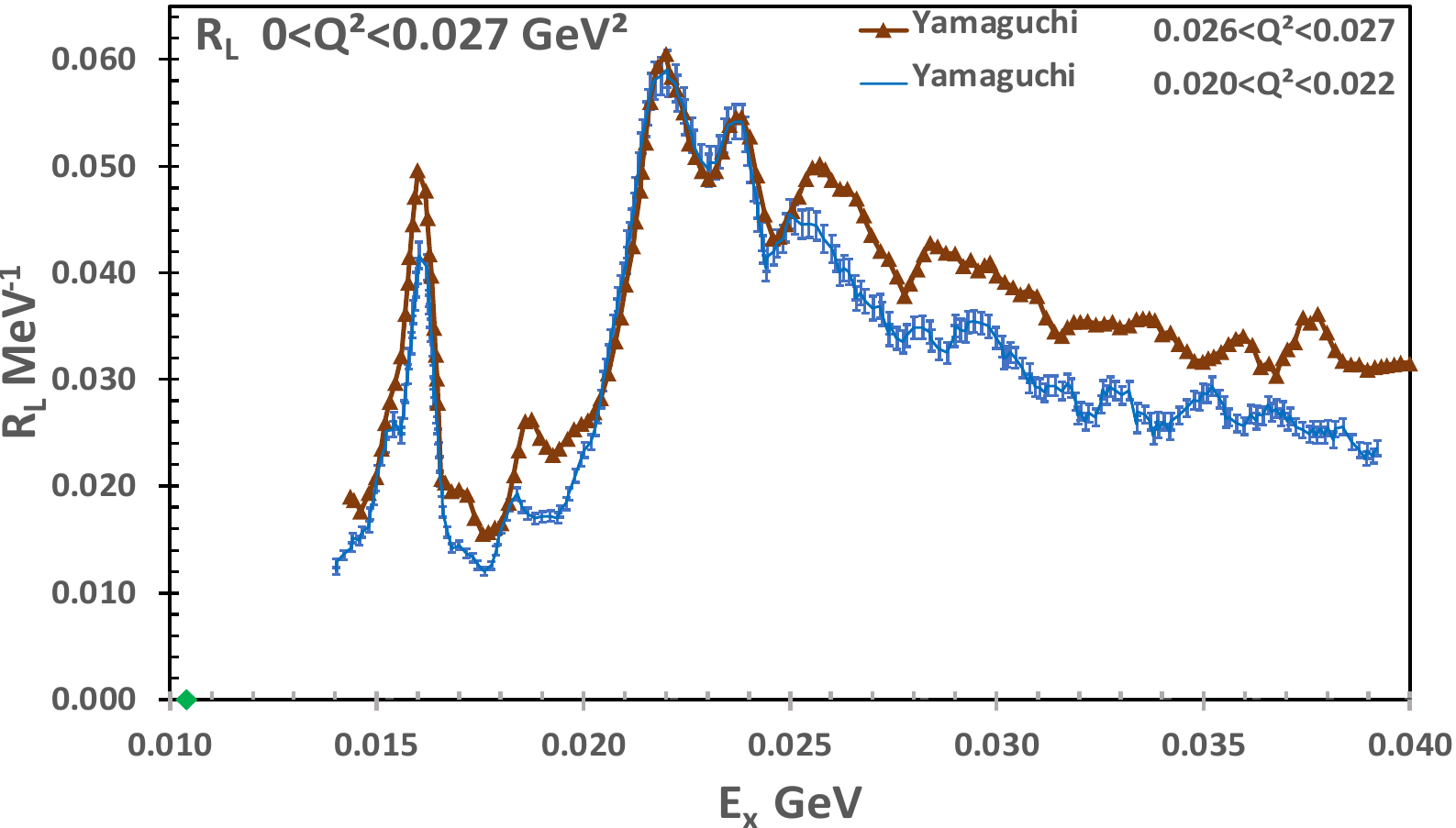}
\includegraphics[width=3.5 in, height=2.6in]  {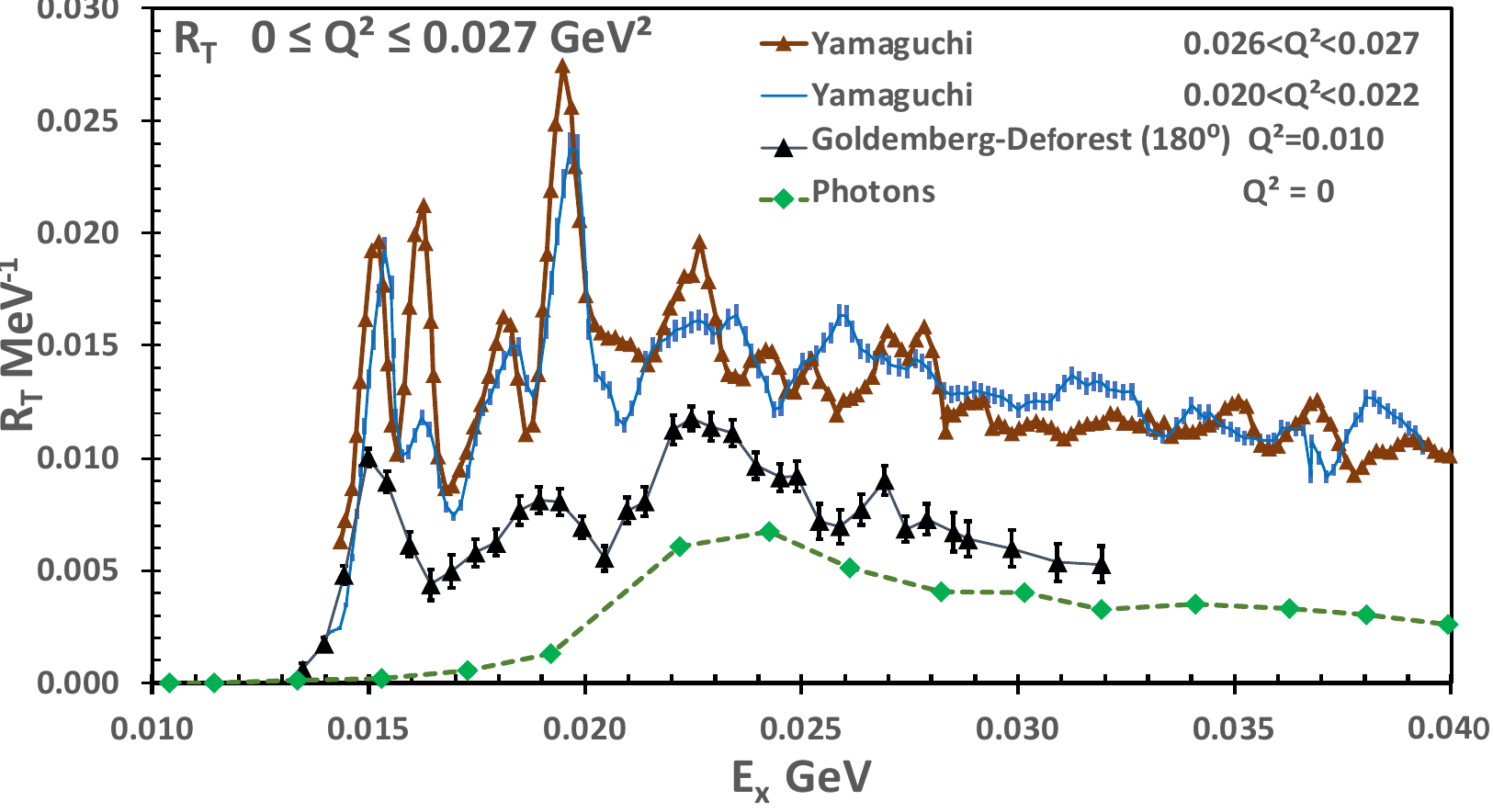}
\includegraphics[width=3.5 in, height=2.6in] {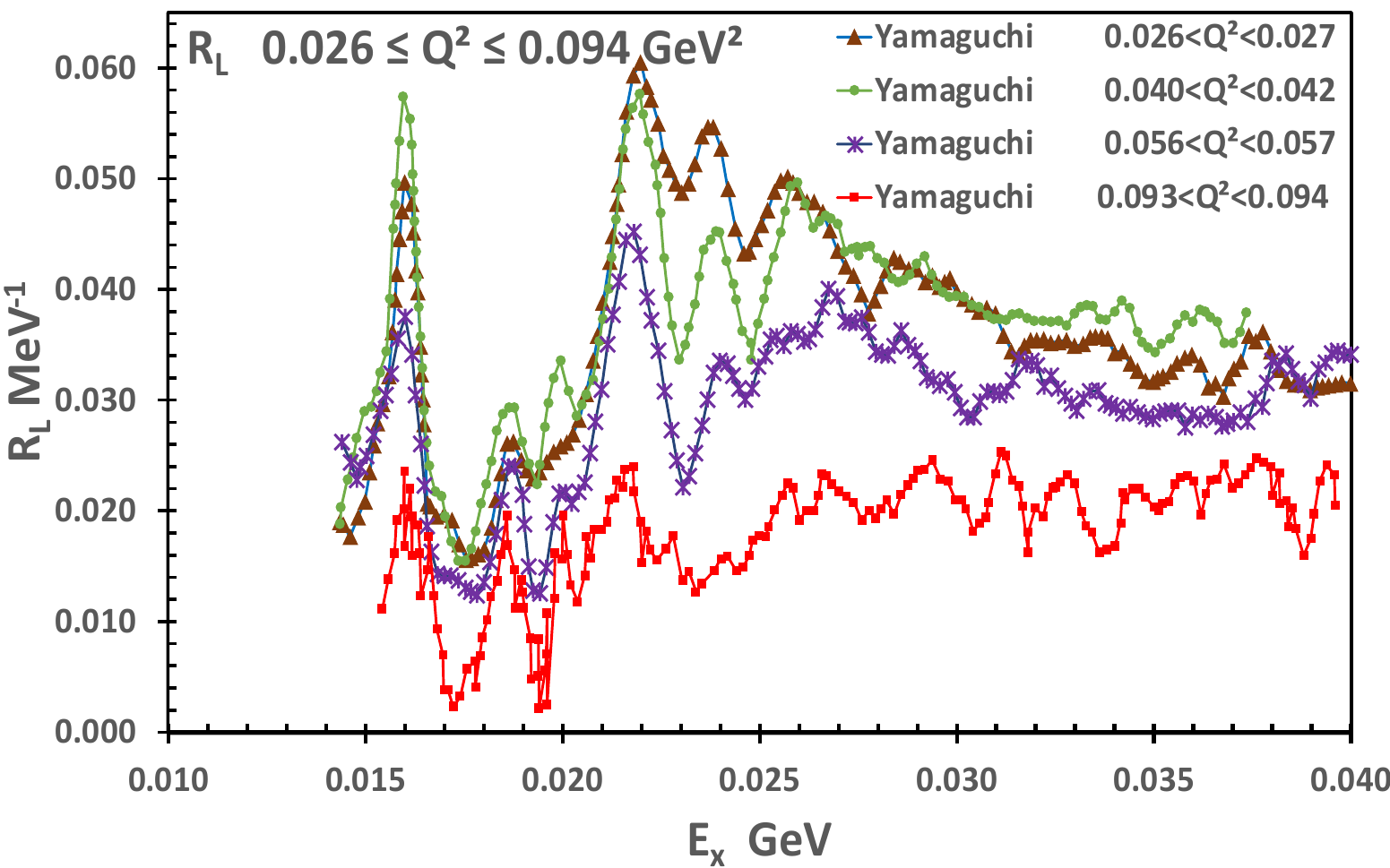}
\includegraphics[width=3.5 in, height=2.6 in] {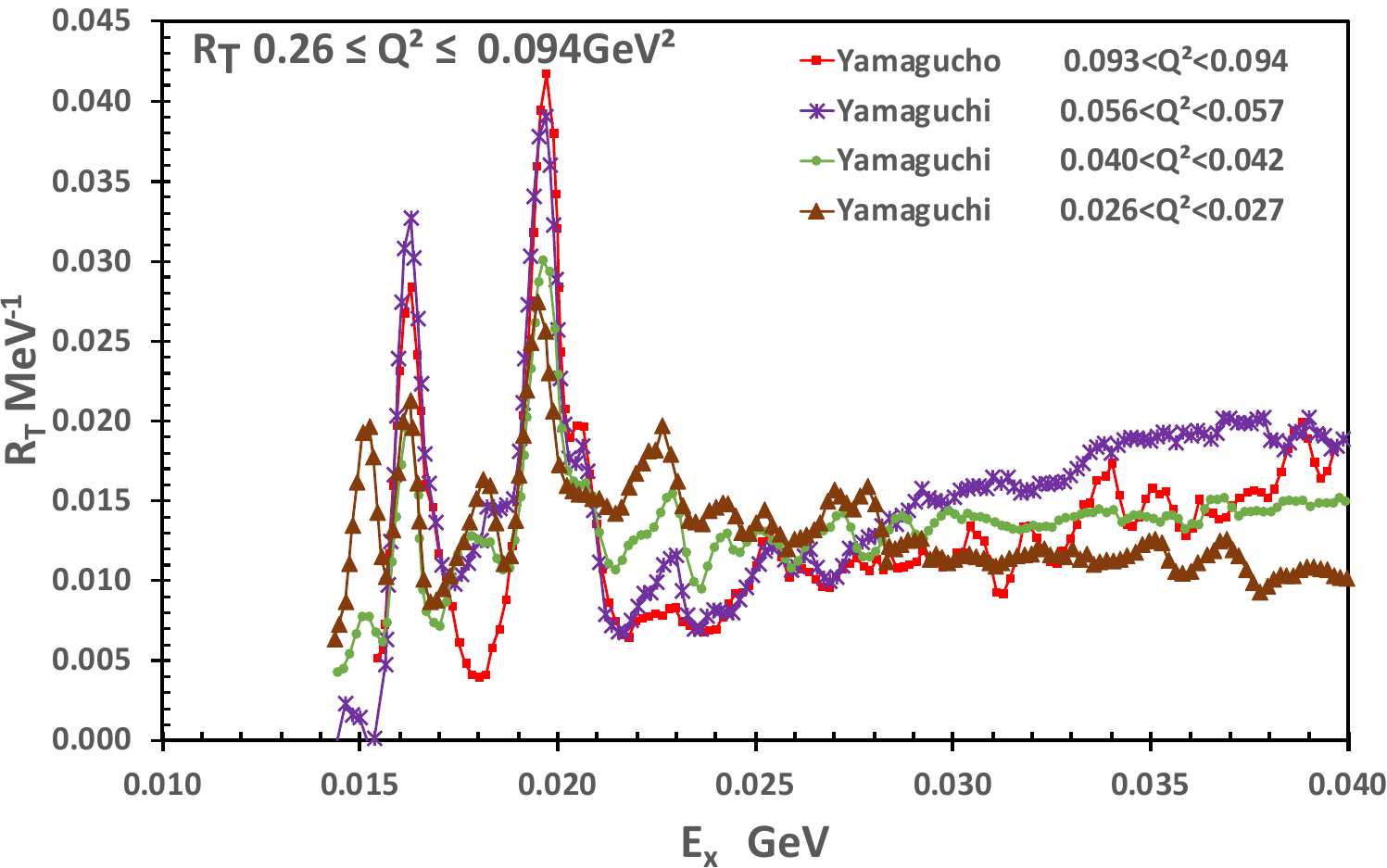}
\caption{ {\bf Nuclear excitation region}: Extractions of ${\cal R}_L$ and ${\cal R}_T$  in the nuclear  excitation region ($16<E_x<40$ MeV) for ${\rm ^{12}C}$.   The  measurements are from Yamaguchi:71~\cite{Yamaguchi:1971ua} except for values of  ${\cal R}_T(Q^2=0)$ which are extracted from photo-absorption data, and values of  ${\cal R}_T(Q^2=0.01)$ which are  extracted  from electron scattering cross section measurements at 180$^{\circ}$ (published by Goldemberg:64~\cite{PhysRev.134.B963} and  Deforest:65~\cite{DEFOREST1965311}). 
 }
\label{RLRTYamaguchi}
\end{figure*}

As shown in the bottom right panel of  Fig. \ref{Ryan_vs_Yama_plot}  the contribution from nuclear excitation is significant at low $Q^2$ (or $\bf q$).   For $E_x<$16 MeV we use ${\cal R}_L$ and ${\cal R}_T$ from   overall fits~\cite{Bodek:2023dsr}  to the nuclear excitation form factors (shown as the red solid line in  Figure \ref{Yamaguchi_RL_RT}). For $16<E_x<40$ MeV (when available)  we use the precise ($\pm 3\%$) Yamaguchi:71 measurements of  ${\cal R}_L$ and ${\cal R}_T$.  In addition, for  ${\cal R}_T$, we  also  use electron scattering data at 180$^{\circ}$~\cite{PhysRev.134.B963, DEFOREST1965311} as shown in the right top panel of   Fig. \ref{RLRTYamaguchi}.  For $E_x>30$ MeV, we extract  ${\cal R}_L$ and ${\cal R}_T$ from our analysis of  all available electron scattering data as described below. 
%
\begin{table}[tbh]
\footnotesize
\begin{center}
\begin{tabular}{|c|c|c||c|c|c|c|} \hline
Center	&	$Q^2$	&$Q^2$	&Center& ${\cal R}_T(\nu={\bf q} )$&	$\bf q$&$\bf q$\\
$Q^2$	&	low	&	high&  {\bf q}&From $\gamma ^{12}C$&low	&	high	\\
\hline
0($\gamma ^{12}C$)&	0	&	0	&    &  &&\\
%

\bf  0.010	&	0.004	&	0.015	& {\bf 0.100}&0.0016$\pm$0.0004	&	0.063	&	0.124	\\
\hline
\bf 0.020    &	0.015	&		0.025              &               {\bf 0.148}&0.0021$\pm$0.0007&	0.124	&	0.158	\\
\bf 0.026	&	0.025	&	0.035	&{\bf 0.167}	&0.0028$\pm$0.0006&	0.158	&	0.186	\\
\bf 0.040	&	0.035	&	0.045	 &{\bf 0.205}	&0.0071$\pm$0.0007&  0.186	&	0.223	\\
\bf 0.056	&	0.045	&	0.070& {\bf 0.240}	&0.0134$\pm$0.0009	&0.223	&	0.270	\\
\bf 0.093	&	0.070	&	0.100	& {\bf 0.300}	&0.0270$\pm$0.0006&	0.270	&	0.340	\\
\bf 0.120	&	0.100	&	0.145	&{\bf 0.380}	&0.0324$\pm$0.0005&	0.340	&	0.428	\\
\bf 0.160	&	0.145	&	0.209	&{\bf 0.475}	&0.0276$\pm$0.0005&	0.428	&	0.523	\\
%
\bf 0.265	&	0.206	&	0.323	&{\bf 0.570}	&0.0262$\pm$0.0008&	0.523	&	0.609	\\
\hline
\bf0.380	&	0.322	&	0.438	& {\bf 0.649}	&0.0290$\pm$0.0006&	0.609	&	0.702	\\
\bf 0.500	&	0.438	&	0.650	& {\bf 0.756}	&0.0299$\pm$0.0010	&0.702	&	0.878	\\
\bf 0.800	&	0.650	&	1.050	& {\bf 0.991}	&0.0371$\pm$0.0003&	0.878	&	1.302	\\
\bf 1.250	&	1.050	&	1.500	&  {\bf 1.619}	&0.0414$\pm$0.0010&	1.302	&	1.770	\\
\bf 1.750	&	1.500       &	2.000	&  {\bf 1.921}	&0.0479$\pm$0.0010&	1.770	&	2.067	\\
\bf 2.250	&	2.000	&	2.500	&  {\bf 2.213}	&0.0542$\pm$0.0020&	2.067	&	2.357	\\
\bf 2.750	&	2.500	&	3.000	&  {\bf 2.500}	&0.0603$\pm$0.0030&	2.357	&	2.642	\\
\bf 3.250	&	3.000	&	3.500	&  {\bf 2.783}	&0.0664$\pm$0.0030&	2.642	&	2.923	\\
\bf 3.750	&	3.500	&	4.000	& {\bf 3.500}      &0.0817$\pm$0.0030&    2.923	        &	4.500	\\
%
%
\hline
\hline
\end{tabular}
\caption{A summary of the 18 bins in $Q^2$ (in GeV$^2$) and the 18 bins in $\bf q$ (in GeV).  The  ${\cal R}_T$ (in MeV$^{-1}$) values for $\nu=\bf q$ are
extracted from  photo-absorption cross sections.
}
\label{tab:Q2bins}
\end{center}
\end{table}

%

\begin{table}[tbh]
\begin{center}
\begin{tabular}{|c|c|c|c|} \hline
   &Data Set           &  Normalization  & Error\\
\hline    
1&Barreau:83~\cite{Barreau:1981rr,Barreau:1983ht,Barreau:1983A}     &    0.9919       &    0.0024      \\
2&O'Connell:87~\cite{OConnell:1987ag}  &     0.9787        &     0.0086      \\
3&Sealock:89~\cite{Sealock:1989nx}    &          1.06        &  0.1000    \\
4&Baran:88~\cite{Baran:1988tw}   &          0.9924      &  0.0046   \\
5&Bagdasaryan:88~\cite{Bagdasaryan:1988hp} &          0.9878       &  0.0083         \\
6&Dai:19~\cite{Murphy:2019wed}             &           1.0108         &   0.0053  \\
7&Arrington:96\cite{Arrington:1995hs} &                 0.9743      &  0.0133   \\ 
8&Day93~\cite{Day:1993md}             &               1.0071     &0.0033    \\
9&Arrington:98~\cite{Arrington:1998ps} &             0.9888     & 0.0034       \\
10&Gaskell:21~\cite{Arrington:2021vuu,Seely:2009gt}     &       0.9934        &    0.0051         \\
11&Whitney:74~\cite{Whitney:1974hr,Moniz:1971mt}     &         1.0149       &  0.0153      \\
12&E04-001-2005:24~\cite{Alsalmi:2019sie,JUPITER:2025uny}      &                   0.9981    & 0.0067    \\ 
13&E04-001-2007:24~\cite{Alsalmi:2019sie,JUPITER:2025uny}       &    1.0029        &   0.0070  \\ 
14&Gomez:74~\cite{Gomez:1993ri, ResArchive}     &           1.0125         &    0.0149     \\
15&Fomin:10~\cite{Fomin:2010ei,JeffersonLabHallCE94-110:2004nsn}         &    1.0046        &   0.0031    \\
16&Yamaguchi:71~\cite{Yamaguchi:1971ua}   &   1.0019         &    0.0029     \\
17&Ryan:84~\cite{Ryan:1984yp} (180$^{\circ}$)&             1.10  &  0.0130       \\
\hline
18& Czy\.z:63~\cite{Czyz:1963zz} (not used)&                      1.000&  0.2000 \\
19& Bounin:63~\cite{Bounin:1963,Lovseth:1968zz}&                         1.150 &  0.2300 \\
21& Antony-Spies70~\cite{Antony-Spies:1970jjs,Yamaguchi:1971ua} (not used) &                   0.95  &  0.25     \\
22& Goldemberg64~\cite{PhysRev.134.B963}(180$^{\circ}$)&                          1.100  &  0.1000 \\
23& Deforest:65~\cite{DEFOREST1965311}(180$^{\circ}$)&                    0.9       &   0.1000  \\\hline
24 & Mainz:2024\cite{Mihovilovic:2024ymj} (not used)&                    1.03      &   0.02\\
25 & CLAS-e4nu:2023\cite{Isaacson:2023} &                   0.85     &   0.02\\
26 & Garino::1992\cite{Garino:1992ca}   (not used)&                   1.00     &   0.02\\
27&  Ricco:1968:\cite{Ricco:1968bsq}  (not used)&            -            &  - \\    
30& Donnelly:68~\cite{Donnelly:1968wsh,Donnelly:1970rk} (not used)  &          -     &   - \\
31 &Zeller:73~\cite{Zeller:1973ge,Heimlich:1974rk}   (not used)       &        -    &     -    \\ \hline
	&	Ahrens:75~\cite{Ahrens:1975rq}(photoabsorption)	&	-	&	-	\\
	&	(not-included-in-fit)	&		&		\\
	&	Carrasco:89~\cite{Carrasco:1989vq}(photoabsorption)	&	-	&	-	\\
	&	Bianchi:95~\cite{Bianchi:1995vb}(photoabsorption)	&		-	&	-	\\
	&	Bezic:69~\cite{Bezic:1969ura}(photoabsorption)	&		-	&	-	\\
	&	(not-included-in-fit)	&		&		\\
\hline
\hline
\end{tabular}
\caption{ A summary table of the $\carbon$ data sets used in the Christy-Bodek 2024  universal fit and in this analysis. 
Shown are  the  normalization factors  and uncertainties. Data sets 18--23 ( have larger normalization uncertainty)  are early data at very low ${\bf q}$.
Data  sets  which are inconsistent with other data sets are identified as not used.
}
\label{datasets}
\end{center}
\end{table}

\section{Inclusive Electron-Nucleon Scattering}
 In terms of the incident electron energy, $E_0$, the 
scattered electron energy $E'$, and the scattering angle $\theta$, the absolute value of the exchanged 4-momentum  squared in electron-nucleon scattering  is given by
\begin{equation}
Q^2 = -q^2 =  4E_0E'{\sin}^2 \frac{\theta}{2}.
\end{equation}
The mass of the undetected hadronic system (nucleon and pions)  is
\begin{equation}
W^2 = M^2 + 2M\nu -Q^2,  
\end{equation}
and the square of the magnitude of  3-momentum transfer vector ${\bf q}$ is
\begin{equation}
{\bf q}^2 = Q^2 +\nu^2.
\end{equation}
Here $M$ is the mass of the proton  and $\nu = E_0-E' $.  In these expressions we have neglected 
the electron mass which is negligible for the kinematic regions investigated in this paper. 

For scattering from a nuclear target such as carbon, the excitation energy $E_x$ is given by $E_x=\nu-\nu_{elastic}$  where
\begin{equation}
\nu_{elastic}= E_0-\frac{E_0}{1+2E_0{\sin}^2 \frac{\theta}{2}/M_A}=\frac{Q^2_{elastic}}{2M_A},
\end{equation}
where  $Q^2_{elastic}$ is the $Q^2$ for elastic scattering from a carbon nucleus for incident energy $E_0$ and
scattering angle~$\theta$ 
and  $M_A$ is the mass of the nuclear target. For carbon $M_{A}=12 u =11.178$ GeV (1 u = 931.502 MeV).
The excitation energy is 
\begin{equation}
E_x=\nu-\nu_{elastic}
\end{equation}

Alternatively, one can describe nuclear excitations in terms of the mass of the excited carbon nucleus.
\begin{equation}
W^2_{A} = M^2_A + 2M_A\nu -Q^2,  
\end{equation}

\subsection{Description in terms of longitudinal and transverse virtual photon cross sections} 
This description is often used in the nucleon resonance region.  In the one-photon-exchange approximation, the spin-averaged cross section for inclusive electron-nucleon (or electron-nucleus) scattering can be expressed in terms of the  photon helicity coupling as
\begin{equation}
\frac{d\sigma}{d\Omega dE^{'}} = \Gamma\left[\sigma_T(W^2,Q^2) + \epsilon \sigma_L(W^2,Q^2)\right],
\label{eq:cs1}
\end{equation}
where $\sigma_T$ ($\sigma_L$) is the cross section  for photo-absorption of purely transverse (longitudinal) polarized photons,
\begin{equation}
\Gamma = \frac{\alpha E^{'}(W^2 - M_N^2)}{(2 \pi)^2 
Q^2 M E_0 (1 - \epsilon)}
\end{equation}
is the flux of virtual photons, $\alpha=1/137$ is the fine structure constant,  and
\begin{equation}
\label{epsilonEQ}
\epsilon = \left[1 + 2\left(1+\frac{\nu^2}{Q^2}\right) 
{\tan}^2 \frac{\theta}{2}\right]^{-1}
\end{equation}
is the relative flux of longitudinal virtual photons (sometimes referred to as the virtual photon polarization). Since $\Gamma$ and $\epsilon$ are purely kinematic factors,
it is convenient to define the reduced cross section 
\begin{equation}
\sigma_r = {1 \over \Gamma} \frac{d\sigma}
{d\Omega dE^{'}} = \sigma_T(W^2,Q^2) + \epsilon \sigma_L(W^2,Q^2).
\label{sig_reduced}
\end{equation}
All the hadronic structure information is,  therefore, contained in $\sigma_T$ and $\sigma_L$, which are 
only dependent on $W^2$ and $Q^2$.   In the $Q^2=0$ limit $\sigma_T(\nu,Q^2)$ should be equal to the measured  photo-absorption cross section~\cite{Bezic:1969ura,Ahrens:1975rq,Carrasco:1989vq,Bianchi:1995vb,Muccifora:1998ct}
  $\sigma_\gamma(\nu)$
for real photons (here $\nu$ is the energy of the photon).
\subsection{Description in terms of structure functions}
This description is primarily used in the inelastic continuum region.
In the one-photon-exchange approximation, the spin-averaged cross section for inclusive electron-nucleon (or electron nucleus) scattering can be expressed in terms of two structure functions as follows
\begin{eqnarray}
\frac{d\sigma}{d\Omega dE^{'}}& =& \sigma_M   [{\cal W}_2(W^2,Q^2) + 2 \tan^2(\theta/2) {\cal W}_1(W^2,Q^2)] \nonumber\\
\sigma_M &= & \frac{\alpha^2 \cos^2(\theta/2) }{[2 E _0 \sin^2(\theta/2)]^2}  = 
 \frac{4\alpha^2E^{\prime 2}}{Q^4} \cos^2(\theta/2)
\label{eq:cs2}
\end{eqnarray}
where $\sigma_M$ is the Mott cross section. 
The ${\cal F}_1$ and ${\cal F}_2$ structure functions are related to ${\cal W}_1$ and ${\cal W}_2$ by ${\cal F}_1 = M {\cal W}_1$ and ${\cal F}_2=\nu {\cal W}_2$.   The structure functions are typically expressed as functions of  $Q^2$ and $W^2$, or alternatively
 $\nu$ and $x=Q^2/(2M\nu)$.  At large $Q^2$ and $\nu$ (deep inelastic region)  $x$ is the fractional momentum  of the nucleon carried by the struck quark.  At low $Q^2$ there are target mass corrections. 
 The target mass  scaling variable ~\cite{Nachtmann:1973mr,Georgi:1976ve,Barbieri:1976rd} 
 is  $\xi_{TM}$ where, 
\begin{eqnarray}
\label{eq:xitm}
\xi_{TM}  &=& \frac{Q^2}
        {M\nu [1+\sqrt{1+Q^2/\nu^2}]}.
\end{eqnarray}

 The quantity $  R_{\sigma LT}(x,Q^2)$ is defined as the ratio of the longitudinal to transverse virtual photo-absorption cross sections, 
  $\sigma_L/\sigma_T$, and is related to the structure functions by
\begin{equation}
 R_{\sigma LT}(x,Q^2)
   = \frac {\sigma_L }{ \sigma_T}
   = \frac{{\cal F}_L }{ 2x{\cal F}_1},
\end{equation}
where ${\cal F}_L$ is called the longitudinal structure function. 

The structure functions are expressed in terms of $\sigma_L$ and
$\sigma_T$ as follows:
\begin{eqnarray}
K &=& \frac{Q^2(1-x)}{2Mx} = \frac{2M \nu - Q^2 }{2M}, \\
 {\cal F}_1 &=& \frac{M K }{ 4 \pi^2 \alpha} \sigma_T, \\
 {\cal F}_2 &=& \frac{\nu K (\sigma_L + \sigma_T)}{4 \pi^2 \alpha (1 + 
 \frac{Q^2 }{4 M^2 x^2} )}, \\
 {\cal F}_L(x,Q^2) &=& {\cal F}_2 \left(1 + \frac{4 M^2 x^2 }{ Q^2}\right) - 2x{\cal F}_1,
\end{eqnarray}
or
\begin{equation}
2x{\cal F}_1 = {\cal F}_2 \left(1 + \frac{4 M^2 x^2 }{ Q^2}\right) -  {\cal F}_L(x,Q^2).
\label{eq:fl-rel}
\end{equation}

In addition, $2x{\cal F}_1$ is given by
\begin{eqnarray}
\label{eq2xF1}
2x{\cal F}_1 (x,Q^{2}) &=& {\cal F}_2 (x,Q^{2}) 
\frac{1+4M^2x^2/Q^2}{1+ R_{\sigma LT}(x,Q^2)},
\end{eqnarray}
or equivalently
\begin{eqnarray}
\label{W2W1}
\label{eqW1}
{\cal W}_1 (x,Q^{2}) &=& {\cal W}_2(x,Q^{2})
 \times  
\frac{1+\nu^2/Q^2}{1+ R_{\sigma LT}(x,Q^2)}. 
\end{eqnarray}
\subsection{Description in terms of response functions}
\label{RLRTdescription}
This description is primarily used in the nuclear excitation and  QE regions.
The electron scattering differential cross section  is written in terms of longitudinal  (${\cal R}_L({Q^2},\nu)$) and transverse (${\cal R}_T({Q^2},\nu)$) nuclear response functions~\cite{Caballero:2009sn} as  
\begin{eqnarray}
\frac{d\sigma}{d \nu d\Omega}&=& \sigma_M [ A {\cal R}_L ({Q^2},\nu) + B {\cal R}_T ({Q^2}, \nu)],
\end{eqnarray} 
where  $\sigma_M$ is the Mott cross section (\ref{eq:cs2}),
 $A=(Q^2/{\bf q}^2)^2$ and $B = \tan^2(\theta/2) +Q^2/2{\bf q}^2$. 
  
 The relationships between the nuclear response functions, structure functions and virtual photon absorption cross sections are:  
 \begin{eqnarray} 
 {\cal R}_T({\bf q},\nu)&=& \frac{2 {\cal F}_1({\bf q},\nu)}  {M} =\frac{K}{2\pi^2\alpha}\sigma_T,\\
 {\cal R}_L ({\bf q},\nu)&=&\frac{{\bf q}^2}{Q^2} \frac{ {\cal F}_L({\bf q},\nu)} {2 M x} =\frac{{\bf q}^2}{Q^2}\frac{K}{4\pi^2\alpha}\sigma_L,
  \label{eq:respcross}
   \end{eqnarray}
  where ${\cal R}_L({\bf q},\nu)$ and ${\cal  R}_T({\bf q},\nu)$ have units of $M^{-1}$.
%
Consequently, 
\begin{equation}
 R_{\sigma LT}(x,Q^2)
   =\frac{2Q^2}{{\bf q}^2} \frac{ {\cal R}_L({\bf q},\nu)}{ {\cal R}_T({\bf q},\nu)}.
   \label{RSLST}
\end{equation}

%
 The square of the electric and magnetic form factors for elastic scattering and nuclear excitations are obtained by the integration of the measured response functions over $\nu$ for each nuclear state.  It should be noted that when  form factors for 
 {\it nuclear excitations} are extracted from electron scattering
  data an additional factor of $Z^2$ has traditionally been included in the definition of $\sigma_M$ (where $Z$ is the atomic number of the nucleus).
 However, to keep consistency between the treatment of QE scattering and  nuclear excitations we chose {\it not} to include the $Z^2$ factor in the definition of $\sigma_M$ in this analysis.
    %
\begin{figure}
\includegraphics[width=3.45in, height=2.in] {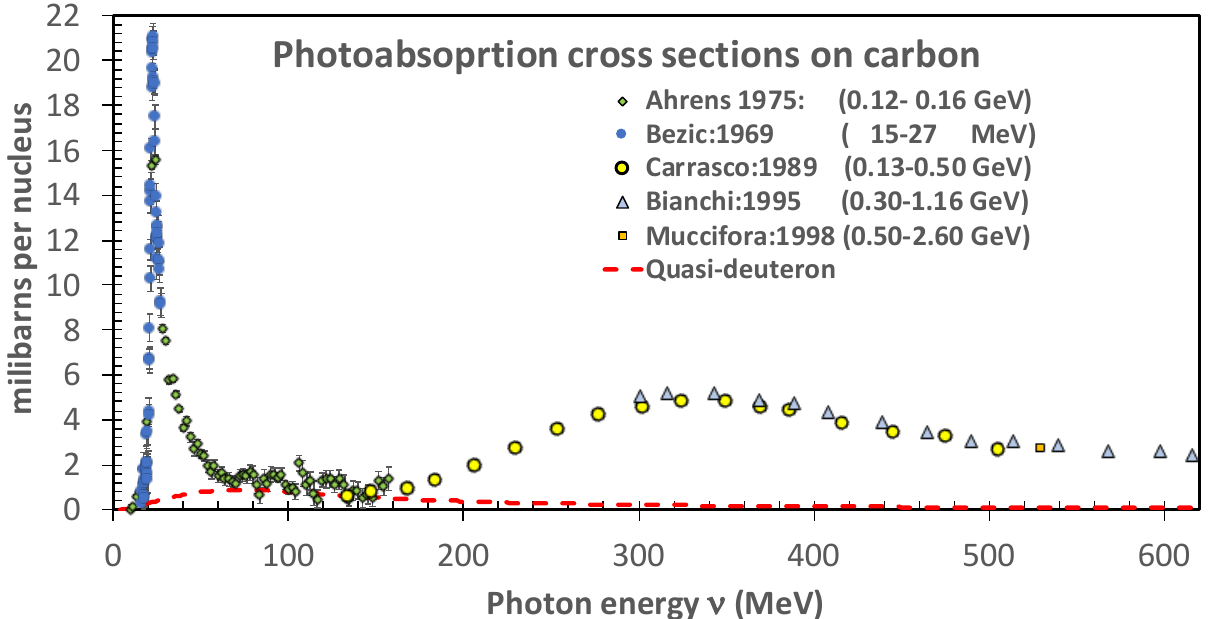}
\includegraphics[width=3.55in, height=2.in] {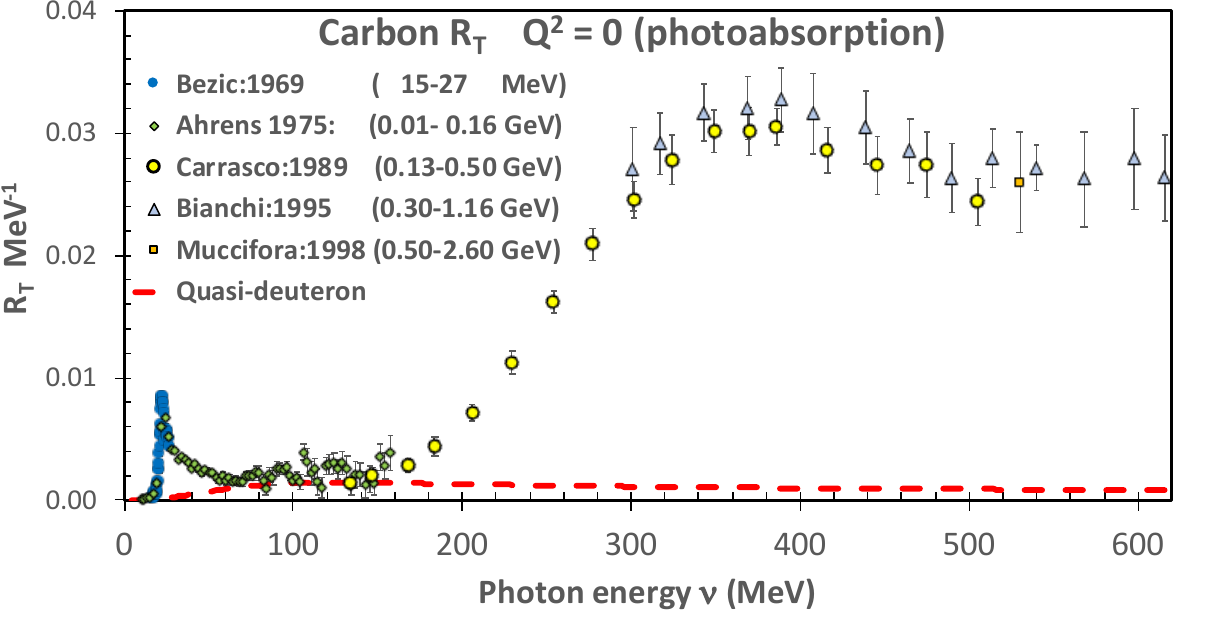}
\caption{Top: ${\rm ^{12}C}$ photo-absorption cross section as a function of photon energy $\nu$
from 0 to 0.6 GeV.
Bottom: The response function ${\cal R}_T ({Q^2=0},\nu)$ extracted from
photo-absorption cross sections using eq. \ref{eq:respphoto}.  
}
\label{photo_cross_figure}
\end{figure}

 \subsubsection{Extraction of response functions from photo-absorption  cross sections}
 At $Q^2=0$  the {\it virtual} photo-absorption cross section is equal to the photo-absorbtion cross section
 for {\it real} photons. 
  Therefore 
\begin{eqnarray} 
   \label{photoRLRT}
    {\cal R}_T(Q^2=0,\nu) &=& \frac{\nu}{2\pi^2\alpha}\sigma_\gamma(\nu),
\label{eq:respphoto}    
\end{eqnarray} 
where $\sigma_\gamma(\nu)$ is the photo-absorption cross section  for real  photons of energy $\nu$. 

As real photons can only have transverse polarization, there are no longitudinal photons at $Q^2=0$.  Therefore,  ${\cal R}_L(Q^2=0, \nu)$
cannot be derived from photo-absorption cross sections and can only be extracted from the universal fit to the  electron scattering data,
\begin{eqnarray} 
   \label{photoRL}
    {\cal R}_L(Q^2=0,\nu)&=&\frac{\nu^3}{4\pi^2\alpha} \lim_{Q^2 \to 0}\frac{\sigma_L(Q^2)}{Q^2}.\end{eqnarray} 
  %
However, the value of ${\cal R}_L(Q^2=0,\nu)$ in this region is not relevant since the longitudinal photo-absorption cross section $\sigma_L(Q^2=0)$ is zero and gives no contribution to the electron scattering cross section.

The measured photo-absorption cross section~\cite{Bezic:1969ura,Ahrens:1975rq,Carrasco:1989vq,Bianchi:1995vb,Muccifora:1998ct} on $^{12}$C is shown on the top panel of Fig~\ref{photo_cross_figure}.
The dashed red line is the expectation from the Quasi-deuteron model \cite{Plujko:2018uum}(scattering from short-correlation neutron-proton pairs).
The Values of ${\cal R}_T(Q^2=0,\nu)$ extracted via Eq~\ref{eq:respphoto} is shown in the bottom panel.
In  electron scattering $Q^2$=${\bf q}^2+\nu^2$ must be $ \geq 0$. Therefore,  for a fixed $\bf q$ the maximum $\nu$ ($\nu^{max}$) occurs when  $\nu = \bf q$ and 
\begin{eqnarray} 
   {\cal R}_T({\bf q,}\nu=\bf q) &=&\frac{\bf q}{2\pi^2\alpha}\sigma_\gamma(\bf q). 
\end{eqnarray}

In photo-absorption  the cross section  is dominated by the Giant Dipole Resonance,  quasi-deuterons \cite{Tavares:1992} and pion production processes 
(the cross section for the quasielastic process is zero).

  We use a form of the Rosenbluth method introduced by Jourdan~\cite{Jourdan:1996np, Jourdan:1995np} 
   to separate ${\cal R}_L({Q^2},\nu)$ and ${\cal  R}_T({Q^2},\nu)$ as follows:
\begin{eqnarray}
\Sigma ({Q^2},\nu)&=&H\frac{d\sigma}{d \nu d\Omega}\\
&=&\epsilon {\cal R}_L ({Q^2},\nu) + \frac{1}{2}\left(\frac{{\bf q}}{Q}\right)^2 {\cal R}_T ({Q^2}, \nu), \nonumber\\
\end{eqnarray}
with
\begin{eqnarray}
H&=&\frac{1}{\sigma_M} {\epsilon}\left(\frac{{\bf q}}{Q}\right)^4\nonumber\\
&=&\frac{{\bf q}^4}{4\alpha^2E^{'2}} \label{H2}
\frac{1}{\cos^2(\theta/2)+2 ({{\bf q}}/{Q})^2\sin^2(\theta/2).}
\end{eqnarray}
The quantity $\Sigma ({Q^2},\nu)$ is plotted as a function of the virtual photon polarization $\epsilon$,
which varies from 0 to 1 as the scattering angle $\theta$ ranges from 180 to 0 degrees.  
We use Eq.~(\ref{H2}) for $H$ because it is valid for all scattering angles, including $180^\circ$.
Here,   ${\cal R}_L({Q^2},\nu)$ is the slope, and  $\frac {1}{2}\frac{{\bf q}^2}{Q^2}{\cal R}_T({Q^2},\nu)$ is the intercept of the linear fit.

However, when Coulomb corrections are included, the above expressions are modified and all the parameters are replaced with "effective parameters"
as described below.
 %
  \subsection{Electron scattering at 180$^{\circ}$}
  Including Coulomb corrections the response function  ${\cal R}_T({Q_{eff}^2},\nu)$ can be extracted directly from the electron scattering cross section  at 180$^{\circ}$ using the following expression:
  \begin{eqnarray}
  \label{RT180}
{\cal R}_T ({Q_{eff}^2}, \nu)&=& \left(\frac {E_0}{E_{0,eff}}\right)^2 \frac{{Q_{eff}}^4}{4\alpha^2E_{eff}^{'2}}\frac{d\sigma}{d \nu d\Omega},
\end{eqnarray}
where  the $eff$  subscript denotes  Coulomb corrected quantities described below.
%
\subsection{${\cal R}_L$ and ${\cal R}_T$ in the nuclear excitation region}
As mentioned earlier, in the nuclear excitation region $16<E_x<40$~MeV we use the precise ($\pm 3\%$) Yamaguchi measurements of ${\cal R}_L$ and ${\cal R}_T$ (where available). For $E_x<16$~MeV we extract ${\cal R}_L$ and ${\cal R}_T$ from fits~\cite{Bodek:2023dsr} to the nuclear excitation form factors, and performing a Gaussian smearing with the width determined by the experimental resolution.  At  $Q^2 = 0.01$~GeV$^2$ we  extract  ${\cal R}_T$ from  electron scattering cross sections at $E_0= 55$ and 70~MeV and $\theta=180^{\circ}$ published in  Goldemberg:64~\cite{PhysRev.134.B963} and  electron scattering cross sections at $E_0= 65$~MeV and $\theta$ =180$^{\circ}$ published in  Deforest:65~\cite{DEFOREST1965311}. At $Q^2=0$ we extract ${\cal R}_T$ from measured photo-absorption cross sections. 
  \section{Coulomb Corrections} 
  \vspace{-10pt}
Coulomb corrections to QE and inelastic pion production processes are taken into account  using the "Effective Momentum Approximation"(EMA)~\cite{Aste:2005wc,Gueye:1999mm}.
The approximation is a  simple energy gain/loss method, using a slightly higher incident and scattered electron energies at the vertex than measured in the lab.  The  effective incident energy is $E_{eff}=E_0+V_{eff}$, and the effective scattered energy is $E'_{eff}= E'+V_{eff}$.  

 Assuming a uniform spherical charge distribution in the nucleus (of radius R) the electrostatic potential inside the charged sphere can be defined as:
\begin{equation} 
     V(r)  =   \frac{3Z\alpha}{2R}   - \frac{Z\alpha }{2R}  \left(\frac{r}{R}\right)^2,
\end{equation}
\noindent where $R$ (in units of GeV) is given by
\begin{equation}
      R     = 1.1 A^{1/3} + 0.86 A^{-1/3},
\end{equation}
and the average potential energy is
\begin{eqnarray} 
     V_{eff}= \frac45 V(r=0)
     = \frac{6Z\alpha}{5R},
\end{eqnarray}
where $Z$ and $A$ are the atomic and mass numbers, respectively. 

%
 \begin{figure*}
 \includegraphics[width=7.in, height=8.5 in]{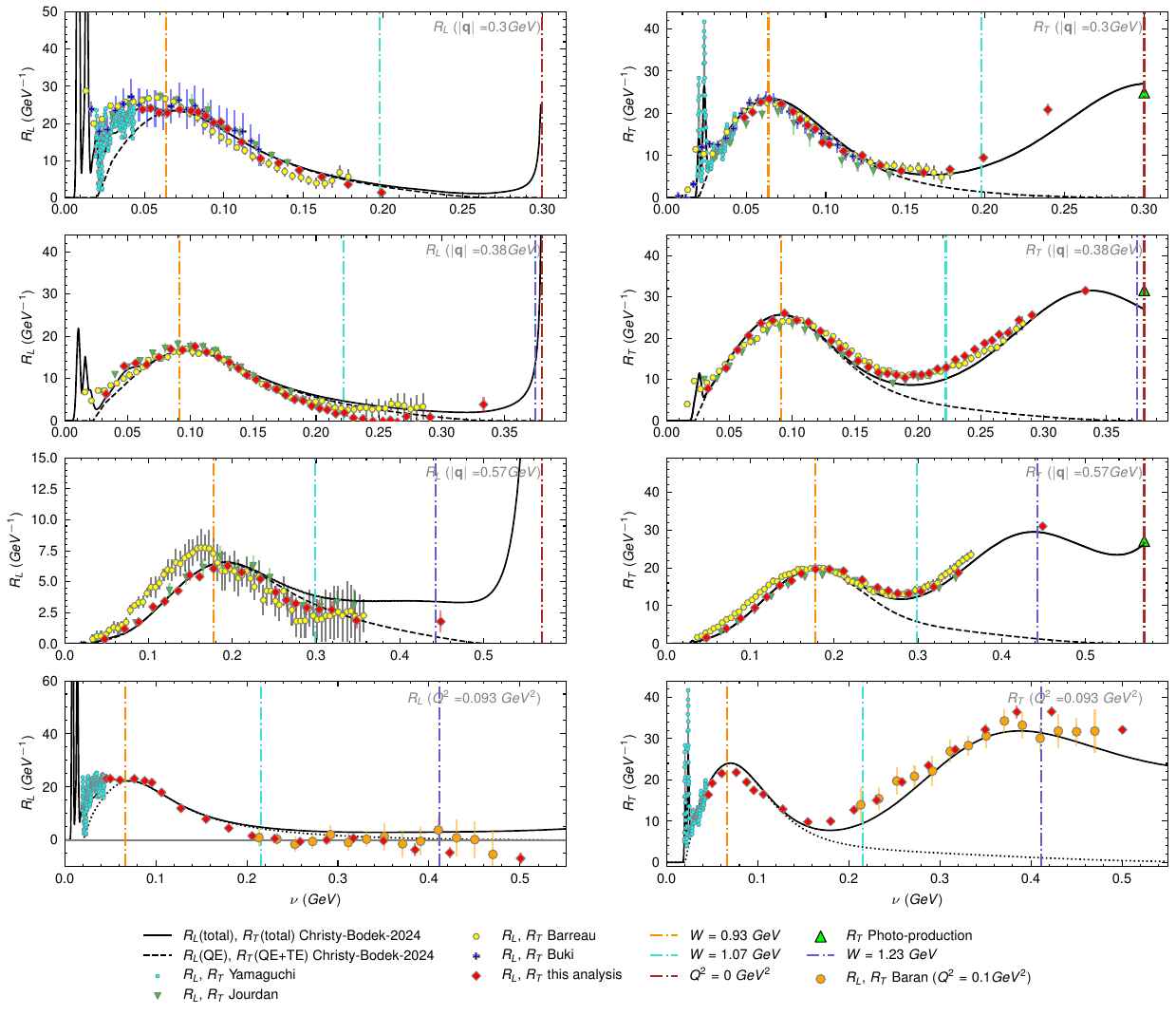}
\caption{{\bf Comparison to previous ${\cal R}_L$  ${\cal R}_T$ extractions:} {\bf  (a) } Comparisons between our extraction  of \rltot and \rttot  for ${\rm ^{12}C}$ in units of GeV$^{-1}$ (red diamonds) to previous extractions by Barreau:82~\cite{Barreau:1983ht},  Jourdan:96~\cite{Jourdan:1996np, Jourdan:1995np}, and Buki:21~\cite{Buki:2021mjo} for {\bf q}=0.3, 0.38 and 0.57 GeV. {\bf (b)}  Comparison of our extraction at $Q^2$=0.093 GeV$^2$ to previous extraction by Baran:1988~\cite{Baran:1988tw} at $Q^2$=0.1 GeV$^2$ (light green circles) in the region of the $\Delta$(1238) resonance.}
\label{Exp_compare}
\end{figure*}

However, in this analysis we use a better determination of the effective potential  extracted from a comparison
  of positron and electron QE scattering cross sections on carbon~\cite{Gueye:1999mm}.
  The experimentally determined $V_{eff}$ for carbon is  $V_{eff}$= 3.1~$\pm$~0.25 MeV.
  
   Including Coulomb corrections (CC)  we define:  
\begin{eqnarray}
      E_{0, eff} &=&  E_0+V_{eff} \\
       E'_{eff} &=&  E'+V_{eff}\nonumber\\
       \nu_{eff}=&\nu  \nonumber \\
      Q^2_{eff}     &=&  4 (E_0+V_{eff}) (E'+V_{eff}) \sin^2(\theta/2)  \nonumber \\
      {\bf q_{eff}}^2 &=&  Q^2_{eff} +\nu^2 \nonumber \\
      W^2_{eff}&=&M^2+2M\nu-Q^2_{eff} \nonumber \\
    E_x^{cc}&=&E_x  \nonumber \\
       W^2_{A-eff}&=&M^2_A+2M_A\nu-Q^2_{eff} \nonumber \\
     \epsilon^{CC} &=&\left[1 + 2\left(1+\frac{\nu^2}{Q_{eff}^2}\right) {\tan}^2 \frac{\theta}{2}\right]^{-1} \nonumber 
\end{eqnarray}
The response functions are
 calculated with   $Q^2=Q^2_{eff}$  and $E'=E'_{eff}$.
In addition, there is a focusing factor $F_{foc}^2=\big[\frac{E_0+V_{eff}}{E_0}\big]^2$ which modifies the Mott cross section.
The modified Mott cross section is  
\begin{eqnarray}
\sigma_{M-eff} &=&  F_{foc}^2 \frac{\alpha^2 \cos^2(\theta/2) }{[2 E_{eff}  \sin^2(\theta/2)]^2}  \nonumber
\end{eqnarray} 
Therefore
\begin{eqnarray}
&&\Sigma^\prime ({Q^2_{eff}},\nu)=H^{CC}\frac{d\sigma}{d \nu d\Omega} \nonumber \\
&=&\epsilon^{CC} {\cal R}_L ({Q_{eff}^2},\nu) + \frac{1}{2}\left(\frac{{\bf q_{eff}}}{Q_{eff}}\right)^2 {\cal R}_T ({Q_{eff}^2}, \nu)\nonumber
\end{eqnarray} 
\begin{eqnarray}
H^{CC}&=&\left(\frac {E_0}{E_0+V_{eff}}\right])^2 \times \\
&&\frac{{\bf q_{eff}}^4}{4\alpha^2E_{eff}^{'2}} \label{H2CC}
\frac{1}{\cos^2(\theta/2)+2 ({{\bf q_{eff}}}/{Q_{eff}})^2\sin^2(\theta/2)}\nonumber
\end{eqnarray} 
%
 %
 \section {Extractions of ${\cal R}_L $ and ${\cal R}_T$ }
From the Christy-Bodek 2024 overall  fit we extract the  relative normalization ($N_i$) of the various data sets as given
in Table \ref{datasets}.   For low excitation energy ($E_x <$ 50 MeV)  bin centering corrections are minimized if we extract the response functions as functions of the final state excited
nuclear mass $W^2_A$,  as described below. Similarly for ($E_x >$ 50 MeV)  bin centering corrections are minimized if we extract the response functions as functions of the square of the final state mass defined in terms of the mass of the nucleon  $W^2$. 

We use the fit to apply $Q^2$  bin centering corrections ($C_i$)
   to all extracted value of $\Sigma_i^\prime$ that are within the $Q_{eff}^2$ range for each $Q_{eff}^2$ and
   of $W_{eff}^2$ bin such that they reflect the values at  $Q_{enter}^2$, $W_{center}^2$, and  $\epsilon_{center}$
   for each cross section measurement,
 \subsection{Bin centering corrections for fix $Q^2$ bins}
 \begin{equation}
C_i=\frac{\epsilon_{center} {\cal R}^{fit}_L  + \frac{1}{2}(\frac{{\bf q_{center}}}{Q_{center}})^2 {\cal R}^{fit}_T}
{\epsilon^{cc}{\cal R}^{fit}_L (W^2_{eff},{Q_{eff}^2)} + \frac{1}{2}(\frac{{\bf q_{eff}}}{Q_{eff}})^2 {\cal R}^{fit}_T ({W^2_{eff},Q_{eff}^2})},\nonumber
\label{bin_centering}
\end{equation}
For excitation energies above 50 MeV we use  $W^2$ based bin centering corrections.
\begin{eqnarray}
&&{\cal R}^{fit}_L ={\cal R}^{fit}_L (W^2_{center},{Q_{center}^2)} \nonumber \\
&&{\cal R}^{fit}_T={\cal R}^{fit}_T (W^2_{center},{Q_{center}^2)}\nonumber \\
 &&\nu_{center}=\frac{W^2_{center}-M^2+Q^2_{center}}{2M}    
   \end{eqnarray}
   and
    \begin{eqnarray}
    \label{W2_Q2_centering}
 && \epsilon_{center} =\left[1 + 2\left(1+\frac{\nu^2_{center}}{Q_{center}^2}\right) {\tan}^2 \frac{\theta}{2}\right]^{-1} \nonumber \\
     \end{eqnarray}
 After the application of the bin centering corrections to all extracted values $\Sigma^\prime_i$ in 
 each $Q_{eff}^2$ bin  we can assume that extracted values of $\Sigma_i$  are at $Q^2_{center}$.  When we apply these
corrections, we keep the final state  mass $W^2$ fixed for $\nu>$~50 MeV, and  only correct for the change resulting
from changing $Q_{eff}^2$ to $Q^2_{center}$, which also results in changing $\nu$ to $\nu_{center}$ and $\epsilon_{cc}$ to $\epsilon_{center}$.

For $Ex<$~50 MeV at fixed values of $Q^2$ the value of $\nu_{center}$ is
 \begin{eqnarray}
  &&\nu_{center}=E_x+ \frac{Q^2_{center}}{2M_A} 
 \end{eqnarray}
 where $M_A$ is the mass of the nuclear target (11.178 GeV for carbon).
   and
    \begin{eqnarray}
    \label{W2_q3_centering}
 && \epsilon_{center} =\left[1 + 2\left(1+\frac{\nu^2_{center}}{Q_{center}^2}\right) {\tan}^2 \frac{\theta}{2}\right]^{-1} \nonumber \\
 &&W^2_{center}=M^2+2M\nu_{center}-Q^2_{center}\nonumber \\
&&{\cal R}^{fit}_L ={\cal R}^{fit}_L (W^2_{center},{Q_{center}^2)} \nonumber \\
&&{\cal R}^{fit}_T={\cal R}^{fit}_T (W^2_{center},{Q_{center}^2)}.
     \end{eqnarray}
For fixed $Q^2$ bins, after  bin centering, all of the Rosenbluth fits are done as a function of  of $\epsilon_{center}$,
and the extracted  ${\cal R}_L $ and ${\cal R}_T $ are plotted as a function of $\nu_{center}$.

\subsection{Bin centering corrections for fixed $\bf q$ bins}
For $E_x>$~50 MeV. we solve the following quadratic equation for $\nu_{center}$.
\begin{eqnarray}
W^2_{center}&=& M^2+2M\nu_{center}-({\bf q}^2_{center}-\nu^2_{center}),\nonumber
\end{eqnarray}
which yields   $$Q^2_{center} = {\bf q}^2_{center}-\nu^2_{center,}$$ and
then follow up with equations \ref{W2_Q2_centering} and \ref{bin_centering}.

For $Ex<$~50 MeV at fixed $\bf q$ beins
We solve the following quadratic equation for $\nu_{center}$.
\begin{eqnarray} 
\nu_{center}&=& E_x +\frac{{\bf q}^2-\nu^2_{center}}{2M_A}
\end{eqnarray}
which yields   $$Q^2_{center} = {\bf q}^2_{center}-\nu^2_{center},$$ 
$$W^2_{center} = M^2+2M\nu_{center}-Q^2_{center},$$ and 
then follow up with equation \ref{W2_q3_centering}.

 \subsection{Comparison to Previous Measurements of \rltot and \rttot}
    A comparison between our extraction of \rltot and \rttot{} (red diamonds) to previous extractions are shown in Fig. \ref{Exp_compare} for the three ${\bf q}$ values used in Jourdan:96~\cite{Jourdan:1996np,Jourdan:1995np}. In the  Barreau:82~\cite{Barreau:1983ht}  analysis the response functions were extracted for only three $\bf q$ values (0.30, 0.40 and 0.55 GeV) using only their own (Saclay) cross section measurements.  A later extraction by Jourdan:96 was also performed for three slightly different values of momentum transfer (${\bf q}=0.30$, 0.38 and 0.57 GeV). In the Jourdan analysis the same Barreau:82  cross sections used,  with additional cross section data from SLAC.  
    
    Our extraction is more extensive because our data sample is much larger than what was available in 1996 (as summarized in Table \ref{datasets}). In Figure  \ref{Exp_compare}(A) we correct the Barreau:82 extractions for the small difference between their $\bf q$ values of 0.40 and 0.55 GeV and the $\bf q$ values of  0.38 and 0.57 GeV, respectively, of Jourdan.  In general, our  \rltot{} and \rttot{} are in better agreement with the Jourdan analysis especially at $\bf q$ of 0.57 GeV.  In all the plots the universal fit for the total (from all processes) \rltot{} and \rttot{} is the solid black line, and the QE component (including  Transverse Enhancement) of the fit is the dashed line.  A comparison to a recent extraction of \rltot{} and \rttot{} at ${\bf q}=0.30$ GeV published in Buki:21~\cite{Buki:2021mjo} (blue circles) using cross sections measured at Kharkov is also shown.  Note that the more recent Kharkov analysis  has larger errors and was performed only for ${\bf q}=0.3$ GeV.  
    
    There is good agreement between  our extraction at $Q^2$=0.093 GeV$2$ and  the extraction by Baran:1988~\cite{Baran:1988tw} at $Q^2$=0.1 GeV$^2$ (light green circles) in the region of the $\Delta$(1238) resonance as seen in Figure  \ref{Exp_compare}(B).
    
    \subsection{Systematic Errors}
    We estimate an overall systematic error of 4\% in the extracted values of the longitudinal and transverse response functions.  This error is
    anti-correlated between  \rltot{} and \rttot{}.  

  %
\section{Comparison to theoretical predictions of \rltot and \rttot}
\begin{figure*}
\includegraphics[width=7.0 in, height=8.in] {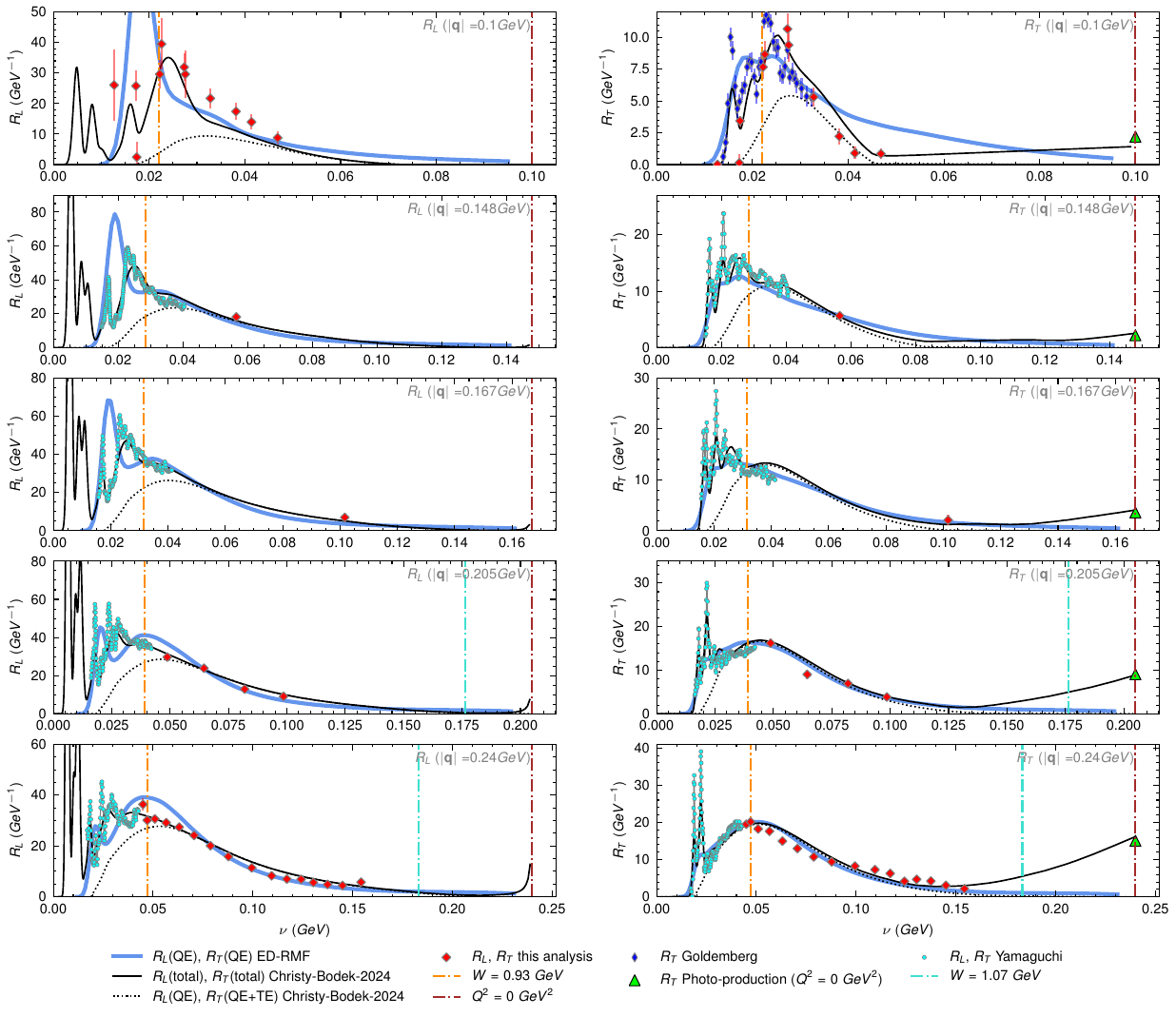}
\caption{{\bf Comparison to ED-RMF}: Our extractions of ${\cal R}_L$ and ${\cal R}_T$ for ${\rm ^{12}C}$ for  $\bf q$ values of 0.10, 0.148, 0.167, 0.205, and  0.240 GeV versus $\nu$.  In the nuclear excitation region we  also show  measurements from Yamaguchi:1971~\cite{Yamaguchi:1971ua}, and  ${\cal R}_T({\bf q}=0.01)$ GeV extracted  from cross sections at 180$^{\circ}$ (Goldemberg:64~\cite{PhysRev.134.B963} and  Deforest:65~\cite{DEFOREST1965311}). 
The values of   ${\cal R}_T({\bf q}=\nu)$ are extracted from photo-absorption data. In all the plots the  universal fit for the total  (from all processes)  \rltot and \rttot is the solid black line and  the QE component  (including  Transverse Enhancement) of the universal fit  is the dotted line. 
 The  blue lines are the predictions of the ED-RMF~\cite{Franco-Munoz:2022jcl,Franco-Munoz:2023zoa} calculations  (which include nuclear excitations).
 }
\label{ED-RMF_q3_A}
\end{figure*}

\begin{figure*}
\includegraphics[width=7.0 in, height=8.3in] {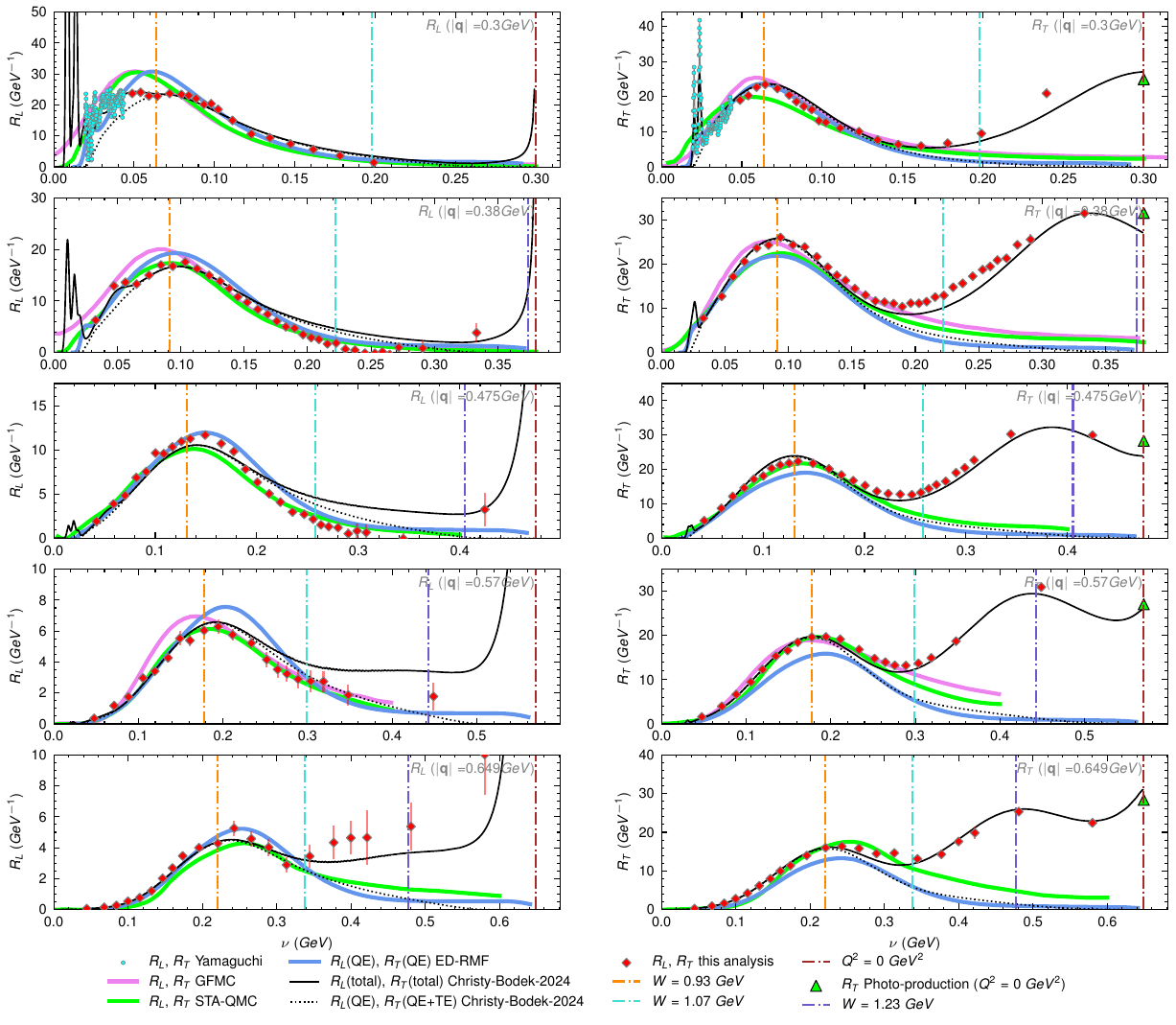}
\caption{{\bf Comparison to ED-RMF, STA-QMC and GFMC}: Same as Fig. \ref{ED-RMF_q3_A} for ${\bf q}$ values of   0.30, 0.38, 0.475, 0.57, and 0.649 GeV versus $\nu$.  the  blue lines are the predictions of ED-RMF, the  light green  lines are the predictions of STA-QMC and the   pink lines are the prediction of GFMC. }
\label{ED-RMF_q3_B}
\end{figure*}

\begin{figure*}
\includegraphics[width=7.0 in, height=5.in] {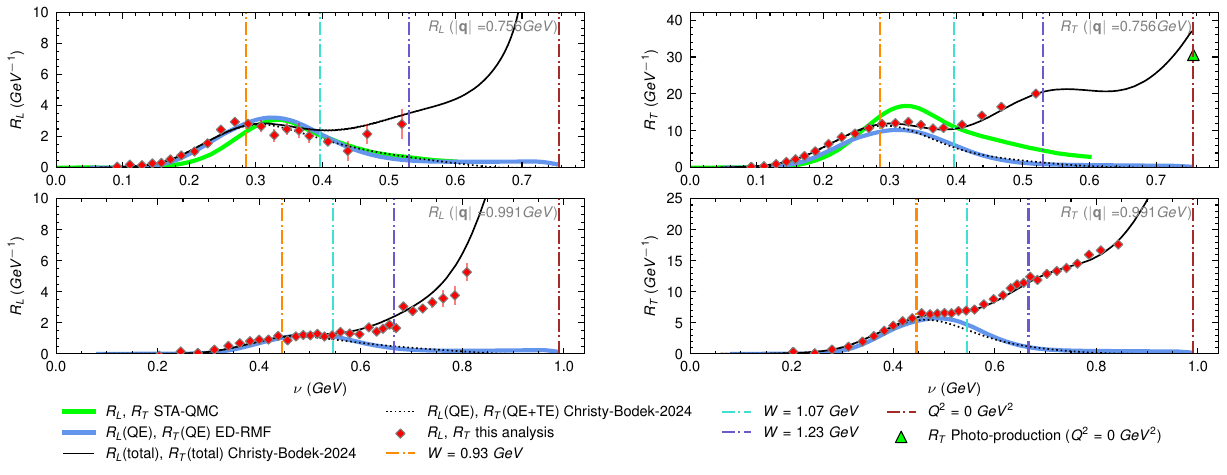}
 \caption{{\bf Comparison to ED-RMF and STA-QMC}: Same as Fig. \ref{ED-RMF_q3_A}  for $\bf q$ values  of 0.756 and  0.991 GeV versus $\nu$. The  blue lines are the predictions of ED-RMF, and the  light green  lines are the predictions of STA-QMC (for $\bf q$=0.756 GeV only).  Although STA-QMC  calculations have been provided for  $\bf q$=0.76 GeV  it is expected that that they will be less accurate in this region where relativistic effects, currently not fully included in STA-QMC, become increasingly more relevant. }
\label{ED-RMF_q3_C}
\end{figure*}

The extracted  \rltot and \rttot (in units of GeV$^{-1}$) are shown as the red diamonds  in Fig.~\ref{ED-RMF_q3_A}-\ref{ED-RMF_q3_C} for 12 values of $\bf q$.  Also shown (where available) are the Yamaguchi:1971~\cite{Yamaguchi:1971ua} measurements of \rltot{} and \rttot  in the nuclear excitation region, and  ${\cal R}_T({\bf q}=0.01)$ GeV extracted  from cross sections at 180$^{\circ}$ (Goldemberg:64~\cite{PhysRev.134.B963} and  Deforest:65~\cite{DEFOREST1965311}). The  values of  \rttot{} at $\nu$ =  $\bf q$ ($Q^2$ = 0) are extracted from  photo-absorption measurements. 

\rltot and \rttot{} extracted from the Christy-Bodek 2024 universal fit (which includes nuclear excitations)  are shown as  solid black lines, and the QE contributions to the fit  (including TE)  are the dashed black lines. 
%
%

%

The dot-dashed vertical orange line is $W=0.93$ GeV, the dot-dashed vertical light green line is $W=1.07$ GeV, the dot-dashed vertical dark blue line is $W=1.23$ GeV and the dot-dashed vertical brown line is where $\nu={\bf q}$ ($Q^2=0$ photo-absorption).

It is noteworthy to mention that: 
 \begin{enumerate} 
 \item  The $\Delta$(1232) nucleon resonance is not visible in \rltot plots and is only seen in \rttot~  because the excitation of the $\Delta$ is primarily transverse.
 \item At low $\nu$ and low $\bf q$ (or $Q^2$) the  contributions of nuclear excitation cross sections are significant.
 \item The QE peak is not visible at high values of  $\bf q$ (or $Q^2$) and the  cross section is dominated by Fermi  motion smeared inelastic pion production processes.
 \item At low $\bf q$ and $Q^2=0$ ($\nu=\bf q$)  the quasielastic cross section is zero and the cross section 
is dominated by the Giant Dipole Resonance,  quasi-deuterons \cite{Tavares:1992} and pion production processes.
The predictions for   \rttot  $Q^2=0$ are zero for models which do not include these processes.
 \end{enumerate} 
In the subsections that follow  we compare our measurements of \rltot and \rttot 
to the following theoretical predictions:
\begin{enumerate} 
\item``Green's Function Monte Carlo'' (GFMC)\cite{Lovato:2016gkq,Lovato:2023raf} which models the contribution of 1-body and two body-currents to the   response functions for single nucleon (1p1h) and two nucleon (2p2h) final states. GFMC predictions are only available for $0.3\le{\bf q}\le 0.57$ GeV.
\item ``Energy Dependent-Relativistic Mean Field'' (ED-RMF)\cite{Franco-Munoz:2022jcl,Franco-Munoz:2023zoa} which models the contribution of 1-body and two body-currents to the response functions for the  single nucleon (1p1h) final state only, but includes nuclear excitations which decay to a single nucleon final state (for both electron and neutrino processes). 
\item  An improved  superscaling model  (SuSAv2)\cite{Gonzalez-Jimenez:2014eqa, Megias:2016lke,Gonzalez-Rosa:2022ltp,Gonzalez-Rosa:2023aim} which uses relativistic mean field to  model  response functions for  the single nucleon (1p1h) final state,  and  a separate Relativistic Fermi Gas (RFG) model\cite{RuizSimo:2016rtu}  for two nucleon (2p2h) final states (for both electron and neutrino processes).  
\item   "Short Time Approximation Quantum Monte Carlo"  (STA-QMC)\cite{Andreoli:2024} which models the contribution of 1-body and two body-currents to the  response functions for single nucleon (1p1h) and two nucleon (2p2h) final states (for both electron and neutrino scattering processes).  Currently,  the STA-QMC  predictions are valid for $0.3\le {\bf q}\le 0.65$ GeV.
\item "Correlated Fermi Gas" (CFG)\cite{Bhattacharya:2024win} which models  response functions for single nucleon (1p1h) and two nucleon (2p2h) final states.
\item  The electron scattering mode of    \nuwro{}\cite{NuWro} which for the  electron scattering mode models only the single nucleon final state. Two nucleon final states are included in the neutrino scattering mode. 
\item   \achilles \cite{Isaacson:2023} A CHIcago Land Lepton Event Simulator which models  the contribution of 1-body and two body-currents to the 
response functions  for the single nucleon final state only (for both electron and neutrino scattering processes). Currently, the   \achilles~ predictions are  only valid for $\bf q>$0.5 GeV.
 \end{enumerate}


Among these theoretical predictions for $\carbon$ \nuwro~ is currently used in neutrino experiments,  and ED-RMF has recently been included in an update of
the NEUT neutrino MC generator.  Therefore, in addition to comparing these theoretical predictions to the data, we also  include a more extended description of  these two  theoretical approaches.  

The  improved SuSAv2 predictions for the  1p1h and 2p2h channels has been  implemented in   \genie~  for electrons and neutrinos \cite{Dolan:2019bxf, electronsforneutrinos:2020tbf}, 
 and  implementation in {{\sc{neut}}}  is underway.  In addition,   the SuSAv2-inelastic model is currently being implemented in  \genie~for electrons and neutrinos.   The SuSAv2 2p2h channel has been  implemented (only for neutrinos) in \nuwro{},~  the  \achilles~generator is being developed for use in future neutrino experiments, and STA-QMC been implemented\cite{Barrow:2020mfy} in \genie~   for  $\rm He^{4}$ (though an hadron tensor interface\cite{GENIE:DocDB137}).
%

\subsection{Comparison to the predictions of GFMC, STA-QMC, and ED-RMF}
The  extracted \rltot and \rttot are compared  to predictions of GFMC in  Fig.~\ref{ED-RMF_q3_B}, STA-QMC  in Fig.~\ref{ED-RMF_q3_B}-\ref{ED-RMF_q3_C}) and  ED-RMF in Fig.~\ref{ED-RMF_q3_A}-\ref{ED-RMF_q3_C}). 
The three  theoretical  calculations include  contributions from both 1-body and 2-body (1b+2b)  currents.    
In these three calculations there is enhancement of the calculated transverse response function \rttot. The  transverse enhancement (TE) primarily originates from the interference \cite{Carlson:2001mp,Simons:2023,Andreoli:2024} between 1b and 2b currents {\it which result in  1p1h. (1 nucleon 1 hole)  final states}.  The contribution of 2b currents to \rttot which results in 2p2h  (2 nucleons 2 holes) final states is not included ED-RMF but are included in GFMC and STA-QMC.   Note that final states with pions is not included in any of the three  calculations (pion production is only included in the improved version of SuSAv2).

 The GFMC  predictions are  shown as the  pink lines in Fig.~\ref{ED-RMF_q3_B}.  These calculations are rather complex and require a large number of CPU hours, especially at low $\bf q$. Consequently the response functions below $\bf q$=0.3  GeV and above $\bf q$=0.57 are not available for GFMC.  In addition,  there are  large uncertainties in the predictions near threshold (small $\nu$).
 

 The STA-QMC predictions for $\carbon$ (shown as the  light green lines in Fig.~\ref{ED-RMF_q3_B}-\ref{ED-RMF_q3_C})  are provided for $0.3 \le {\bf q} \le 0.76$ GeV.  By its  nature STA-QMC  only works for $\bf q>$ 0.3 GeV.   The STA-QMC predictions for    $0.3 \le {\bf q} \le 0.6$ GeV are  in agreement with our measurements.   Presently,  STA-QMC does not fully account for relativistic effects at the vertex. Therefore,  although STA-QMC  calculations have been provided for up to  $\bf q$=0.76 GeV (shown in Fig. \ref{ED-RMF_q3_C}), it is expected that that they will be less accurate in  the high $\bf q$ region where relativistic effects become increasingly more relevant.  The  STA-QMC calculations are based on the same many-nucleon Hamiltonian and electromagnetic currents as GFMC.
Since  STA-QMC does not have any knowledge of the correct threshold behavior or low-energy properties of the system, the correct threshold behavior is imposed as a constraint\cite{Pastore:2019urn}.

%
The  ED-RMF theoretical predictions are shown as the  blue lines.  These predictions (which are available at  all values of $\bf q$) include contributions to 1p1h final states from QE scattering as well as from nuclear excitations. 
 At higher values of $\bf q$ and higher values of $\nu$ the ED-RMF predictions for \rttot are below the data. This is expected because  the  ED-RMF  theoretical calculations for  \rttot do not include 2-body current processes with 2p2h  final states or Fermi smeared pion production processes.  
The STA-QMC calculation includes both single and two nucleon final states but is valid over a  more restricted kinematic  range and does not account for nuclear excitations.  The calculations of both ED-RMF and  STA-QMC  are  directly applicable to the same kinematic regions for neutrino scattering.

\subsubsection{The ED-RMF theoretical formalism}
Because the ED-RMF approach has recently been implemented in the NEUT neutrino generator, we include additional details on the ED-RMF approach. In the  ED-RMF  approach the hadronic current operator includes one- and two-body current contributions (see details in ref. \cite{Franco-Munoz:2022jcl}).
In this 1p1h calculation,  only the contributions from transitions where one nucleon below the Fermi level (hole)  is promoted to a continuum state above the Fermi level (particle)  by the current operator are included. Matrix elements of this current operator in between  hole-particle combinations are computed in an unfactorized fashion to compute the cross-section. The momentum distribution of the bound nucleons in the initial state are obtained by  solving the Diract equation with the Relativistic Mean Field as in ref. \cite{Sharma:1993it}.  The occupations and energies are chosen to be consistent with a representation of a semi-phenomenological spectral function similar to the Rome spectral function for $^{12}C$.  A continuous missing-energy profile and occupations is employed  as in  ref. \cite{Franco-Patino:2022tvv}, i.e. 3.3 nucleons for the 1p3/2 and 1.8 nucleons for the 1s1/2 shell for  both protons and neutrons. The remaining 0.8 protons and neutrons, are ascribed to the non-discrete content of the spectral function, represented by a high missing-energy and momentum contribution.

For the final state, the  momentum distributions for the particle states in the continuum are computed  by solving the Dirac equation with the energy-dependent relativistic mean-field (ED-RMF) potential, which has no imaginary part so no flux is lost. The ED-RMF is the same potential used in the bound state but multiplied by a function that weakens it for increasing energy (for see details ref. \cite{Gonzalez-Jimenez:2019qhq,Nikolakopoulos:2019qcr}).
At low energies of the final nucleon (up to about 20 MeV in the continuum) the ED-RMF potential is the same potential employed to compute the momentum distributions for the initial bound nucleons and at larger energies it approaches the behavior of a phenomenological optical potentials. This ensures orthogonality between the initial and final states at these low energies and thus the reliability  of the cross-section computed in this region. 
In this way, the ED-RMF calculation also preserves the contribution (and resonant character)  from single-channel resonances  corresponding to excitations appearing slightly above the single-nucleon emission threshold of the 12C final system, which subsequently decay by single nucleon emission.

\subsubsection{Comparisons to ED-RMF  in the nuclear excitation region}
%
 \begin{figure*}
\includegraphics[width=7in, height=8.5in]{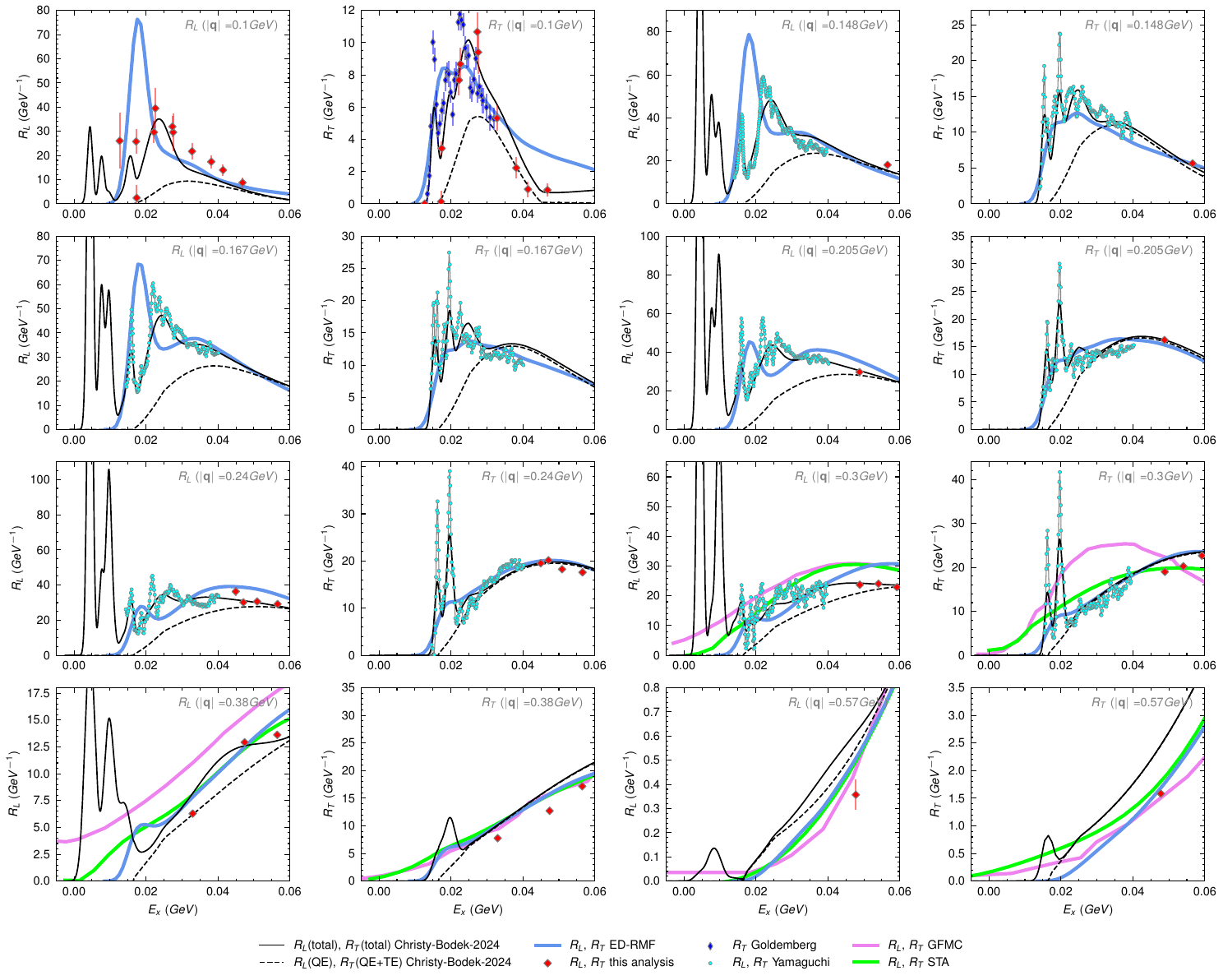}
\vspace{-10pt} 
\caption{ {\bf Nuclear excitation Region for fixed q bins vs $E_x$}: Comparison of measurements of \rltot and \rttot in the nuclear excitation region to the predictions of  ED-RMF ( blue lines),  GFMC ( pink lines) and STA-QMC ( light green lines).
The GFMC and STA-QMC  predictions are only available  for {\bf q}  values of 0.3, 0.38 and 0.57 GeV.}
\vspace{-10pt} 
\label{C_theory_Ex}
\end{figure*}
As seen in Fig. Fig.~\ref{ED-RMF_q3_A} - Fig.~\ref{ED-RMF_q3_B} the contribution of nuclear excitation to the response functions is significant for $\bf q$ values less than 0.3 GeV.  In this region,  for excitation energies above  20 MeV the ED-RMF predictions are in reasonable agreement (on average) with the measurements.   Fig. \ref{C_theory_Ex} is an expanded view of the nuclear excitation region, where the excitations of the nuclear targets can be seen above the
  QE contribution (the dashed black line).  Here \rltot and \rttot are shown versus excitation energy $E_x$. 
  The ED-RMF calculations require  one nucleon in the continuum in the final state, thus they show strength only above the one-nucleon separation energy (16 MeV for a proton and 18.7 MeV for a neutron in $\carbon$). 

Above the one-nucleon separation threshold and small energy transfers, where the nucleon in the continuum has less than 10 MeV, typically, the ED-RMF calculation exhibits single-channel resonances of the RMF potential employed to compute the continuum wave functions for the nucleon in the final state.   At these low energies of the nucleon, one expect these resonances of the potential to correspond to  actual resonances of the target nucleus, with the correct cross sections, if the potential is realistic enough, in a phenomenological sense. The ED-RMF calculations are performed with some moderate binning in energy, which contributes to the averaged smoothed look of the   ED-RMF predictions. As seen in  Fig. \ref{C_theory_Ex}  for $E_x$ above the nucleon separation energy the ED-RMF predictions in the nuclear excitation region for $\bf q$ values less than 0.3 GeV are reasonable (on average).  This indicates that the ED-RMF potential used is a reasonable approximation of the actual nuclear potential.  Since the same potential is used in the ED-RMF calculations of the response functions in neutrino scattering, the resulting response functions in the nuclear excitation region for neutrinos would probably also be a reasonable approximation to the actual nuclear excitations in neutrino scattering.  None of the other models used in neutrino MC generators to date include the contribution of  nuclear excitations.

As mentioned earlier, the  ED-RMF response functions incorporate the average contribution from nuclear excitations which decay via single nucleon emission. Decay modes which decay to other channels would not, however, contribute to ED-RMF responses. For example, decays with $\alpha$  particles, $\gamma$'s or deuterons in the final state are not included. 
 The ED-RMF predictions in the nuclear excitation region above 20 MeV are expected to be  reasonable because  the predominant decay modes of the nuclear excitations are to single nucleon final states. 
 
 For example,  the relative branching ratios for the decays of nuclear excitations between 12 and 21 MeV to protons and $\alpha$ particles have been measured\cite{Neuschaefer:1985zz}. The  decays of the  12.71 and 16.11 MeV excitations (which are below the proton and neutron separations energy)  are primarily via the  the emission of  $\alpha$ particles, while for the  20.62 MeV excitation the branching ratio to decay via proton emission is a factor of 2 higher than the decay via $\alpha$ emission.   
Note that  Christy-Bodek 2024 fit to inclusive electron scattering data describes all nuclear excitations irrespective of their decay modes.

 %

\begin{figure*}
\includegraphics[width=7.0 in, height=8.in] {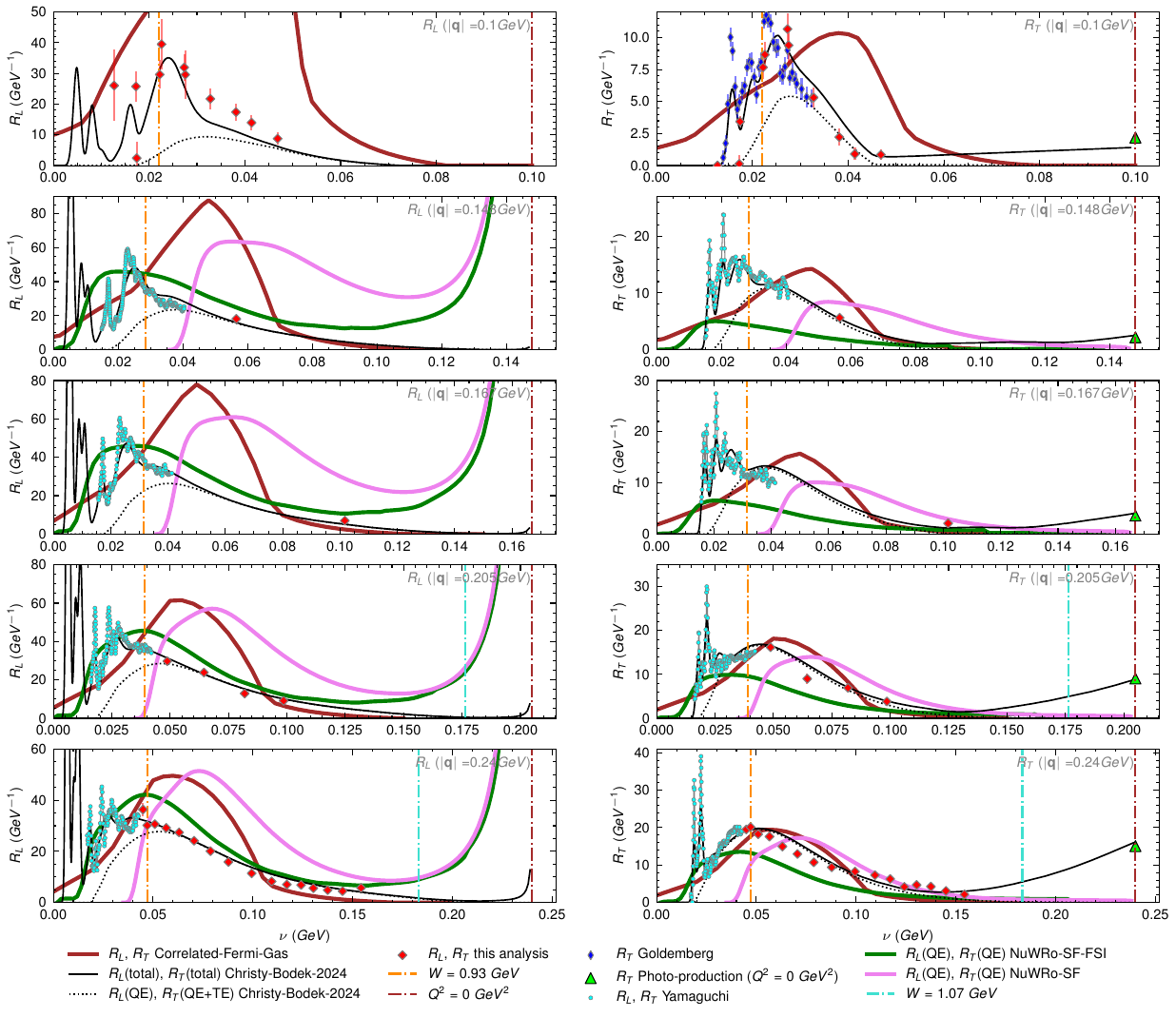}
\caption{ {\bf Comparison to CFG,  and NuWro for fixed q bins:}  
Our extractions of ${\cal R}_L$ and ${\cal R}_T$ for  $\bf q$ values of 0.10, 0.148, 0.167, 0.205, and  0.240 GeV versus $\nu$.  In the nuclear excitation region we  also show  measurements from Yamaguchi:1971~\cite{Yamaguchi:1971ua}, and  ${\cal R}_T({\bf q}=0.01)$ GeV extracted  from cross sections at 180$^{\circ}$ (Goldemberg:64~\cite{PhysRev.134.B963} and  Deforest:65~\cite{DEFOREST1965311}). 
The values of   ${\cal R}_T({\bf q}=\nu)$ are extracted from photo-absorption data.  The   universal fit for the total  (from all processes)  \rltot and \rttot is the solid black line and  the QE component  (including  Transverse Enhancement) of the universal fit  is the dotted line. 
The solid brown line is  the prediction  of the "Correlated Fermi Gas" (CFG)~\cite{Bhattacharya:2024win} calculation.The  solid pink line is the prediction of the  \nuwro{}  Monte Carlo Generator using a spectral function formalism (labeled \nuwro{}-SF) , and  the  solid green line is the predictions of \nuwro{}-SF and accounting for the effect of  final state interaction (labeled \nuwro{}-SF-FSI). (Note that \rttot in \nuwro{} is underestimated because the electron scattering mode does not include an MEC model). }
\label{RLRT_q3_A}
\end{figure*}

\begin{figure*}
\includegraphics[width=7.0 in, height=8.3in] {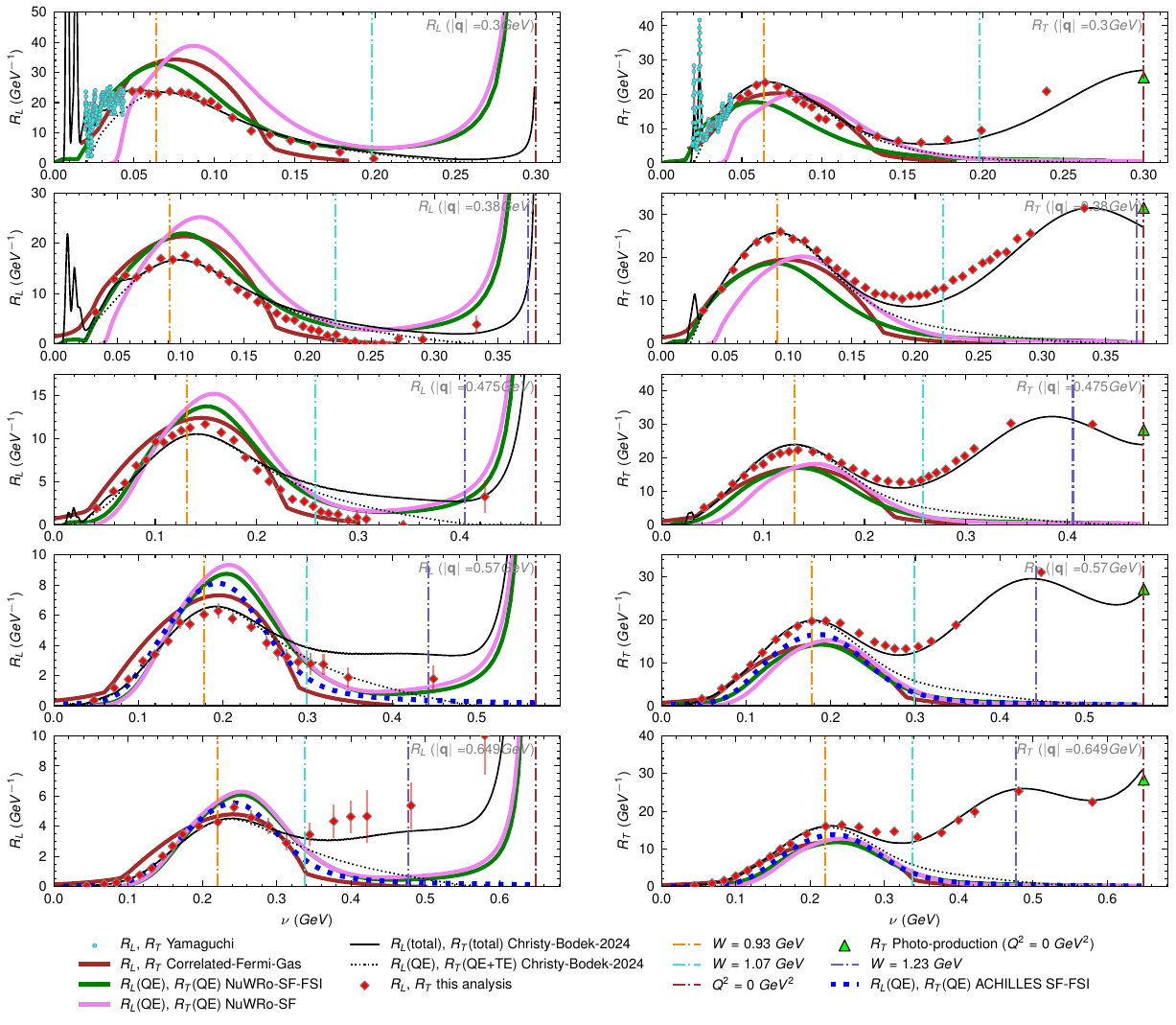}
\caption{{\bf Comparison to CFG, \nuwro{} and  \achilles~ for fixed q bins:}  Same as Fig. \ref{RLRT_q3_A} for ${\bf q}$ values of   0.30, 0.38, 0.475, 0.57, and 0.649 GeV versus $\nu$. 
 The the solid blue squares are the predictions of \achilles~ (for   $\bf q$ = 0.57 and 0.659 GeV). (Note that \rttot in \nuwro{} is underestimated because the electron scattering mode does not include an MEC model).}
\label{RLRT_q3_B}
\end{figure*}
%
\begin{figure*}
\includegraphics[width=7.0 in, height=8.3in] {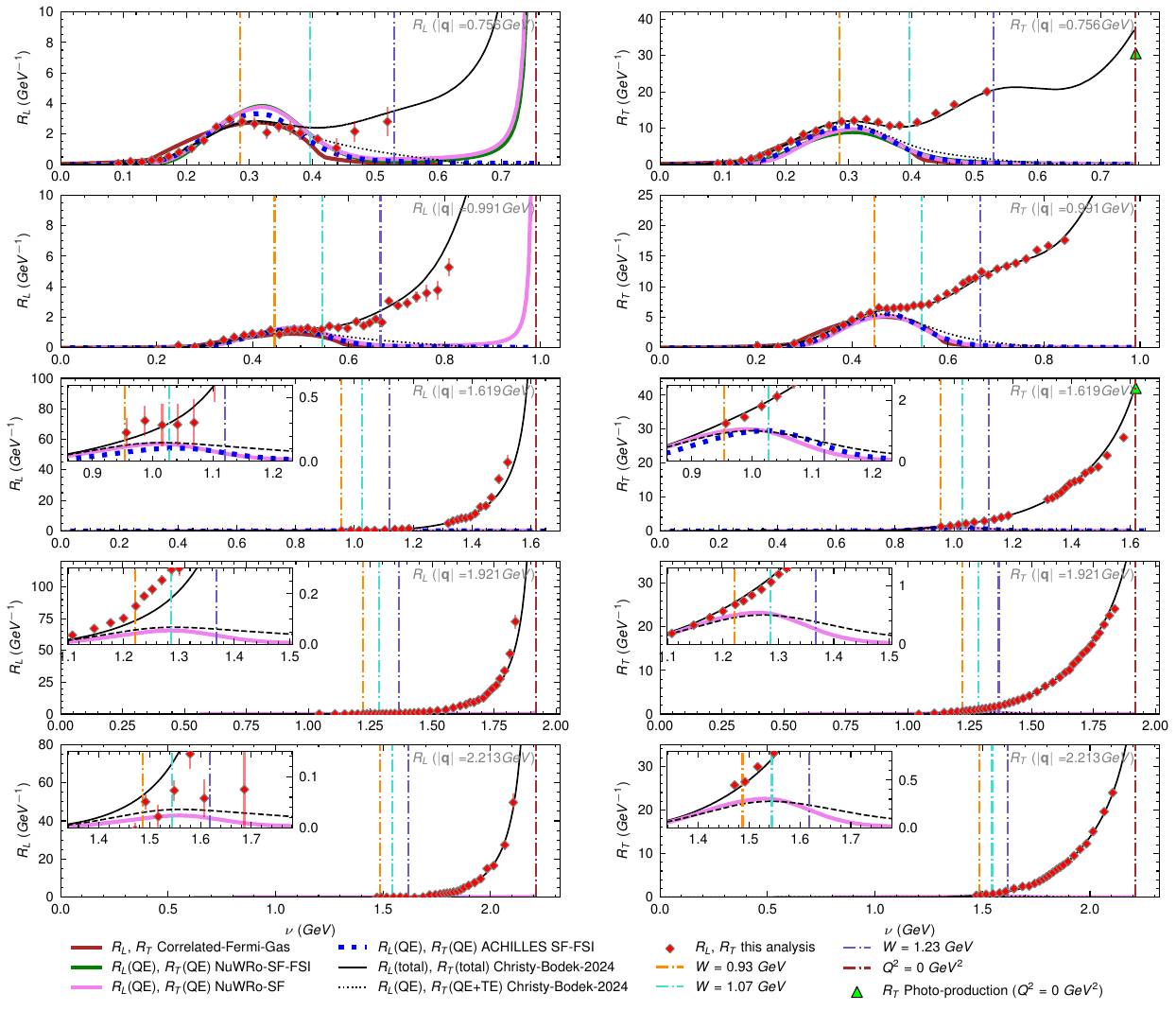}
 \caption{{\bf Comparison to CFG, NuWro and  ACHILLESS for fixed q bins:}  Same as Fig. \ref{RLRT_q3_A}  for $\bf q$ values  of 0.756, 0.991, 1.619 1.92,  and 2.213 GeV versus $\nu$. 
 The the solid blue squares are the predictions of \achilles~ (for   $\bf q$ = 0.756, 0.991 and 1.619 GeV).(Note that \rttot in \nuwro{} is underestimated because the electron scattering mode does not include an MEC model).}
\label{RLRT_q3_C}
\end{figure*}

\begin{figure*}
\includegraphics[width=7.0 in, height=7.in] {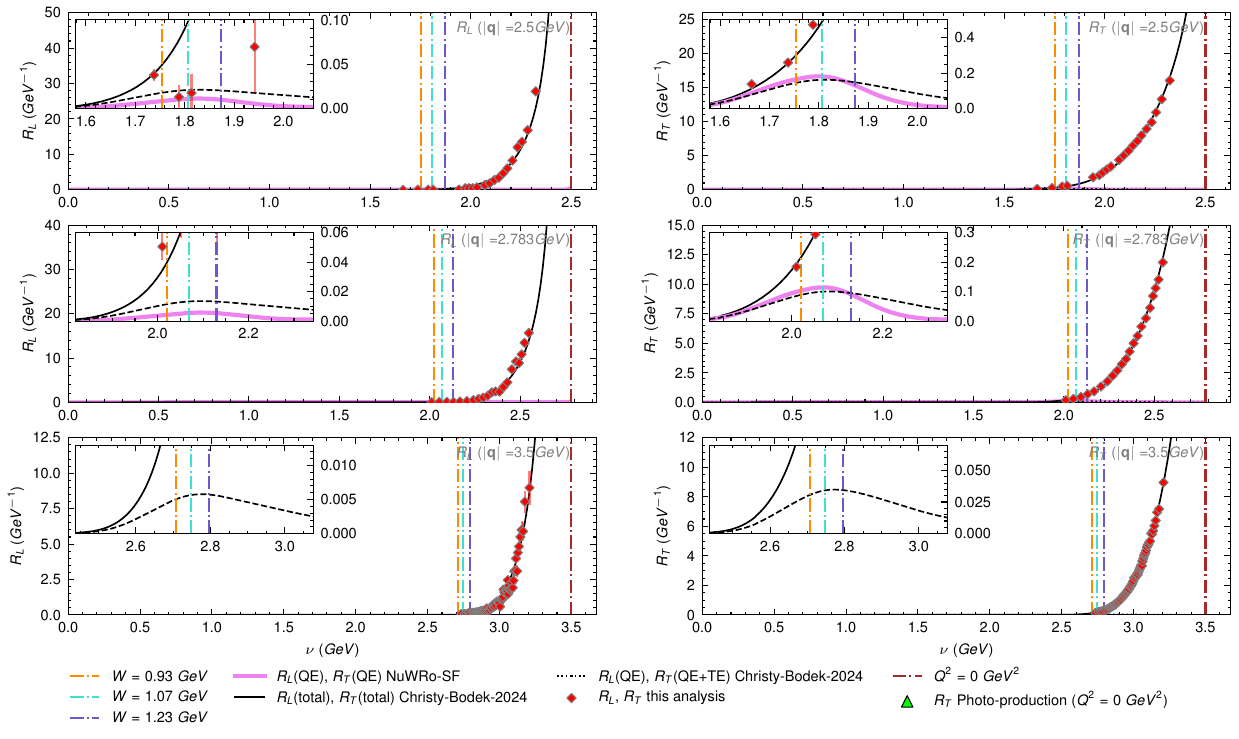}
 \caption{{\bf Comparison to  NuWro for fixed q bins:} Same as Fig. \ref{RLRT_q3_A}  for $\bf q$ values of $\bf q$ values of 2.5, 2.763,  and 3.5 GeV versus  $\nu$.  }
\label{RLRT_q3_D}
\end{figure*}

\subsection{Comparison to \nuwro{},  CFG and  ACHILLES}

In this section we compare the extracted \rltot and \rttot~for 18 fixed values of $\bf q$ (shown in Fig.\ref{RLRT_q3_A} - \ref{RLRT_q3_D}) 
to the predictions of the \nuwro{}-SF  (spectral function) calculation and   the predictions  of   \nuwro{}-SF-FSI (spectral function calculation with Final State Interaction). In addition, we compare the data to the predictions of the Correlated Fermi Gas (CFG), 
 and to the predictions of \achilles~(which are only valid for $\bf q>$0.5 GeV). Currently, the   \nuwro{} generator is being used in neutrino experiments,  and  \achilles~is being developed for use in neutrino experiments. Therefore, these two calculations also described in more detail below.

 \subsubsection{The  \nuwro{} Monte Carlo Generator}
\label{nuwro_sec}

The \nuwro{} Monte Carlo generator~\cite{NuWro} is designed to simulate neutrino-nucleus interactions in the few-GeV energy region, relevant for accelerator-based experiments. It has been primarily developed by the neutrino-theory group at the University of Wroc{\l}aw since 2004~\cite{Sobczyk:2004va,Juszczak:2005wk}.

To model lepton interactions with atomic nuclei, \nuwro{} assumes the validity of the impulse approximation, in which the process of scattering is assumed to involve predominantly a~single nucleon, with the remaining nucleons acting as a spectator system.

Simulating neutrino interactions with nucleons, \nuwro{} accounts for charged-current quasielastic (QE) scattering~\cite{LlewellynSmith:1971uhs}, hyperon production~\cite{Thorpe:2020tym}, single pion production~\cite{Sobczyk:2004va}, and deep-inelastic scattering~\cite{Bodek:2002vp}. In interactions with atomic nuclei, \nuwro{} also takes into account mechanisms involving two-body currents~\cite{Bonus:2020yrd} and coherent pion production~\cite{Berger:2008xs}.

In the electron mode, \nuwro{} currently simulates only QE scattering~\cite{Banerjee:2023hub}, with several options available to model nuclear effects. The results presented in this article are obtained within the spectral function (SF) approach~\cite{Benhar:2005dj} (without and with FSI).

\nuwro{} uses the carbon SF of Ref.~\cite{Benhar:1994hw}, which consistently combines the shell structure determined in coincidence electron-scattering experiments  at Saclay~\cite{Mougey:1976sc} with the results of theoretical calculations for infinite nuclear matter at different densities~\cite{Benhar:1989aw}, by employing the local-density approximation.

Happening inside the nucleus, the interaction between a lepton and a nucleon is affected by the spectator system. The surrounding nucleons interact with the struck nucleon, which modifies its energy spectrum and leads to a more complicated energy conservation in the vertex than for a~free nucleon. These effects of final-state interactions (FSI) induce a shift and a broadening of the double differential cross section in a kinematics-dependent manner. In \nuwro{}, they are implemented following Ref.~\cite{Ankowski:2014yfa}. In addition, the effect of Pauli blocking is accounted for by the action of the step function~\cite{Benhar:2005dj}, with the average Fermi momentum $\overline p_F=211$ MeV~\cite{Ankowski:2014yfa}.

In the context of the $^{12}{\rm C}(e,e')$ response functions, it is important to bear in mind that the electron mode of \nuwro{} does not account for mechanisms of interaction {\it other than  QE scattering induced by the one-body current}. The two-body currents  and interference between one-body and 2-body currents are responsible for the enhancement of the transverse response functions. 
Therefore, in electron scattering mode \nuwro{}  is expected to underestimate the transverse response function and overestimate the longitudinal response function in the QE region

In addition, in electron scattering mode  \nuwro{} does not include pion production processes.
Moreover, discrepancies with experimental data are expected to occur for the processes beyond the impulse approximation, such as nuclear excitations and the excitation of the giant dipole resonance, which are not included in the electron scattering or neutrino scattering modes in \nuwro{}.

  
%
\subsubsection{Comparison to   \nuwro{} ~predictions }
 As seen  in Fig. \ref{RLRT_q3_A}  and  \ref{RLRT_q3_D} 
  the predictions of   \nuwro{}  with FSI are in better agreement with the data than the prediction without FSI,  especially at low $\bf q$. (The effect of FSI above $\bf q$ of 0.65 GeV is  small). Therefore,  we focus on  comparisons  to  the predictions of \nuwro{}  with FSI  (\nuwro{}-SF--FSI, solid solid green curve).
 
 As mentioned above,   \nuwro{} 
in electron  scattering mode does not include a model for  the enhancement in  ${\cal R}_T$ from 2-body currents. Therefore,  as expected the predictions of \nuwro{}-SF-FSI  are lower than the data.

The predictions of \nuwro{}-SF-FSI for \rltot at low $\bf q$ are higher than the data, since they do not account for the quenching
of \rlqe at small values of $\bf q$.  However, since \nuwro{}-SF-FSI  also does not include a model for nuclear excitations, the overestimate of \rltot partially compensates for the missing nuclear excitations in the longitudinal channel. 

The direct comparison with the measured values of \rltot and \rttot as well as comparison with the Christy-Bodek 2024 universal fit
can be used to extract corrections (e.g. tuning) that will bring the predictions into better agreement with the data as discussed in section \ref{tuning_generators}.

  \subsection{ Comparison to the prediction of the Correlated Fermi Gas (CFG) }
  
The extraction of the Fermi momentum parameters in the  Relativistic Fermi Gas (RFG) model  from fits to  electron scattering cross sections   on various nuclei (at $\bf q$=0.45 GeV)  were published by Moniz and collaborators in 1971\cite{Moniz:1971mt}.  In 1981, Bodek and Ritchie\cite{Bodek:1980ar,Bodek:1981wr}  added the contribution of two-nucleon correlations to the model.  That  models was developed for analysis of  deep inelastic scattering on nuclear targets at high values of $Q^2$. 

The  recent Correlated Fermi Gas model~\cite{Bhattacharya:2024win} includes two-nucleon correlations in a more sophisticated way.   As seen  in Fig.\ref{RLRT_q3_A} - \ref{RLRT_q3_D} 
 the model  breaks down at low momentum transfers and works  somewhat better at higher momentum transfers.  Here also, the direct comparison with the measured values of \rltot and \rttot as well as comparison with the Christy-Bodek 2024 universal fit can be used to extract corrections (e.g. tuning) that will bring the predictions into better agreement with the data as discussed in section \ref{tuning_generators}.

   \subsection{Comparison to the predictions of  \achilles~ }
   \label{achilles_sec}
  
The   \achilles~ predictions\cite{Noah:2024}     for ${\bf q}$ values of  0.570, 0.649, 0.756, 0.991, and  1.659 GeV
 are shown as solid blues squares in  Fig.\ref{RLRT_q3_B} - \ref{RLRT_q3_C}..
 \subsubsection{The  \achilles~approach}
The   \achilles~ predictions  are computed  in the impulse approximation using a quantum Monte Carlo (QMC) based spectral function.   The contributions from 1-body currents and the interference between 1-body and 2-body currents, both leading to 1-nucleon knockout (1p1h) are included.
 These are computed using the $\rm ^{12}C$ QMC spectral function used in  
\cite{PhysRevC.107.L061301,Simons:2023,Lovato:2023raf,Lovato:2023khk} and are corrected for elastic FSI using a folding function. The interaction mechanisms  which are included contribute almost all of the strength in the QE peak and lower in $\nu$. Not included  are the contribution from two nucleon knockout by 2-body currents or pion production,  which contribute at higher energy transfer.  Pauli blocking is implemented as a simple fermi gas prescription, where the cross section is put to 0 if the final state nucleon momentum is less than 
$k_F$ ($ k_F$ = 225 MeV for $\rm ^{12}C$ is used). The \achilles~ predictions  are only valid for  $\bf q > 0.5$~GeV.
 \subsubsection{Comparison to  \achilles~predictions}
 As seen in Fig.\ref{RLRT_q3_B} and Fig.\ref{RLRT_q3_C} 
the \achilles~ predictions for \rltot and \rttot are  in better agreement with the data than the predictions of  \nuwro{}-SF-FSI.  For  \rttot  this is expected  because \achilles~ also include the contribution of 2-body currents and the interference between 1-body and 2-body currents leading to 
1-nucleon final states. The predictions for \rttot are lower  than the QE+TE component of Christy-Bodek 2024 fit because the transverse enhancement resulting in  2-nucleon final states (2p2h) are not included in the model.
%


\begin{figure*}
\includegraphics[width=7.0 in, height=9.in] {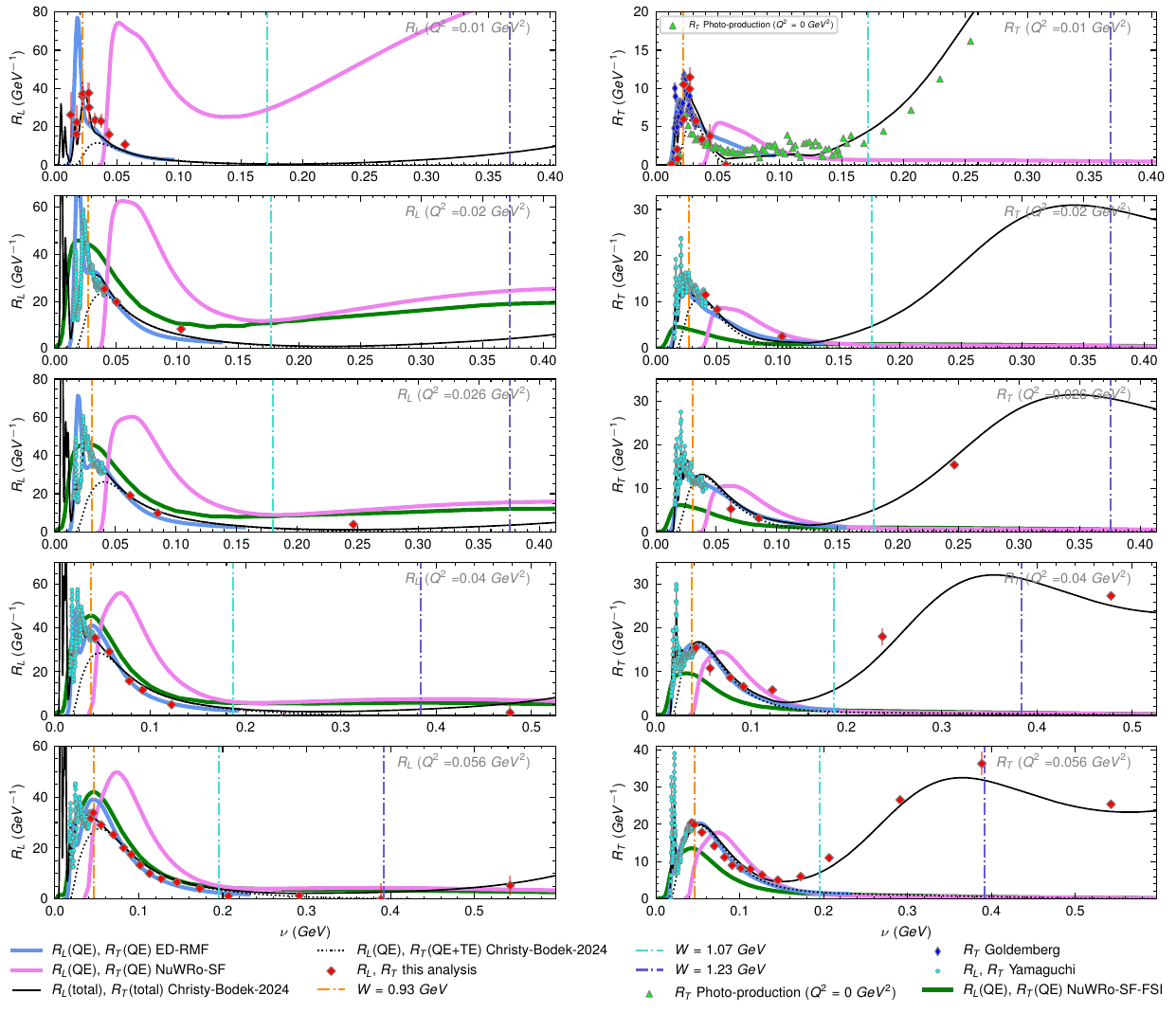}
\caption{ {\bf Comparison to ED-RMF and NuWro for fixed $Q^2$ bins:} Same as Fig. \ref{RLRT_q3_A}  for $Q^2$ values of 0.01 0.02, 0.026, 0.04 and 0.056  GeV$^2$ versus $\nu$. ( \rttot in \nuwro{} is underestimated because the electron scattering mode does not include an MEC model). }
\label{ED-RMF_Q2_A}
\end{figure*}
%
\begin{figure*}
\includegraphics[width=7. in, height=9.in] {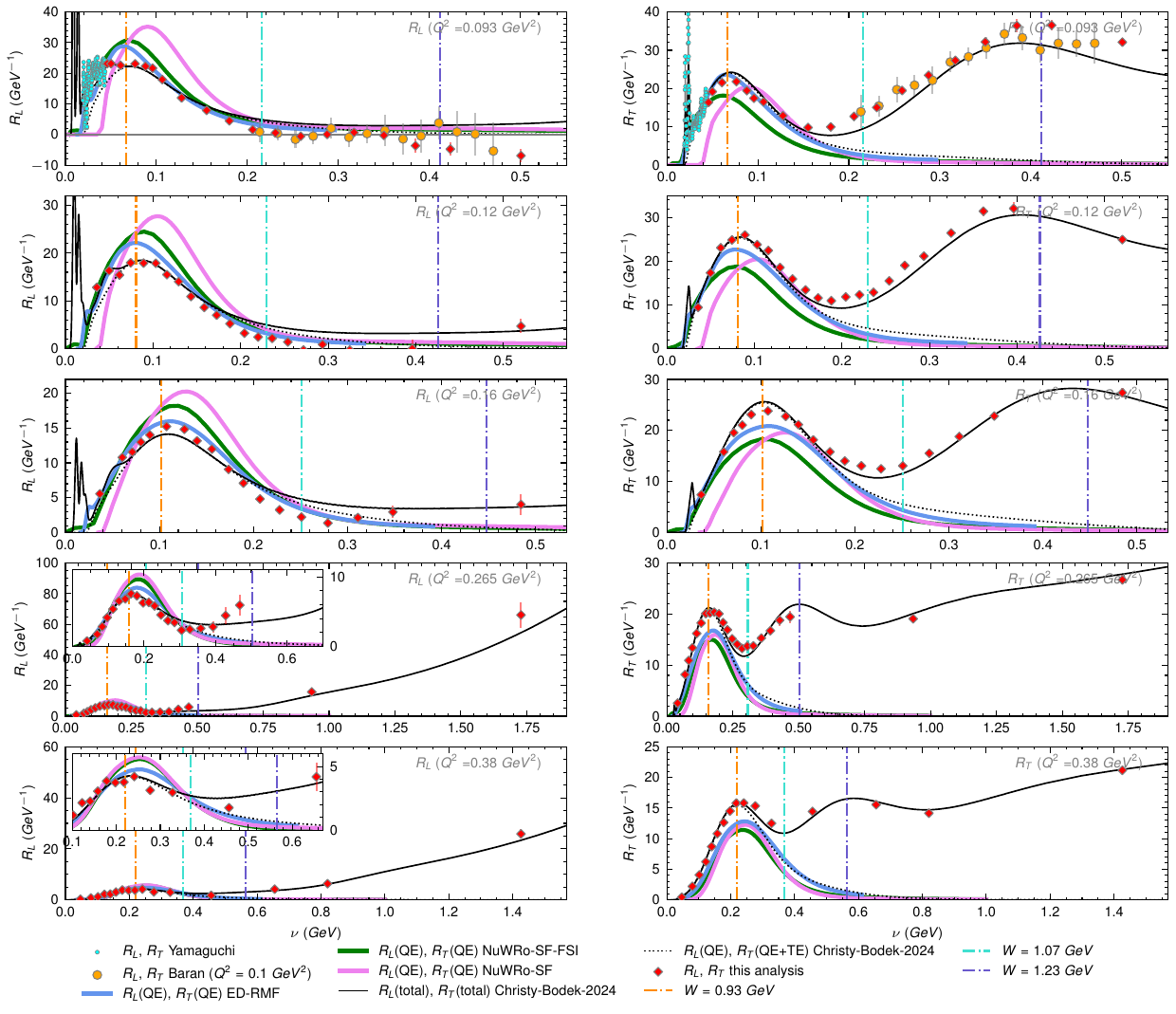}
\caption{ {\bf Comparison to ED-RMF and NuWro for fixed $Q^2$ bins:} Same as Fig. \ref{RLRT_q3_A} for $Q^2$ values of 0.093 0.12, 0.16, 0.265  and 0.38  GeV$^2$ versus $\nu$. (Note that \rttot in \nuwro{} is underestimated because the electron scattering mode does not include an MEC model). }
\label{ED-RMF_Q2_B}
\end{figure*}
%
\begin{figure*}
\includegraphics[width=7. in, height=9.in] {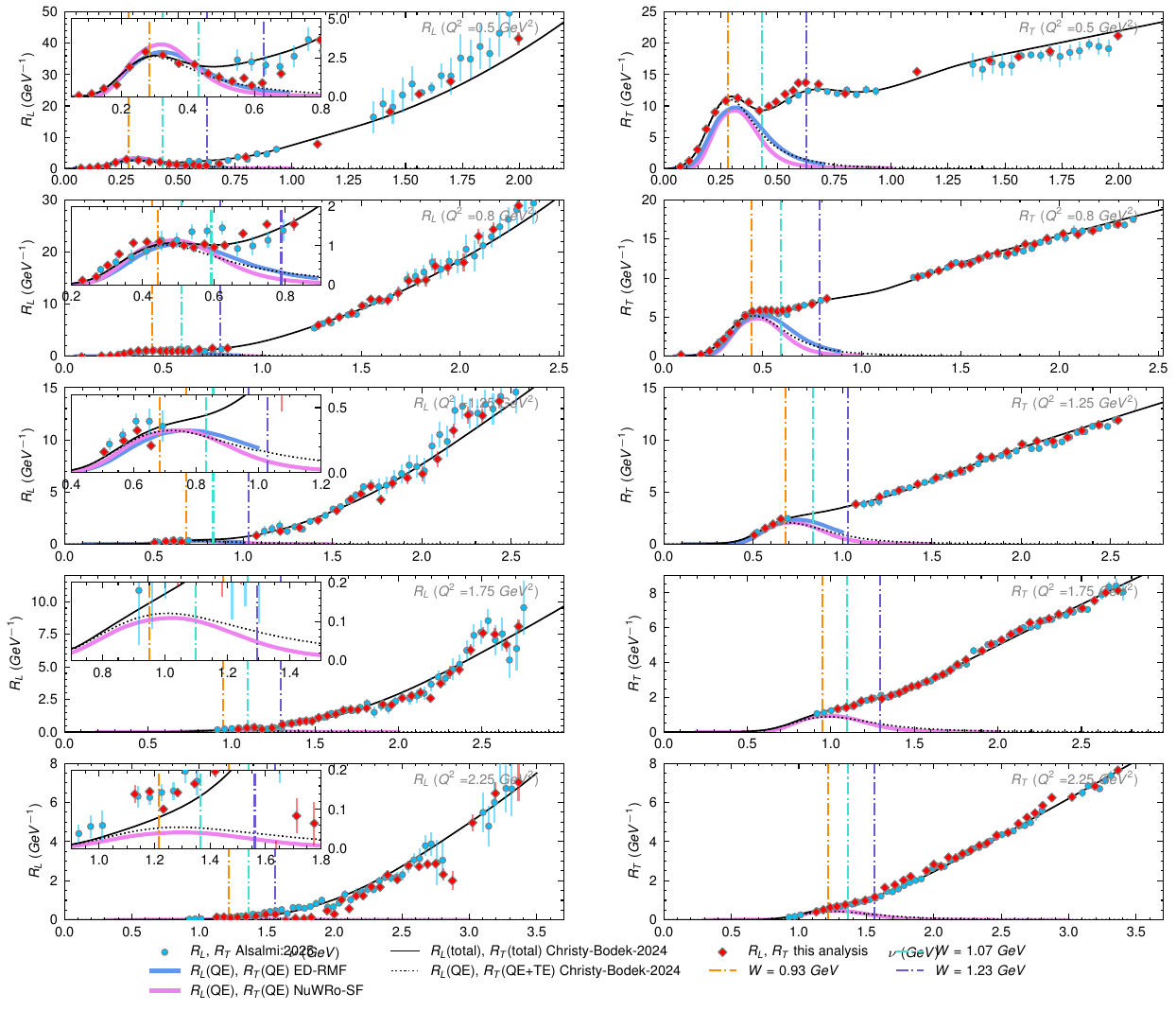}
 \caption{{\bf Comparison to ED-RMF and NuWro for fixed $Q^2$ bins:} Same as Fig. \ref{RLRT_q3_A}  for  $Q^2$ values of 0.50, 0.8, 1.25, 1.75 and 2.25  GeV$^2$ versus $\nu$. (Note that \rttot in \nuwro{} is underestimated because the electron scattering mode does not include an MEC model).  }
\label{ED-RMF_Q2_C}
\end{figure*}

\begin{figure*}
\includegraphics[width=7. in, height=5. in] {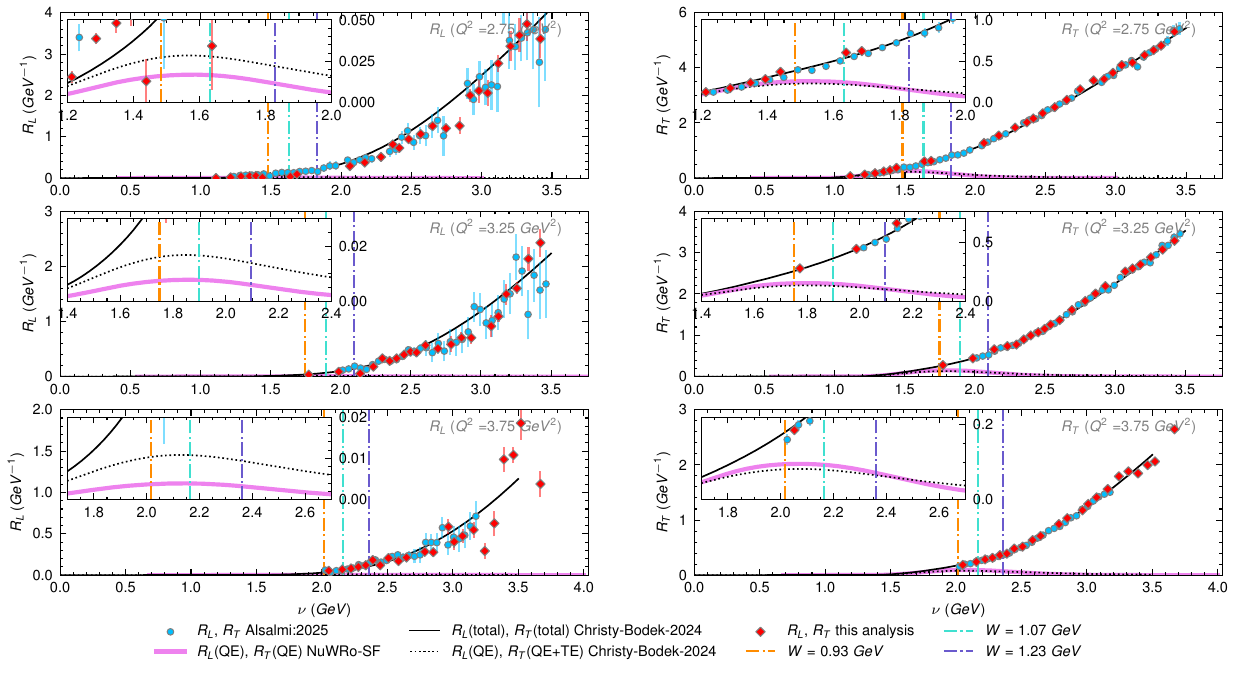}
 \caption{{\bf Comparison to ED-RMF for fixed $Q^2$ bins:} Same as Fig. \ref{RLRT_q3_A}  for  $Q^2$ values of  2.75, 3.25, and 3.75 GeV$^2$.  }
\label{ED-RMF_Q2_D}
\end{figure*}

 \begin{figure*}
\includegraphics[width=7in, height=8.5in]{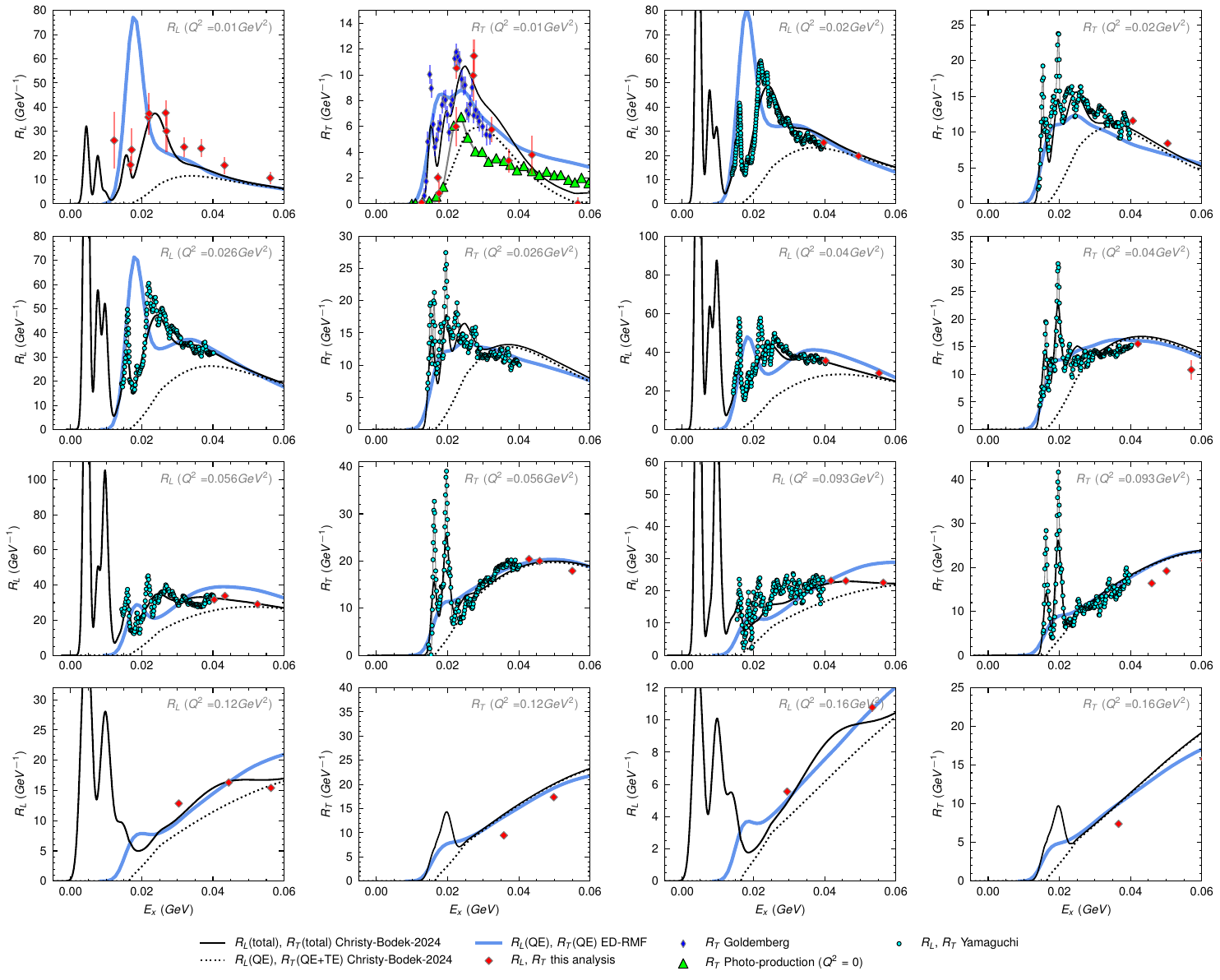}
\vspace{-10pt} 
\caption{{\bf Nuclear excitation Region for fixed $Q^2$ bins vs $E_x$}: Comparison of measurements of \rltot and \rttot  as a function of $E_x$  for  Q$^2$ values 
of 0.01, 0.026, 0.056, 0.12 GeV$^2$ to the predictions of  ED-RMF.  }
\vspace{-10pt} 
\label{C_theory_Ex_Q2}
\end{figure*}

\subsection{Comparisons of \rltot to the predictions of  \nuwro{}-SF-FSI  and ED-RMF  in  18 fixed  $Q^2$ bins}
In the previous sections  we compare  the extracted values of \rltot and \rttot to
the predictions of \nuwro{}-SF-FSI  and ED-RMF for fixed values of $\bf q$. Comparisons versus $\bf q$ are typically used to validate nuclear theory predictions. 

For neutrino scattering the validation
of MC generators at fixed values of $Q^2$ is more relevant.  Therefore, 
in  this  section we present the  comparison of   the extracted values of \rltot and \rttot
to the predictions of both  \nuwro{}-SF-FSI  and ED-RMF  18 fixed values of  $Q^2$  as
shown in Fig.\ref{ED-RMF_Q2_A} - \ref{ED-RMF_Q2_D}. 

As expected,  the    \nuwro{}  predictions for \rttot 
are underestimated   because  the electron mode of \nuwro{}  
does not include the  contribution of Meson Exchange Currents 
(which are included in the neutrino version of \nuwro{}). 

At low $Q^2$ we find that the  \nuwro{}-SF-FSI predictions  are in better
agreement with the data than the predictions of  \nuwro{}-SF. 
At high $Q^2$ we find that  there is little difference between
 the predictions of   \nuwro{}-SF-FSI  and \nuwro{}-SF.  Therefore, we only show the predictions of
  \nuwro{}-SF-FSI at low $Q^2$ and  \nuwro{}-SF at high $Q^2$.

Fig. \ref{C_theory_Ex_Q2} is an expanded view of the nuclear excitation region, where the excitations of the nuclear targets can be seen above the
  QE contribution (the dashed black line).  Here the response functions are shown versus excitation energy $E_x$. This figure is the same as  Fig. \ref{C_theory_Ex} except that it is for bins in fixed $Q^2$ instead of bins in fixed {\bf q}.
  
As seen in Fig.\ref{ED-RMF_Q2_A} - \ref{C_theory_Ex_Q2} the ED-RMF predictions 
for \rltot are in much better agreement with experiment than the  \nuwro{}-SF-FSI
predictions  especially at low $Q^2$ where  the contribution from nuclear excitations is significant.

    \section{Comparison to the  improved  superscaling model SuSAv2}   
   The improved  superscaling model  (SuSAv2) is described in detail in  references \cite{Megias:2016lke,Gonzalez-Rosa:2022ltp,Gonzalez-Rosa:2023aim}.    
    Among the models that we have investigated, the recently improved  SuSAv2 is the only model  that includes QE, 2p2h-MEC, and pion production processes  including resonance production, 
     Shallow Inelastic Scattering (SIS) and Deep Inelastic  (DIS) processes, 
    (but does not include nuclear excitations).  In this publication we  focus on the comparison of the  SuSAv2 response functions   for  the  QE-1p1h  and 2p2h-MEC processes.  Investigation of pion production processes will be reported in a future publication.
    
    The SuSAv2 predictions were  compared  to few  electron scattering cross section measurement in  references \cite{Megias:2016lke,Gonzalez-Rosa:2022ltp,Gonzalez-Rosa:2023aim}. Here we present the  first comparison of the SuSAv2 predictions  to measurements of   \rltot and \rttot  spanning the complete kinematic range of $Q^2$ and $\nu$ of interest to neutrino experiments. These comparisons are shown for the 18 values of fixed  $\bf q$ (Figures \ref{SuSAv2_q3_A}-\ref{SuSAv2_q3_D}) and the 18 values of fixed  $Q^2$  (Figures \ref{SuSAv2_Q2_A}-\ref{SuSAv2_Q2_D}) in Appendix A. 
    
     In all the plots the  universal fit for the total  (from all processes)  \rltot and \rttot is the solid black line and  the QE component  (including  Transverse Enhancement) of the universal fit  is the dotted line. The  predictions of SuSAv2 are the dashed   pink 
lines for  QE-1p1h, and solid  pink lines for  the sum of  QE-1p1h  and MEC-2p2h processes. The  blue lines are the predictions of ED-RMF~\cite{Franco-Munoz:2022jcl,Franco-Munoz:2023zoa}  for  QE-1p1h (including nuclear excitations).   Since ED-RMF does not account for 2p2h final states, we also show the sum of the ED-RMF prediction for QE-1p1h  and the SuSAv2  MEC-2p2h model  to account for  the 2p2h contribution to the response functions  (solid green line).

\subsection{SuSAv2 QE(1p1h)}
    In the original SuSA model, which is  based on a fit to the  electron longitudinal  response functions, the longitudinal and transverse  scaling functions are equal, thus resulting in an underestimation in the  transverse  channel.  In the improved  SuSAv2 the  Relativistic Mean Field (RMF)  approach is used  to obtain the theoretical longitudinal and transverse  scaling functions.  The SuSAv2 RMF transverse  scaling function is somewhat  larger than the RMF  longitudinal scaling function, thus resulting in a transverse enhancement of the QE-1p1h cross section.  However differences between SuSAv2-QE-1p1h  and ED-RMF-QE-1p1h are expected at  small  {\bf q}, $\nu$  and $Q^2$  where scaling violations present in  the  RMF theory (and also nuclear excitations) as well as the effect of 2-body currents leading to 1p1h in the ED-RMF approach are not fully incorporated in the SuSAv2 approach. 
    
    We find that  SuSAv2-QE-1p1h predictions for  \rlqe  (dashed pink lines) are  somewhat  higher than our measurements  (especially at small {\bf q}), and the predictions for \rtqe are  somewhat  lower than our measurements (especially at small {\bf q}). Although ED-RMF and SuSAv2 cross sections are similar apart from the very low-energy region, some important differences appear when studying the \rlqe  and \rtqe channels separately. These differences can be mainly ascribed to the lack of 2-body currents leading to 1p1h in the SusAv2 approach.      When comparing theory predictions to experimental electron scattering cross sections, such differences may be hard to observe because an   overestimation of \rlqe  can be partially compensated by an underestimation of \rtqe unless the comparisons include  both cross section data at very small angles and cross section data at very large angles.
    
     A more significant issue is that  for low $\bf q$ and low $Q^2$ ($\bf q <$ 0.3 GeV and $Q^2<$ 0.1 GeV$^2$) the  SuSAv2-QE-1p1h predictions for   \rlqe and  \rtqe are shifted to low values of $\nu$  which results in unphysical cross sections (at negative excitation energies)  at small  $\nu=0$. 
     
     In summary,, additional  corrections and tuning of  the SuSAv2-QE-1p1h  calculations are needed to achieve better agreement with our measurements.

\subsection{SuSAv2  MEC-2p2h}

In the Meson-Exchange-Current 2p2h model (MEC-2p2h)  incorporated in SuSAv2  the main contribution in  electron scattering is transverse and the contribution in the longitudinal channel is small.  This is not the case for neutrinos where the 2p2h longitudinal  axial contribution is important.  In electron scattering the  enhancement in the QE peak region originating from the 1p1h RMF transverse enhancement is in general bigger than the 2p2h contribution. The  2p2h channel becomes more important in the dip region between the QE and $\Delta$(1238) peaks.  The  sums of  the predictions of SuSAv2-QE-1p1h and  SuSAv2--MEC-2p2h are shown as the solid pink lines in the plots in Appendix A.  In general
the inclusion of the 2p2h contribution results in better agreement with our measurements in the dip region. However, at  high values of {\bf q}  and $Q^2$ 
(e.g. $\bf q >$1.6 GeV and $Q^2 >$ 1.25 GeV$^2$)
 the  SuSAv2 MEC-2p2h model predictions for the {\it longitudinal} response functions have unphysical peaks.  
 This could be because the  parameterization of the 2p2h RFG-based nuclear responses from Ruiz-Simo et al.\cite{RuizSimo:2016rtu} 
 included in this implementation of the SuSAv2-MEC model is valid only for  ${\bf q}< 2$ GeV and $\nu<2$ GeV (although it can be extended to higher kinematics).

Note that the   SuSAv2-MEC-2p2h implemented  in  \genie~  uses interpolation methods and therefore  is  more accurate than the parameterization formulae used in this implementation. This can introduce important differences at low kinematics (${\bf q}<$ 0.3 GeV) and also at very high kinematics (${\bf q} >$1.5 GeV).

\subsection{Resonance }

  Since
the ED-RMF-QE-1p1h predictions are in better agreement with our measurements than the predictions of  SuSAv2-QE-1p1h, we also show the  sum of  the predictions of  ED-RMF-QE-1p1h and  the predictions of  SuSAv2-MEC-2p2h  as the solid {\it green}  lines in the plots in Appendix A.  Combining the ED-RMF-QE-1p1h predictions with the SuSAv2-MEC-2p2h  predictions provides a better description of the response function for  both single nucleon (from QE and nuclear excitations) and two nucleon final states.  
 \begin{table*} [tbh]
\begin{center}
\footnotesize
\begin{tabular}{|c|c|c|c|c|c|c|c|     } \hline
 Model	&	Currents	&	Final State	&	Available for 	&	RL	&	RT	&	Large $\bf q$	& Small $\bf q$\\
   \hline 
                  ED-RMF	&	1b,2b	&	1p1h, nucl. exci.	&	$all ~\bf q$	&	best 1p1h  	&	best 1p1h 	&	best 1p1h & best 1p1h\\	
  QMC       &          & 1b+2b &  {\bf best model} &  2p2h small  &   need 2p2h     & need 2p2h  & need 2p2h\\
                  &          & no 2p2h & {\bf for 1p1h}  &  {\bf use SuSAv2}  &   {\bf use SuSAv2}&{\bf  use SuSAv2}&{\bf use SuSAv2}\\  \hline \hline
                        SuSAv2	&	1p1h	&	1p1h  &	$all ~\bf q$	& low $\nu$ unphysical  	& low $\nu$ unphysical	&  OK		& {\bf needs RL }  \\	
RMF Scaling
      &  only 1b       & only 1b&1p1h no 1b-2b & low {\bf q} model-high  & low {\bf q}  model-low     & &{\bf quench}\\  
        function                   & 2p2h        &  no nucl exc. & interference & {\bf needs 1p1h}& {\bf needs 1p1h }       &  & {\bf needs RT  }\\ 
                          &  1b+2b     &{\bf 2p2h OK} & {\bf 2p2h OK} & {\bf quench or}& {\bf enhancement }       &  & {\bf enhancement }\\ 
                                                    &      & &  &use ED-RMF?& use ED-RMF?      &  & use ED-RMF?\\ \hline \hline
  %
                  STA-QMC	&	1b,2b	&	1p1h + 2p2h.	&	$0.3 \le {\bf q}< \le .65$	&	OK  	&	OK	&	relativistic 	&analytic extrapol.  \\	
         &          &  no nucl exc. &  & $0.3 \le {\bf q}< \le .65$ &  $0.3 \le {\bf q} \le 0.65$     & corr. needed  & needed\\  \hline \hline
      \nuwro{} SF-FSI  &1b        & 1p1h & all {\bf q} & RL   high   &RT  low           & needs  & needs\\  
  e-mode  spectral &          & no nucl exc.   & & needs   &   needs       &  2p2h& 2p2h \\  
 function     &          & no 2p2h  &  & quenching   &     enhancement     &model & model\\ \hline
                          \achilles~ 	&	1b,2b	&	1p1h 	&	${\bf q}>$0.5	&	OK  	&	OK	&needs& need other\\	
  QMC   spectral   &          &  no 2p2h. &   & need &   need    &  2p2h model &low {\bf q}  models \\
   function    &          &  no nucl exc. &   & 2p2h model  &   2p2h model     & for RL RT&for 1p1h, 2p2h \\  \hline       \hline
           GFMC	&	1b,2b	&	1p1h+ 2p2h	&	$0.3 \le {\bf q} \le 0.57$	&	low  $\nu$ unphysical  	&	OK	&	CPU intensive	&CPU intensive \\	
  QMC      &          & no nucl. exc. & &  {\bf q}=0.57 high &   $0.3 \le {\bf q}< \le .57$    & {\bf not possible} & {\bf not possible }\\  \hline
   CFG   &1b         & 1p1h+2p2h  & all {\bf q} &   poor   & poor          & poor  & poor\\  
  Correlated       &         &no nucl. exc.&  & agreement   &agreement          & agreement & agreement \\ 
   Fermi Gas      &         & &  &    &       &  & \\ 
\hline
\end{tabular}
\caption{A summary of comparisons  of $\carbon$ \rltot and \rttot to theoretical  predictions ({\bf q} units are in GeV).}
\label{summary}
\end{center}
\end{table*}  
\label{tuning_generators}
Because our extractions of \rltot and \rttot from all available data electron scattering data span a large range of $\bf q$ and $Q^2$,  the comparisons can be used extract  $\bf q$ dependent corrections  to nuclear models and   neutrino/electron MC generators to achieve better agreement our measurement s (as has been done in the Christy-Bodek  2024 fit).  For example,
in the QE and nuclear excitation region:  
\begin{enumerate} 
 \item  The locations of the QE  in $\nu$ are sensitive to the removal energy (energy shift) which shifts the QE peak to higher values of $\nu$)  and also a $\bf q$ dependent   final state optical potential~\cite{Bodek:2018lmc}  which shifts the QE  peak to lower values of $\nu$ at low $\bf q$. Tuning these parameters in  SuSAv2 would bring the position of the QE peaks into better agreement with  the measurements at  low $\bf q$ and $Q^2$.   In addition,  introducing a {\bf q}-dependence in the RMF-based SuSAv2 scaling functions (which are roughly constant for ${\bf q}>$ 0.3 GeV, but not for smaller values
 of {\bf q}) can be used to incorporate into SuSAv2 the  the low-energy  RMF nuclear effects that break the scaling behavior at low {\bf q}, which will modify the shape and  magnitude of the SuSAv2  QE predictions at that region.
  %
  \item  The  magnitude of \rltot  can tuned be via  a $\bf q$ dependent longitudinal quenching factor,  and
 the amplitude of \rttot can be tuned via an $\bf q$ dependent additive transverse enhancement contribution  (as done in the Christy-Bodek 2024 fit\cite{Bodek:2022gli}). Including and tuning such  parameters in SuSAv2  would bring the magnitude of \rlqe and \rtqe predictions  into better agreement with the measurements.
  \item  In  ED-RMF  nuclear excitations  (which decay to single nucleons in the final state) occur via the  interactions of final state nucleons with the Energy Dependent Mean Field.  In the other calculations,  a  
model describing nuclear excitations (with excitation energies {\it above} the proton separation energy) can be added. 
  \item  A model describing nuclear excitations (with excitation energies {\it below} the proton separation energy) which decay via $\gamma$ ray emission is
  included in the inclusive Christy-Bodek 2024 fit~\cite{Bodek:2023dsr}. At present none of the theoretical predications include nuclear excitations  which decay via $\gamma$ emission.
  \end{enumerate}


%
  \section{Summary}
  
We have done a  global  extraction of the ${\rm ^{12}C}$ {\it longitudinal}   (${\cal R}_L$) and {\it transverse} (${\cal R}_T$) nuclear electromagnetic response functions from  an analysis of all available inclusive electron scattering cross section on $^{12}C$. The response functions are extracted for a large range of energy transfer $\nu$,  spanning the nuclear excitation, quasielastic,  resonance and inelastic continuum and over a large range of $Q^2$. We extract  ${\cal R}_L$ and ${\cal R}_T$ as a  function of $\nu$ for fixed values of both $Q^2$ (0 $\le Q^2\le3.5$ GeV$^2$) and  $\bf q$ (0.1 $\le {\bf q} \le3.75$ GeV).

Since  the extracted response functions cover a large range of  $Q^2$ and $\nu$, they can be readily used to validate nuclear models as well as  validate and tune Monte Carlo generators for electron and neutrino scattering experiments.   Alternatively, the Christy-Bodek 2024  global fit to all electron scattering data on $\carbon$ can also be used to validate nuclear models and also validate and tune Monte Carlo generators for electron and neutrino scattering experiments over a larger region of $\nu$.  In this paper we focus on  comparison to models in the nuclear excitation region,  and for  single nucleon  (QE-1p1h) and two nucleon  (2p2h) final states.

Because the mass of the muon is much larger than the mass of the electron there are kinematic regions in electron-neutrino charged current scattering which are not accessible in the scattering of muon-neutrinos.  Therefore, in neutrino oscillation experiments  {\it the modeling of the cross sections at low momentum transfer and low energy transfer is important for accounting for the  difference between  muon-neutrino and electron-neutrino cross sections.} This is the region where the contribution of  nuclear excitations is significant. At present, only ED-RMF accounts for nuclear excitations which decay to  a single nucleon in the  final state (but not excitations which decay via gamma emission).

We have compared measurements in the QE region to the predictions of several theoretical calculations as summarized in Table \ref{summary}.  In particular  ED-RMF,  STA-QMC  and SuSAv2  have the added benefit that the calculations are also  directly applicable to the same kinematic regions for neutrino scattering.  We find that the SuSAv2-QE-1p1h predictions for  \rlqe   are  somewhat higher than our measurements and the predictions for \rtqe are a  somewhat  lower than our measurements.  In addition, at small momentum transfers the SuSAv2-QE-1p1h  predictions for  \rlqe and  \rtqe are shifted to lower values of $\nu$  which result in unphysical cross sections at small  $\nu$.  These differences can be remedied via the introduction of additional tuning parameters in the theory.

Among all the QE theoretical predictions,   ED-RMF  provides  the best  description of  \rltot and \rttot for QE scattering (for 1p1h final states)  over the largest kinematic range in $Q^2$ and $\nu$.  These predictions (which are available at  all values of $\bf q$) include contributions to 1p1h final states from QE scattering as well as from nuclear excitations. 
 At higher values of $\bf q$ and higher values of $\nu$ the ED-RMF predictions for \rttot are below the data. This is expected because  the  ED-RMF  theoretical calculations for  \rttot do  not include processes with two nucleons  or pions in  the final states. Combining the ED-RMF-QE-1p1h  predictions with the SuSAv2-MEC-2p2h  predictions provides  a better description  of both single nucleon (from QE and nuclear excitations) and two nucleon final states.  In our next publication (in which the resonance region will be investigated) we plan to include  ED-RMF-1-pion production model.
 
 The STA-QMC calculation includes both single and two nucleon final states but  is  valid over a  more restricted kinematic range and does not account for nuclear excitations.  Extension of the theory to higher and lower values of momentum transfer are currently under way.

 
In future communications we plan to report on the comparisons  of response functions (including the nucleon resonance region) for  $\rm ^{12}C$,  $\rm ^{40}Ca$   and  $\rm ^{56}Fe$ to the predictions of   ED-RMF, SuSAv2 and the {{\sc{genie}}} MC generator.  


    The extracted  ${\cal R}_L$ and ${\cal R}_T$ values (shown in the figures) are available as supplemental materials~\cite{supp}. In addition, the values of 
    the Christy-Bodek 2024  fit to  ${\cal R}_L$ and ${\cal R}_T$ over a larger range of $\nu$ (and also listing separately the contributions of nuclear excitations, quasielastic scattering, transverse enhancement, and pion production processes) are available as supplemental materials.
    
      \section{Acknowledgments}
      We thank Noah A. Steinberg for providing the predictions from the \achilles~ MC generator.  We thank Lorenzo Andreoli for providing the predictions from STA-QMC and Saori Pastore for clarification of the predictions of STA-QMC.
      We thank Sam Carey for providing the predictions from CFG.  We thank Tania Franco Munoz for providing the predictions from ED-RMF,  and  Jose Manuel Udias and Raul Gonzalez-Jimenez   for clarification of the predictions of ED-RMF.
      
 Research supported  in part by   the Office of Science, Office of Nuclear Physics under contract DE-AC05-06OR23177 (Jefferson Lab) and by  the U.S. Department of Energy under University of Rochester grant number DE-SC0008475.  The work of A. M. Ankowski is partly supported by the Polish National Science Centre under grant UMO-2021/41/B/ST2/02778. The work of Julia Tena Vidal is supported by a European Research Council ERC grant (NeutrinoNuclei, 101078772). 
G. D. Megias' work is supported by the Spanish Ministerio de Ciencia, Innovación y Universidades and ERDF (European Regional Development Fund) under contracts PID2020-114687GB-100 and PID2023-146401NB-I00, by the European Union’s Horizon 2020 Research and Innovation Programme under the  HORIZON-MSCA-2023-SE-01 JENNIFER3 grant agreement no.101183137, by the Junta de Andalucia (grant No.~FQM160) and by University of Tokyo ICRR’s Inter-University Research Program FY2024 (Ref.~2024i-J-001)
 

 \begin{appendices}
  
 \section{Comparison to SuSAv2 in bins of  $\bf q$ and $Q^2$}

\begin{figure*}
\includegraphics[width=7.0 in, height=8.in] {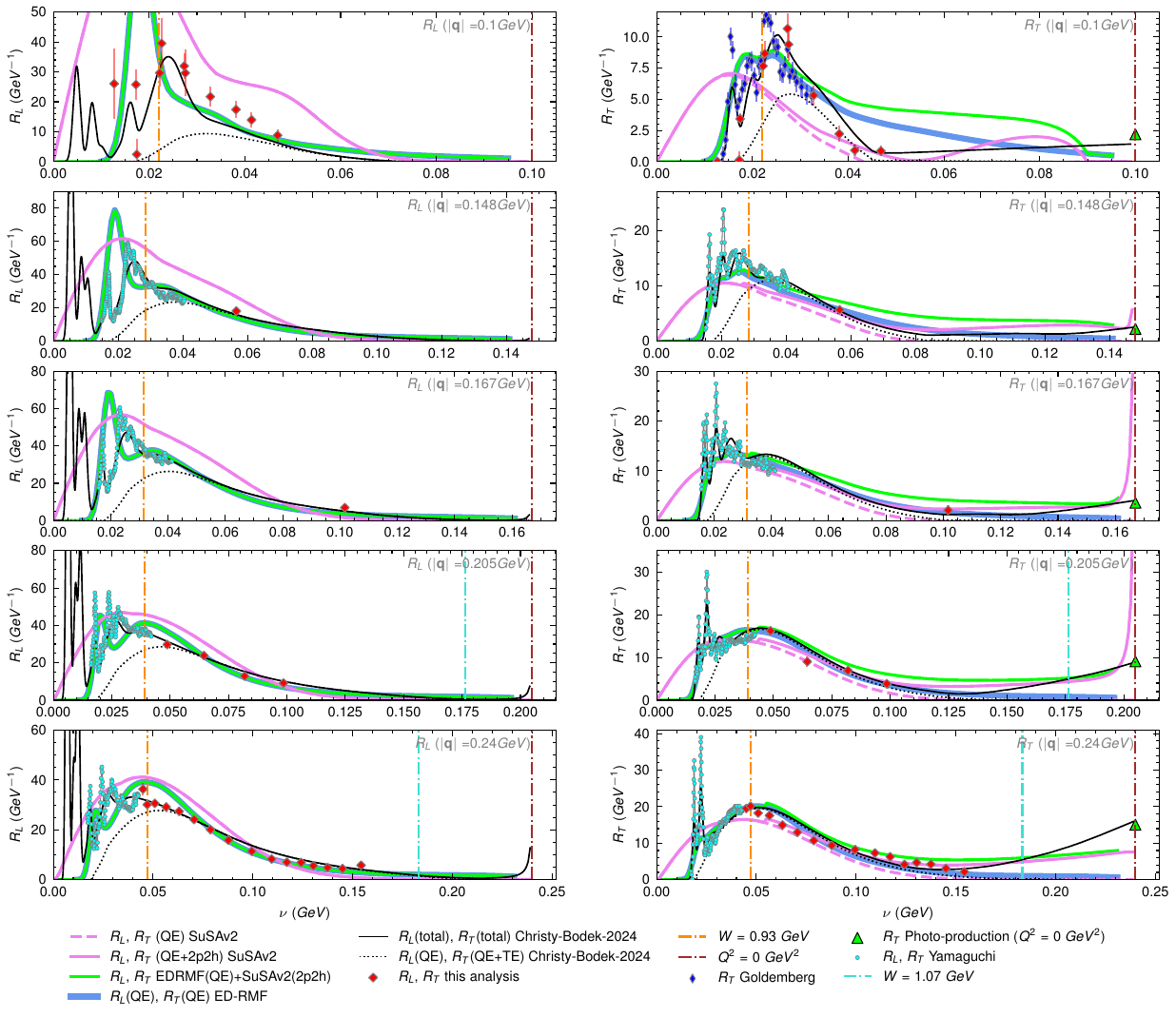}
\caption{{\bf Comparison to ED-RMF and  SuSAv2 for fixed bins in q}: Our extractions of ${\cal R}_L$ and ${\cal R}_T$ for ${\rm ^{12}C}$ for  $\bf q$ values of 0.10, 0.148, 0.167, 0.205 and  0.240 GeV versus $\nu$. In all the plots the  universal fit for the total  (from all processes)  \rltot and \rttot is the solid black line and  the QE component  (including  Transverse Enhancement) of the universal fit  is the dotted line. The  predictions of SuSAv2 \cite{Megias:2016lke,Gonzalez-Rosa:2022ltp,Gonzalez-Rosa:2023aim}  are the dashed   pink 
line for  QE-1p1h, and solid  pink lines for the sum of QE-1p1h and MEC-2p2h processes.  The  blue lines are the predictions of the ED-RMF~\cite{Franco-Munoz:2022jcl,Franco-Munoz:2023zoa}  (QE-1p1h including nuclear excitations).   Since ED-RMF does not account for 2p2h final states, we also show the sum of the ED-RMF prediction for QE-1p1h  and the SuSAv2  MEC model prediction for the 2p2h contribution  (solid green line).
 }
\label{SuSAv2_q3_A}
\end{figure*}

\begin{figure*}
\includegraphics[width=7.0 in, height=8.3in] {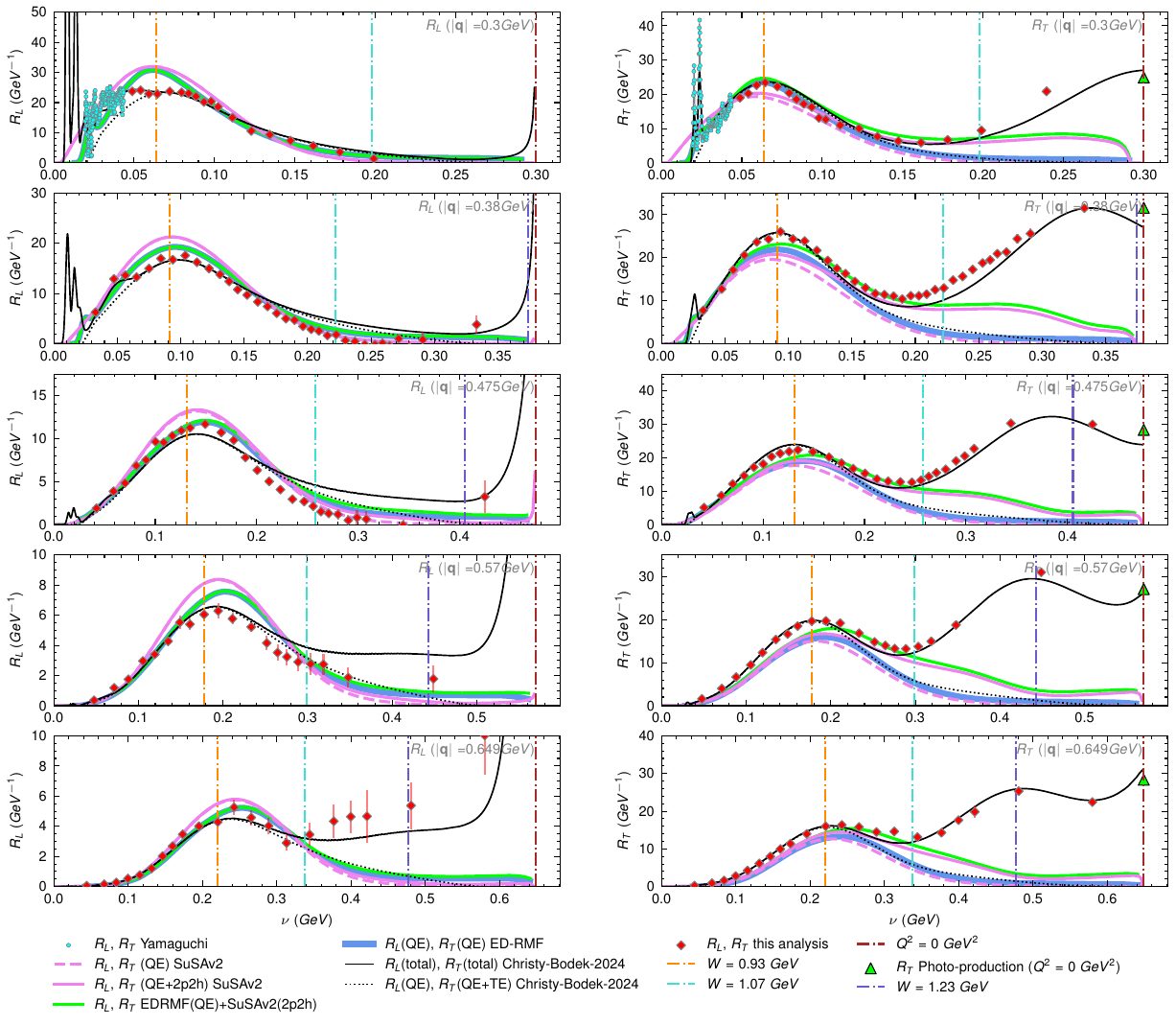}
\caption{{\bf Comparison to ED-RMF and  SuSAv2 for fixed bins in q}:  Same as Fig. \ref{SuSAv2_q3_A} for ${\bf q}$ values of   0.30, 0.38, 0.475, 0.57 and 0.649 GeV versus $\nu$.  The  predictions of SuSAv2 \cite{Megias:2016lke,Gonzalez-Rosa:2022ltp,Gonzalez-Rosa:2023aim}  are the dashed   pink 
line for  QE-1p1h, and solid  pink lines for the sum of QE-1p1h and MEC-2p2h processes.  The  blue lines are the predictions of ED-RMF~\cite{Franco-Munoz:2022jcl,Franco-Munoz:2023zoa} (QE-1p1h including nuclear excitations). Since ED-RMF does not account for 2p2h final states, we also show the sum of the ED-RMF prediction for QE-1p1h  and the SuSAv2  MEC model prediction for the 2p2h contribution  (solid green line).
 }
\label{SuSAv2_q3_B}
\end{figure*}

\begin{figure*}
\includegraphics[width=7.0 in, height=8.in] {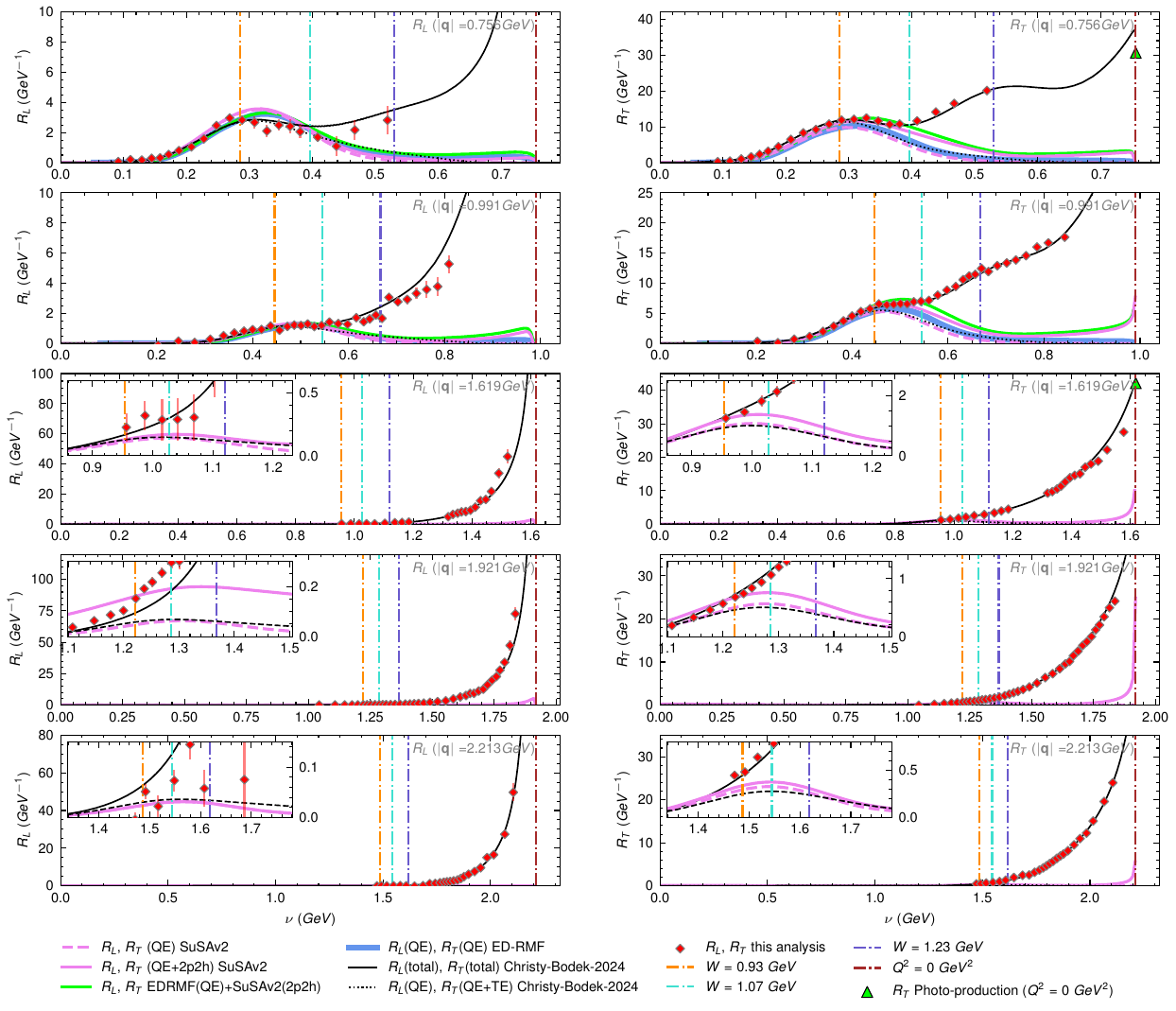}
 \caption{{\bf Comparison to ED-RMF and SuSAv2 for fixed bins in q}: Same as Fig. \ref{SuSAv2_q3_A} for $\bf q$ values  of 0.756, 0.991, 1.619 and 1.921 GeV versus $\nu$. 
 The  predictions of SuSAv2 \cite{Megias:2016lke,Gonzalez-Rosa:2022ltp,Gonzalez-Rosa:2023aim}  are the dashed   pink 
line for  QE-1p1h, and solid  pink lines for the sum of QE-1p1h and MEC-2p2h processes.  The  blue lines are the predictions of ED-RMF~\cite{Franco-Munoz:2022jcl,Franco-Munoz:2023zoa} (QE-1p1h including nuclear excitations). Since ED-RMF does not account for 2p2h final states, we also show the sum of the ED-RMF prediction for QE-1p1h  and the SuSAv2  MEC model prediction for the 2p2h contribution  (solid green line).
 }
\label{SuSAv2_q3_C}
\end{figure*}

\begin{figure*}
\includegraphics[width=7.0 in, height=5.in] {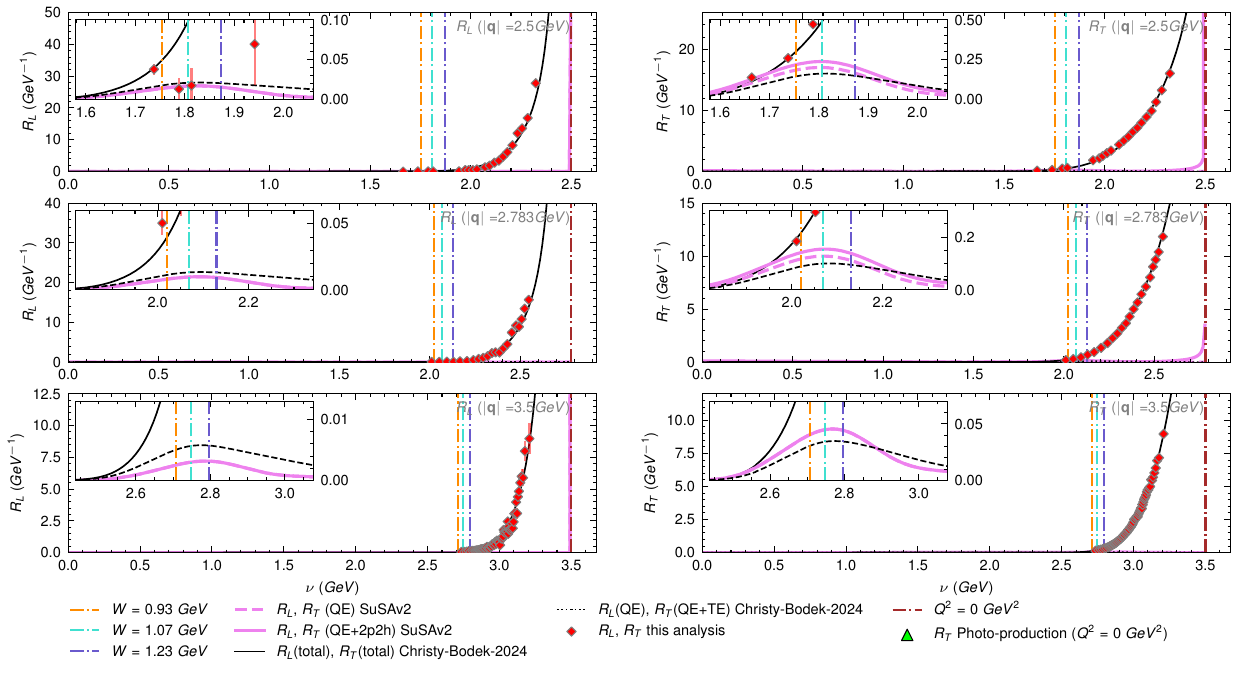}
 \caption{{\bf Comparison to ED-RMF and SuSAv2 for fixed bins in q}: Same as Fig. \ref{SuSAv2_q3_A} for $\bf q$ values  of 2.4,  2.783, and 3.5 GeV versus $\nu$. 
The  predictions of SuSAv2 \cite{Megias:2016lke,Gonzalez-Rosa:2022ltp,Gonzalez-Rosa:2023aim}  are the dashed   pink 
line for  QE-1p1h, and solid  pink lines for the sum of QE-1p1h and MEC-2p2h processes.  The  blue lines are the predictions of ED-RMF~\cite{Franco-Munoz:2022jcl,Franco-Munoz:2023zoa} (QE-1p1h including nuclear excitations). Since ED-RMF does not account for 2p2h final states, we also show the sum of the ED-RMF prediction for QE-1p1h  and the SuSAv2  MEC model prediction for the 2p2h contribution  (solid green line).
 }
\label{SuSAv2_q3_D}
\end{figure*}

\begin{figure*}
\includegraphics[width=7.0 in, height=8.in] {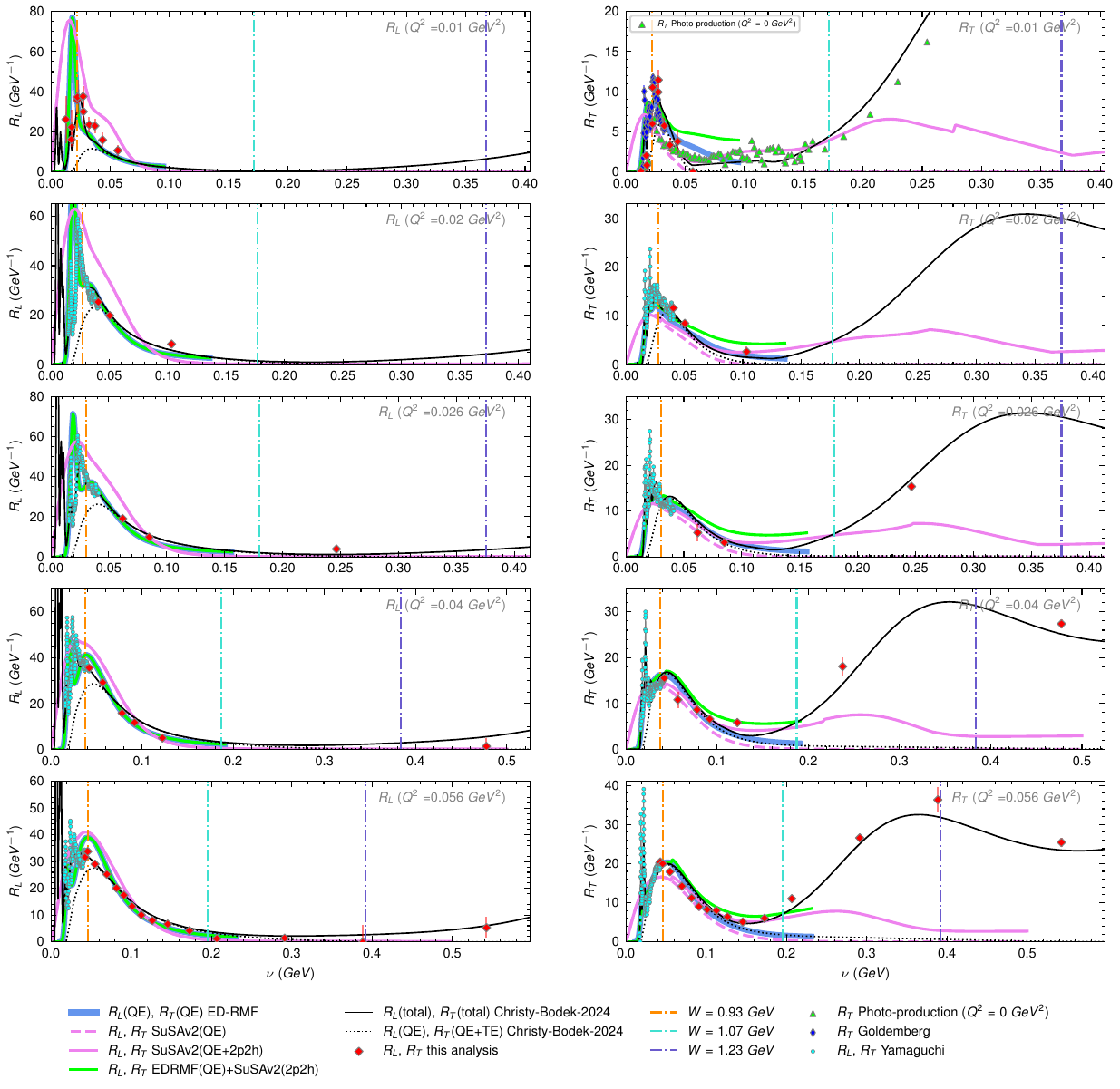}
\caption{ {\bf Comparison to ED-RMF and SuSAv2 for fixed $Q^2$ bins:} Same as Fig. \ref{SuSAv2_q3_A}  for $Q^2$ values of 0.01 0.02, 0.026, 0.04 and 0.056  GeV$^2$ versus $\nu$.
 The  predictions of SuSAv2 \cite{Megias:2016lke,Gonzalez-Rosa:2022ltp,Gonzalez-Rosa:2023aim}  are the dashed   pink 
line for  QE-1p1h, and solid  pink lines for the sum of QE-1p1h and MEC-2p2h processes.  The  blue lines are the predictions of ED-RMF~\cite{Franco-Munoz:2022jcl,Franco-Munoz:2023zoa}  (QE-1p1h including nuclear excitations). Since ED-RMF does not account for 2p2h final states, we also show the sum of the ED-RMF prediction for QE-1p1h  and the SuSAv2  MEC model prediction for the 2p2h contribution  (solid green line).
 }
\label{SuSAv2_Q2_A}
\end{figure*}

\begin{figure*}
\includegraphics[width=7. in, height=8.in] {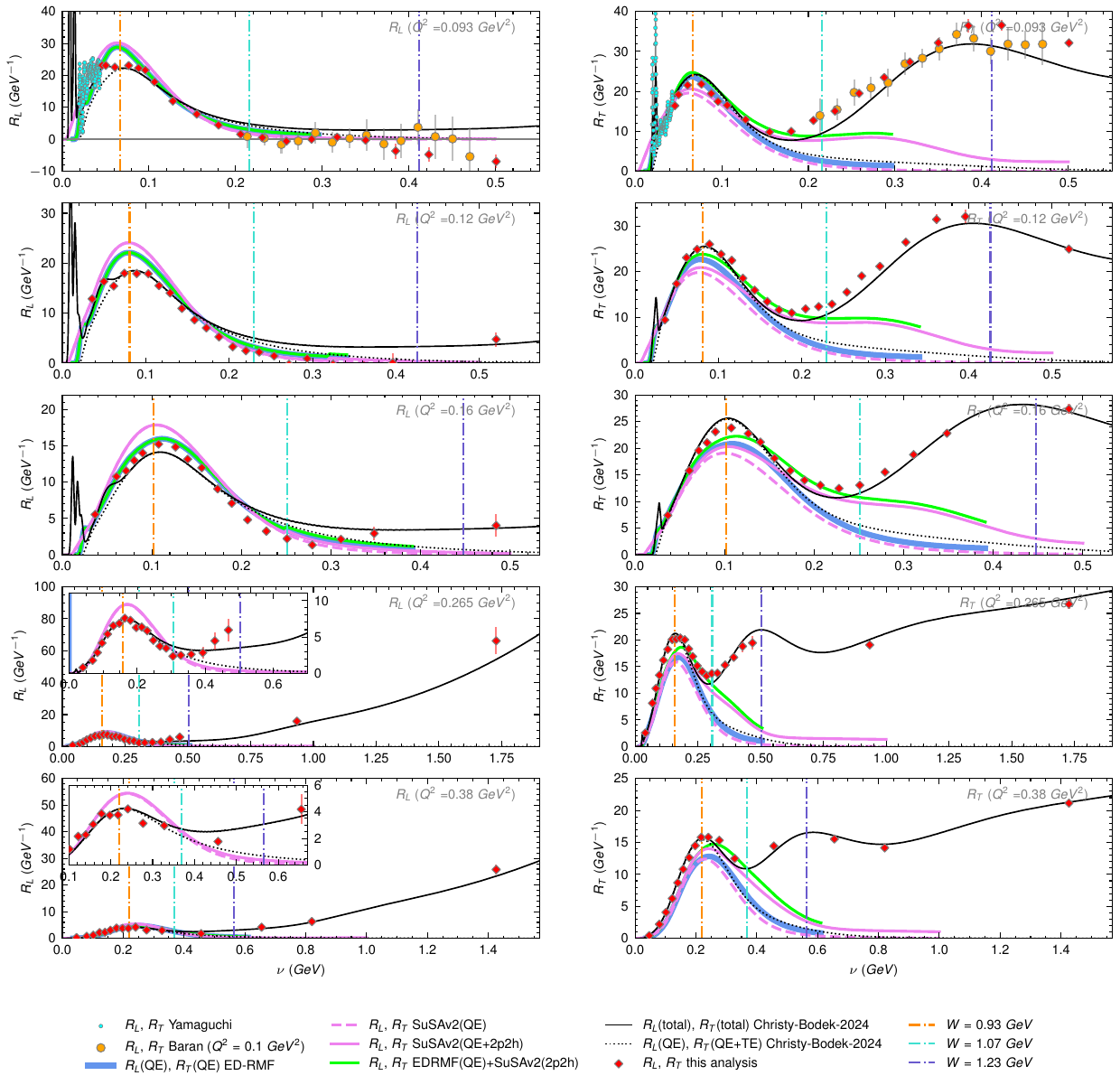}
\caption{ {\bf Comparison to ED-RMF and  SuSAv2 for fixed $Q^2$ bins:} Same as Fig. \ref{SuSAv2_Q2_A} for $Q^2$ values of 0.093 0.12, 0.16, 0.265  and 0.38  GeV$^2$ versus $\nu$. 
The  predictions of SuSAv2 \cite{Megias:2016lke,Gonzalez-Rosa:2022ltp,Gonzalez-Rosa:2023aim}  are the dashed   pink 
line for  QE-1p1h, and solid  pink lines for the sum of QE-1p1h and MEC-2p2h processes.  The  blue lines are the predictions of ED-RMF~\cite{Franco-Munoz:2022jcl,Franco-Munoz:2023zoa}  (QE-1p1h including nuclear excitations).  Since ED-RMF does not account for 2p2h final states, we also show the sum of the ED-RMF prediction for QE-1p1h  and the SuSAv2  MEC model prediction for the 2p2h contribution (solid green line).
}
\label{SuSAv2_Q2_B}
\end{figure*}

\begin{figure*}
\includegraphics[width=7. in, height=8.in] {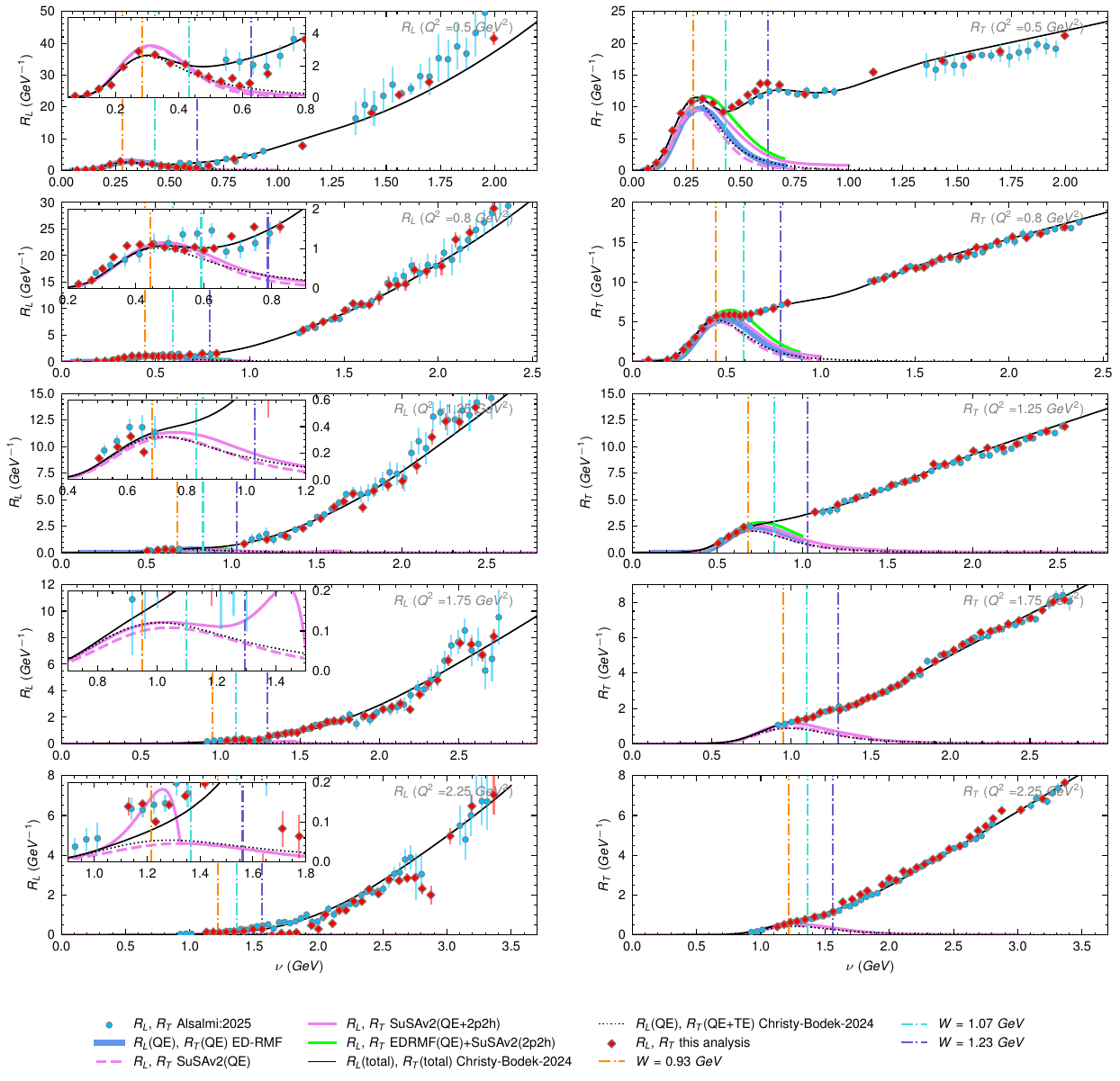}
 \caption{{\bf Comparison to ED-RMF and SuSAv2 for fixed $Q^2$ bins:} Same as Fig. \ref{SuSAv2_Q2_A} for  $Q^2$ values of 0.50, 0.8, 1.25, 1.75 and 2.25  GeV$^2$ versus $\nu$. The  predictions of SuSAv2 \cite{Megias:2016lke,Gonzalez-Rosa:2022ltp,Gonzalez-Rosa:2023aim}  are the dashed   pink 
line for  QE-1p1h, and solid  pink lines for the sum of QE-1p1h and MEC-2p2h processes.  The  blue lines are the predictions of ED-RMF~\cite{Franco-Munoz:2022jcl,Franco-Munoz:2023zoa}  (QE-1p1h including nuclear excitations). Since ED-RMF does not account for 2p2h final states, we also show the sum of the ED-RMF prediction for QE-1p1h  and the SuSAv2  MEC model prediction for the 2p2h contribution  (solid green line).
}
\label{SuSAv2_Q2_C}
\end{figure*}

\begin{figure*}
\includegraphics[width=7. in, height=5.in] {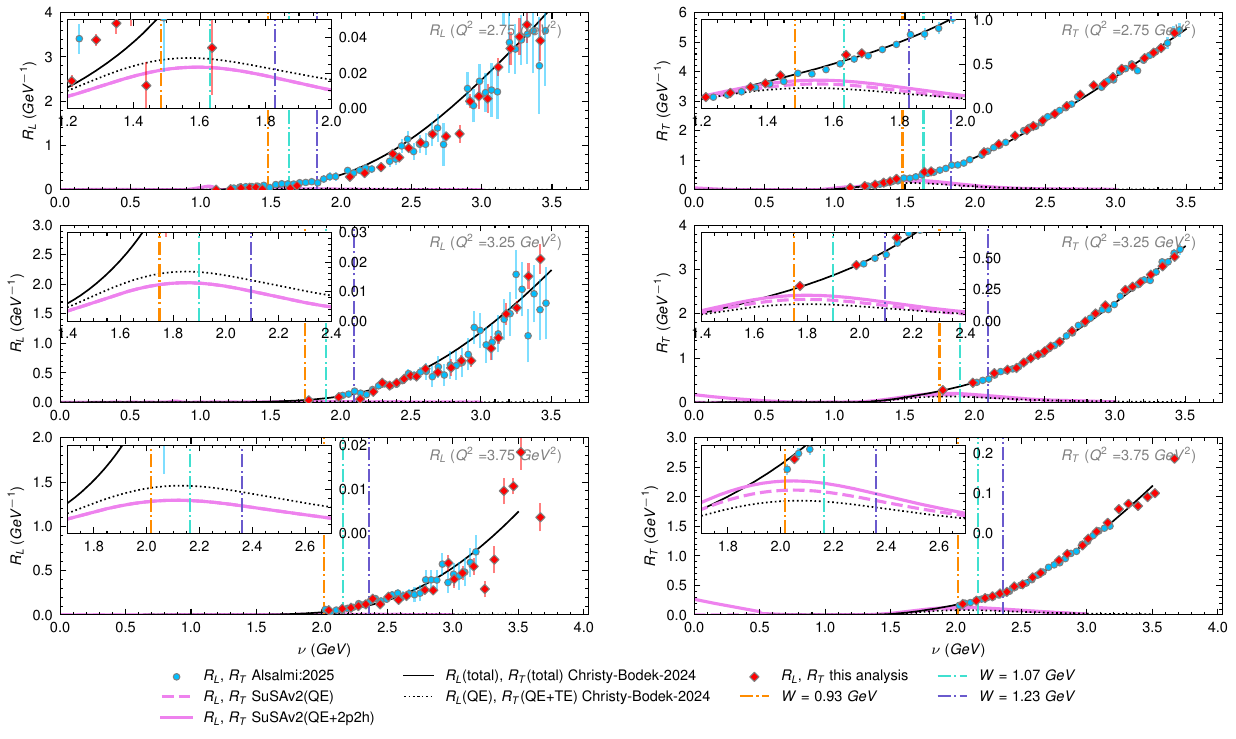}
 \caption{{\bf Comparison to ED-RMF and SuSAv2 for fixed $Q^2$ bins:} Same as Fig. \ref{SuSAv2_Q2_A} for  $Q^2$ values of 2.75, 3.25 and 3.75  GeV$^2$ versus $\nu$. The  predictions of SuSAv2 \cite{Megias:2016lke,Gonzalez-Rosa:2022ltp,Gonzalez-Rosa:2023aim}  are the dashed   pink 
line for  QE-1p1h, and solid  pink lines for the sum of QE-1p1h and MEC-2p2h processes.  The  blue lines are the predictions of ED-RMF~\cite{Franco-Munoz:2022jcl,Franco-Munoz:2023zoa}  (QE-1p1h including nuclear excitations). Since ED-RMF does not account for 2p2h final states, we also show the sum of the ED-RMF prediction for QE-1p1h  and the SuSAv2  MEC model prediction for the 2p2h contribution (solid green line).
}
\label{SuSAv2_Q2_D}
\end{figure*}

Comparisons to the predictions of SuSAv2 are shown in Figures  \ref{SuSAv2_q3_A}-\ref{SuSAv2_Q2_D}. 
\section{Quasi-deuteron contribution}

The Quasi-deuteron contribution  to the  photoproduction ($Q^2=0$)
cross section ($\sigma_{Quasi-D}$) as a function of $\nu=\varepsilon _{\gamma}$  is taken from  \cite{Plujko:2018uum}. 
 For a nucleus with $N$ neutrons, $Z$ protons and mass number $A=N+Z$, the $\sigma_{Quasi-D}$ is equal to
 \vspace{-0.3cm}
\begin{equation}
\label{EQ3}
\sigma _{Quasi-D} (\varepsilon _{\gamma } )=397.8\; \frac{NZ}{A} \frac{(\varepsilon _{\gamma } -2.224)^{3/2} }
{\varepsilon _{\gamma }^{3} }  \phi (\varepsilon _{\gamma } )
\end{equation}
\noindent with $\varepsilon _{\gamma } $ in MeV and $\sigma _{QD} $ in units of mb.
The function $\phi (\varepsilon _{\gamma } )$ accounts for the Pauli-blocking of the excited neutron-proton pair in
the nuclear medium.
\vspace{-0.3cm}
\begin{eqnarray}
&& \phi (\varepsilon _{\gamma } <20\; {\rm MeV})  = {\rm exp}(-73.3/\varepsilon _{\gamma } ), 
~~~~~~~~~~~~~~~~~~~~~~~~~~\nonumber \\
&&  \phi (20<\varepsilon _{\gamma } <140\; {\rm MeV}) =  8.3714\times 10^{-2} \nonumber \\
&&-9.8343\times 10^{-3} \varepsilon _{\gamma } +4.1222\times 10^{-4} \varepsilon _{\gamma }^{2} 
 \nonumber \\
&& -3.4762\times 10^{-6} \varepsilon _{\gamma }^{3} +9.3537\times 10^{-9} \varepsilon _{\gamma }^{4} , \nonumber\\
&& \phi (\varepsilon _{\gamma } >140\; {\rm MeV})={\rm exp}(-24.2/\varepsilon _{\gamma } ).
\end{eqnarray}
 \section{Supplemental Materials}
 
 The following are available as supplemental materials~\cite{supp}
      
     (a) Comparison of the normalized $^{12}C$ electron scattering cross sections to the universal fit (15 Ffgures).
     
     (b) Tables of the  extracted  ${\cal R}_L$ and ${\cal R}_T$ values for bins of  $\bf q$ and bins of $Q^2$.
     
      (c) Tables of the  values of  universal fit values of  ${\cal R}_L$ and ${\cal R}_T$ (the  contributions of nuclear excitations, quasielastic scattering, transverse enhancement, and pion production processes are listed separately).

%

%

%
%

\end{appendices}


%
%
\bibliographystyle{apsrev4-1}
\bibliography{12C_Global_RL_RT_PRD}
\vspace{+25.3in}
 

\section*{Supplementary material}

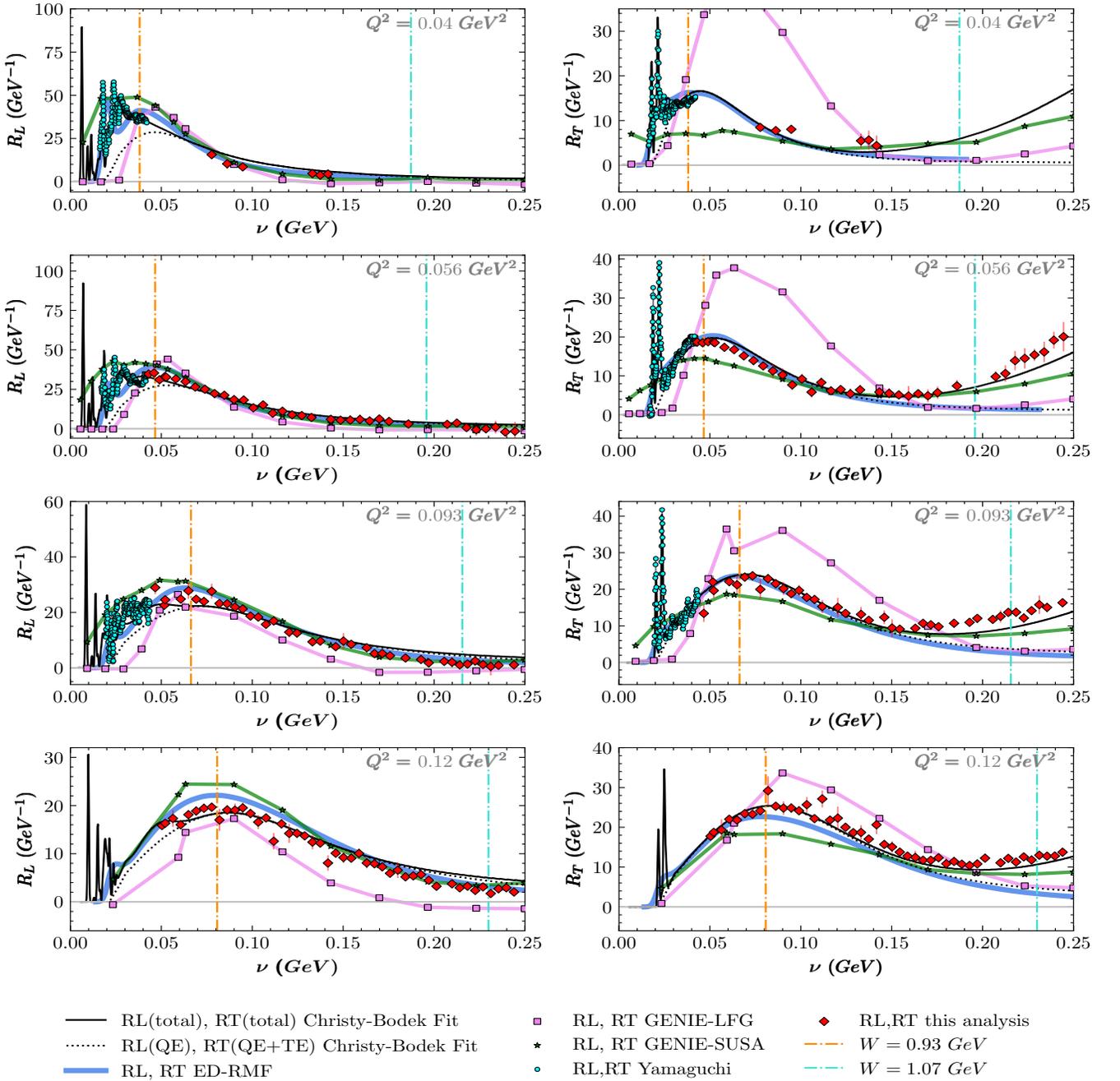

Figure 21: Comparison of the $\mathcal{R}_L(\mathbf{q}, \nu)$ and $\mathcal{R}_T(\mathbf{q}, \nu)$ measurements and Christy-Bodek fit all electron scattering data to the predictions of two versions of GENIE (GENIE-LFG and GENIE-SUSA) in the QE region only for $Q^2$ values of 0.04, 0056, 0.093, and 0.12 GeV$^2$ versus $\nu$. For comparison we also show the predictions of ED-RMF (which has recently been implemented in NEUT).

a large amount of CPU hours, and the extractions can be statistically limited, especially at low $Q^2$. GENIE generates events in the Q2, W space. To avoid the divergence of the cross section at $Q^2$=0, we currently use a Q2 min of 0.02 (comparisons at lower $Q^2$=0 will be reported in a future communication).

Unlike the other calculations, the extracted GENIE pre-

dictions for $\mathcal{R}_L(\mathbf{q}, \nu)$ and $\mathcal{R}_T(\mathbf{q}, \nu)$ include the QE, resonance production and inelastic continuum processes, but nuclear excitations are not modeled.

Here we focus on predictions in the kinematic regions dominated by nuclear excitations, QE and $\Delta(1238)$ production, especially at low values of $Q^2$. Comparisons to GENIE predictions at higher (and lower) values of $Q^2$ (and

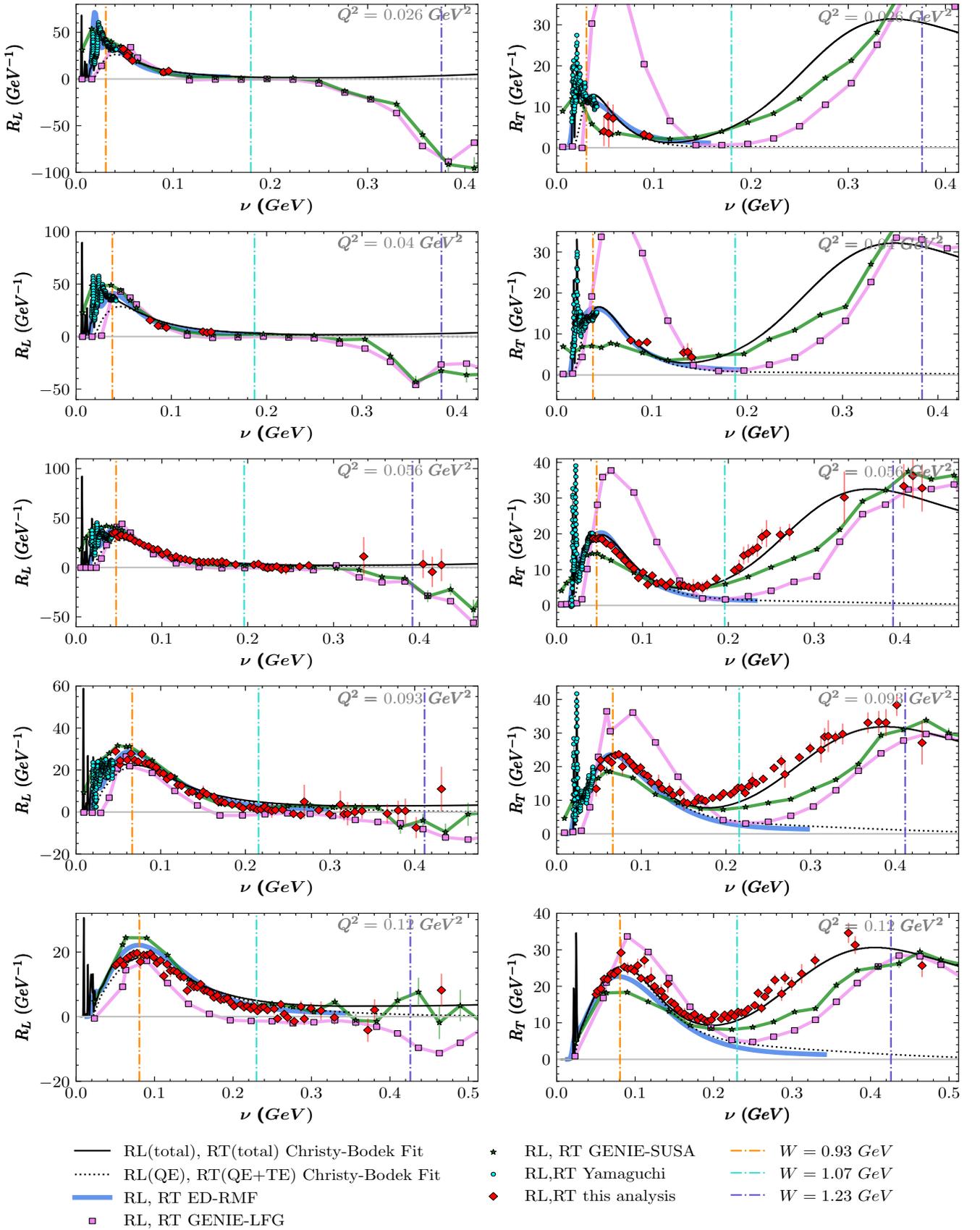

Figure 22: Comparison of the $\mathcal{R}_L(\mathbf{q}, \nu)$ and $\mathcal{R}_T(\mathbf{q}, \nu)$ measurements and Christy-Bodek fit all electron scattering data to the predictions of two versions of GENIE (GENIE-LFG and GENIE-SUSA) in both the QE and $\Delta(1232)$ regions for $Q^2$ values of 0.026, 0.040, 0.056, 0.093, and 0.120 GeV$^2$ versus $\nu$. For comparison we also show the predictions of ED-RMF in the QE region.

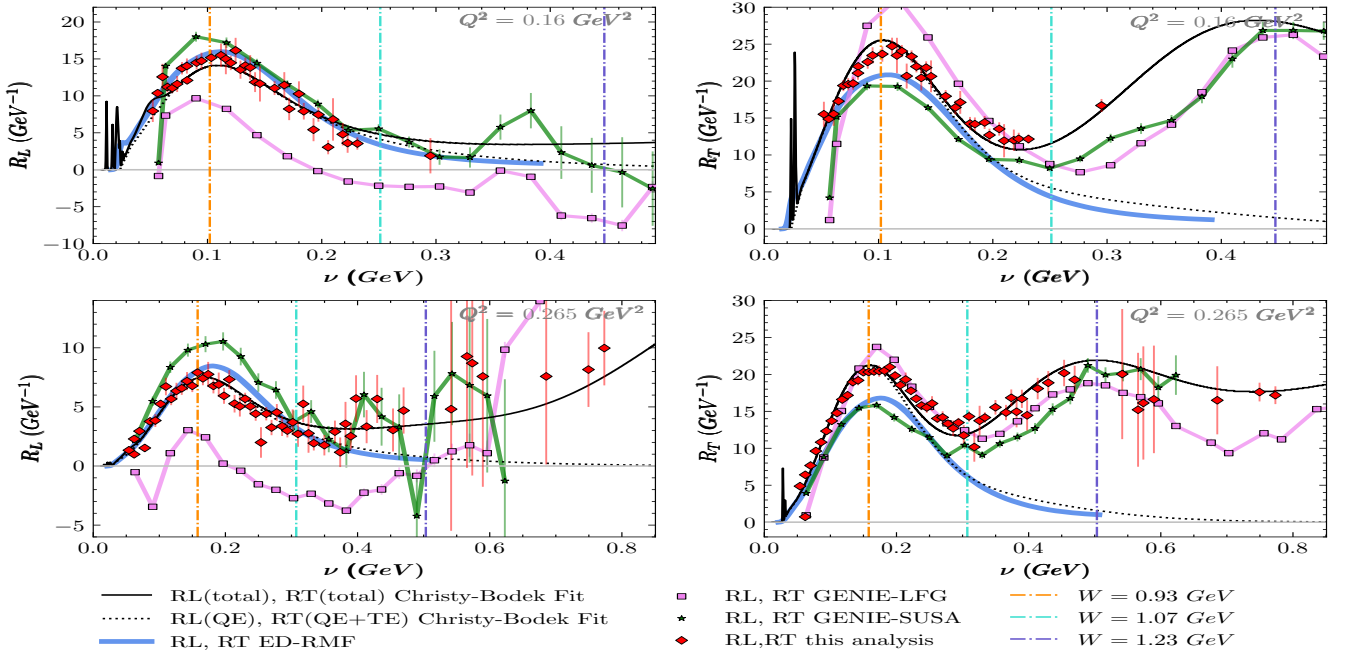

Figure 23: Comparison of the $\mathcal{R}_L(\mathbf{q}, \nu)$ and $\mathcal{R}_T(\mathbf{q}, \nu)$ measurements and Christy-Bodek fit all electron scattering data to the predictions of two versions of GENIE (GENIE-LFG and GENIE-Susa) in both the QE and $\Delta(1232)$ regions for $Q^2$ values of 0.160 and 0.265 GeV$^2$ versus $\nu$. For comparison we also show the predictions of ED-RMF in the QE region.

including the nucleon resonance region) will be presented in a future communication

The GENIE Collaboration provides a range of models for each interaction mechanisms. These are grouped in self consistent model sets, often referred as tunes. In the comparisons shown below the label GENIE-LFG refers to the G18-10a-00-000 tune and the label GENIE-Susa refers to the GEM21-11a-00-000 tune. Both predictions are computed with the latest GENIE version (GENIE-v3.4.2). GENIE-LFG uses a Local Fermi Gas (LFG) for the nucleon momentum distribution for QE scattering and an empirical Meson Exchange Currents (MEC) model[107]. GENIE-Susa(Super Scaling Approach) also uses a Local Fermi Gas, with a modified Susa (superscaling approach) to model QE and MEC (for details of modifications see [108]). Instead of a $\psi'$ distribution extracted from the longitudinal component in electron scattering, the Susa implementation in GENIE uses LFG to provide the initial 4-momentum of the struck nucleon. In both models, momentum and energy conservation are used to compute the kinematics of the out-going nucleons. The inelastic channels are modeled in the same way in GENIE-Susa and GENIE-LFG (as described below). Note that at present GENIE is the only generator that includes the resonance and inelastic continuum in its predictions.

In summary, in both cases the same nuclear ground state, resonance, non-resonance, deep inelastic and FSI models are used. The Local Fermi Gas ground state model is used with a constant removal energy of 25 MeV for Carbon. Resonance production is modeled with

the Berger-Sehgal model which model each resonance separately up to W=1.7 GeV. A non-resonance background is added using the Bodek-Yang model extended to W<1.7 GeV. This non-RES model is scaled by ad-hoc parameters which depend on the multiplicity of the final hadronic system[109]. No interference is accounted for between resonances or with the non-resonance background. Deep-Inelastic-Scattering (DIS) is modeled using the Bodek-Yang model[92, 110] and FSI is modeled using the data-driven hA model [111]. Since no Coulomb effects are accounted for in the GENIE simulations, in our extraction of $\mathcal{R}_L(\mathbf{q}, \nu)$ and $\mathcal{R}_T(\mathbf{q}, \nu)$ we assume $V_{eff}=0$, thus setting Coulomb corrections to zero.

### 2. Comparison of GENIE predictions to data and to the predictions of ED-RMF

Fig. 21 shows the comparison of the $\mathcal{R}_L(\mathbf{q}, \nu)$ and $\mathcal{R}_T(\mathbf{q}, \nu)$ measurements and the Christy-Bodek fit all electron scattering data to the predictions of two versions of GENIE (GENIE-LFG and GENIE-Susa). Here we focus on the QE region and show comparisons for $Q^2$ values of 0.040, 0.056, 0.093, and 0.120 GeV$^2$. For comparison we also show the predictions of ED-RMF (since it has recently been implemented in NEUT).

Note that the Christy-Bodek universal fit describes all electron scattering measurements and therefore, the values $\mathcal{R}_L(\mathbf{q}, \nu)$ and $\mathcal{R}_T(\mathbf{q}, \nu)$ from the fit span a much larger kinematic range than the individually measured

$\mathcal{R}_L(\mathbf{q}, \nu)$ and $\mathcal{R}_T(\mathbf{q}, \nu)$ points. This is because individually measured points require cross sections measurements in both small and large angles to be in the same bin of $Q^2$ and $\nu$. In contrast, the universal fit to all electron scattering data is constrained by data at small and large angles over a wide range of $Q^2$ and $\nu$ to constrain the fit parameters.

Figures 22 and 23 show comparisons of the $\mathcal{R}_L(\mathbf{q}, \nu)$ and $\mathcal{R}_T(\mathbf{q}, \nu)$ measurements and the Christy-Bodek fit all electron scattering data to the predictions of the two versions of GENIE in both the QE and $\Delta(1232)$ regions for $Q^2$ values of 0.026, 0.040, 0.056, 0.093, 0.120, 0.160 and 0.265 GeV$^2$. For comparison we also show the predictions of ED-RMF in the QE region.

### 3. Comparisons to GENIE in the QE region

In the QE region, ED-RMF predictions (blue lines) are in better agreement with the measurements of $\mathcal{R}_L(\mathbf{q}, \nu)$ and $\mathcal{R}_T(\mathbf{q}, \nu)$ than GENIE-LFG (pink lines) or GENIE-SUSA (green lines). In some QE regions the GENIE-LFG predictions for $\mathcal{R}_T(\mathbf{q}, \nu)$ are more than a factor two higher than the data, and the prediction for $\mathcal{R}_L(\mathbf{q}, \nu)$ in the region between the QE peak and the $\Delta(1232)$ is sometimes negative (which is unphysical). In addition, the QE cross section is zero at low values of $\nu$ and in the nuclear excitation region.

In the QE region, the predictions of GENIE-SUSA are closer to the data than the predictions of GENIE-LFG. However, neither is as good as the predictions of ED-RMF especially in the nuclear excitation region. The GENIE-SUSA predictions for $\mathcal{R}_L(\mathbf{q}, \nu)$ are higher than data, and the predictions for $\mathcal{R}_T(\mathbf{q}, \nu)$ are lower than the data (sometimes by a factor of 2). Therefore, the GENIE-SUSA QE cross sections at small $\theta$ are overestimated and the predicted GENIE-SUSA QE cross sections at large $\theta$ are underestimated.

### 4. Comparisons to GENIE in the $\Delta(1232)$ region

Both versions of GENIE use the same Berger-Sehgal model for the $\Delta(1232)$ and in both versions the peak position of the $\Delta(1232)$ in GENIE is at higher $\nu$ than the data. This can be remedied by using an "effective optical potential" [47] for the $\Delta(1232)$.

As seen in the figures the GENIE predictions of $\mathcal{R}_T(\mathbf{q}, \nu)$ in the $\Delta(1232)$ region are larger than data. The electroproduction of the $\Delta(1232)$ is primarily transverse, and the longitudinal contribution is close to zero. In contrast, the GENIE predictions for $\mathcal{R}_L(\mathbf{q}, \nu)$ at the peak of the $\Delta(1232)$ is large and negative (which is unphysical).

Comparisons to GENIE predictions at higher (and lower) values of $Q^2$ (including in the resonance region) will be presented in another paper.

### G. Tuning MC generators

Because our extractions of $\mathcal{R}_L(\mathbf{q}, \nu)$ and $\mathcal{R}_T(\mathbf{q}, \nu)$ from all available data electron scattering data span a large range of $\mathbf{q}$ and $Q^2$, neutrino/electron MC generators can introduce various $\mathbf{q}$ dependent corrections (which can be different for $\mathcal{R}_L(\mathbf{q}, \nu)$ and $\mathcal{R}_T(\mathbf{q}, \nu)$ ) to achieve better agreement with data (as has been done in the Christy-Bodek fit). Examples include:

1. The locations of the QE and $\Delta(1232)$ peaks in $\nu$ are sensitive to the binding energy as well as a $\mathbf{q}$ dependent optical potentials [46, 47], which are different for QE and $\Delta(1232)$ processes.

2. The amplitude of $\mathcal{R}_L(\mathbf{q}, \nu)$ can be changed by a $\mathbf{q}$ dependent longitudinal quenching factor [23].

3. The amplitude of $\mathcal{R}_T(\mathbf{q}, \nu)$ can be changed by a $\mathbf{q}$ dependent additive transverse enhancement contribution (including 2p2h final states) [23].

4. A model describing nuclear excitations (with excitation energies *above* the proton separation energy) should be added [22] (Only ED-RMF accounts for such excitations on average).

5. A model describing nuclear excitations (with excitation energies *below* the proton separation energy) which decay via $\gamma$ ray emission should be added [22]. At present none of the theoretical approaches include these excitations in the predictions.

Since the ED-RMF predictions are in general in good agreement with the data, any correction factors extracted from the comparison to the data would be relatively small. In contrast, the NuWro and GENIE Monte Carlo generator would require more significant tuning corrections.

## VI. SUMMARY

We have done a global extraction of the $^{12}$C Longitudinal ($\mathcal{R}_L$) and Transverse ($\mathcal{R}_T$) nuclear electromagnetic response functions from an analysis of all available inclusive electron scattering cross section on $^{12}C$. The response functions are extracted for a large range of energy transfer $\nu$, spanning the nuclear excitation, quasielastic, resonance and inelastic continuum and over a large range of $Q^2$. We extract $\mathcal{R}_L$ and $\mathcal{R}_T$ as a function of $\nu$ for fixed values of both $Q^2$ ($0 \leq Q^2 \leq 3.5$ GeV$^2$) and $\mathbf{q}$ (0.1 $\leq \mathbf{q} \leq$ 3.75 GeV). In this paper we focus on comparison to models in the nuclear excitation, QE and $\Delta(1232)$ regions.

Since the extracted response functions cover a large range of $Q^2$ and $\nu$, they can be readily used to validate nuclear models as well as validate and tune Monte Carlo

generators for electron and neutrino scattering experiments. Alternatively, the Christy-Bodek global fit to all electron scattering data on $^{12}$C can also be used to validate nuclear models and also validate and tune Monte Carlo generators for electron and neutrino scattering experiments over a larger region of $\nu$.

Because the mass of the muon is much larger than the mass of the electron there are kinematic regions which are not accessible in the scattering of muon-neutrinos. Therefore, in neutrino oscillation experiments *the modeling of the cross sections at low momentum transfer and low energy transfer is important for accounting for the difference between muon-neutrino and electron-neutrino cross sections.* This is the region where the contribution of nuclear excitations is significant. At present, only ED-RMF includes nuclear excitations in the calculations.

We have compared measurements in the QE region to the predictions of several theoretical calculations. We find that the among all the QE theoretical predictions, ED-RMF provides the best description of $\mathcal{R}_L(\mathbf{q}, \nu)$ and $\mathcal{R}_T(\mathbf{q}, \nu)$ for QE scattering (for 1p1h final states) over the largest kinematic range in $Q^2$ and $\nu$.

The ED-RMF predictions include the enhancement of $\mathcal{R}_T(\mathbf{q}, \nu)$ from the interference of 1b and 2b currents (leading to 1p1h final state). However, they do not include enhancement of $\mathcal{R}_T(\mathbf{q}, \nu)$ from 2b currents leading to 2p2h final states. Therefore, a model for the contribution of 2p2h final states to $\mathcal{R}_T(\mathbf{q}, \nu)$ is needed.

We have also investigate the GENIE predictions for the $\Delta(1232)$. All versions of GENIE uses the Berger-Sehgal model model for the $\Delta(1232)$. We find that the peak position of the $\Delta(1232)$ in GENIE is at higher $\nu$ than the data. This can be remedied by including an "effective optical potential" [47] in the Fermi smearing of the single nucleon $\Delta(1232)$ cross sections.

We have also investigated other calculations which at present are only valid in a limited kinematic range. These are of theoretical interest (e.g. GFMC, STA-QMC), but in general are not useful in neutrino MC generators since generator predictions are needed in all regions of $Q^2$ and $\nu$.

Comparisons of GENIE predictions at higher (and lower) values of $Q^2$ (including the resonance region) will be presented in another paper. We plan to do similar analyses for other nuclei including $^{40}$Ca, $^{48}$Ca, $^{56}$Fe, $^{27}$Al, and $^{208}$Pb.

## VII. SUPPLEMENTAL MATERIALS

The extracted $\mathcal{R}_L$ and $\mathcal{R}_T$ values (shown in the figures) are available as supplemental materials [112]. In addition, the values of the Christy-Bodek fit to $\mathcal{R}_L$ and $\mathcal{R}_T$ over a larger range of $\nu$ (and also listing separately the contributions of nuclear excitations, quasielastic scattering, transverse enhancement, and pion production processes) are available as supplemental materials.



\end{document}